\documentclass[12pt,a4paper,oneside]{book}
\usepackage{mystyle}
\usepackage[top=20mm, bottom=20mm, left=40mm, right=30mm]{geometry}

\usepackage{amsmath,amssymb,bbm,graphicx,gensymb}
\usepackage{authblk}
\usepackage{faktor}
\usepackage{caption}
\usepackage{subcaption}
\usepackage[utf8]{inputenc}
\usepackage{amsmath,amsthm,amssymb,bm,graphicx}
\usepackage{comment}
\usepackage{empheq}
\usepackage{adjustbox}
\usepackage{xr}
\usepackage{tikz}
\usepackage{setspace}
\begin{document}
\pagestyle{plain}

\thispagestyle{empty}

\begin{center}

\LARGE{\bf{Geometric and General Relativistic Techniques for Non-relativistic Quantum Systems}}
\\

\vspace{12mm}
\Large{\bf{Aonghus Hunter-McCabe}}\\
\large{\bf{B.A.Int, B.Phil, H.Dip}}\\
\vspace{8mm}

\begin{figure}[ht]
\centering
\includegraphics[width=0.5\linewidth]{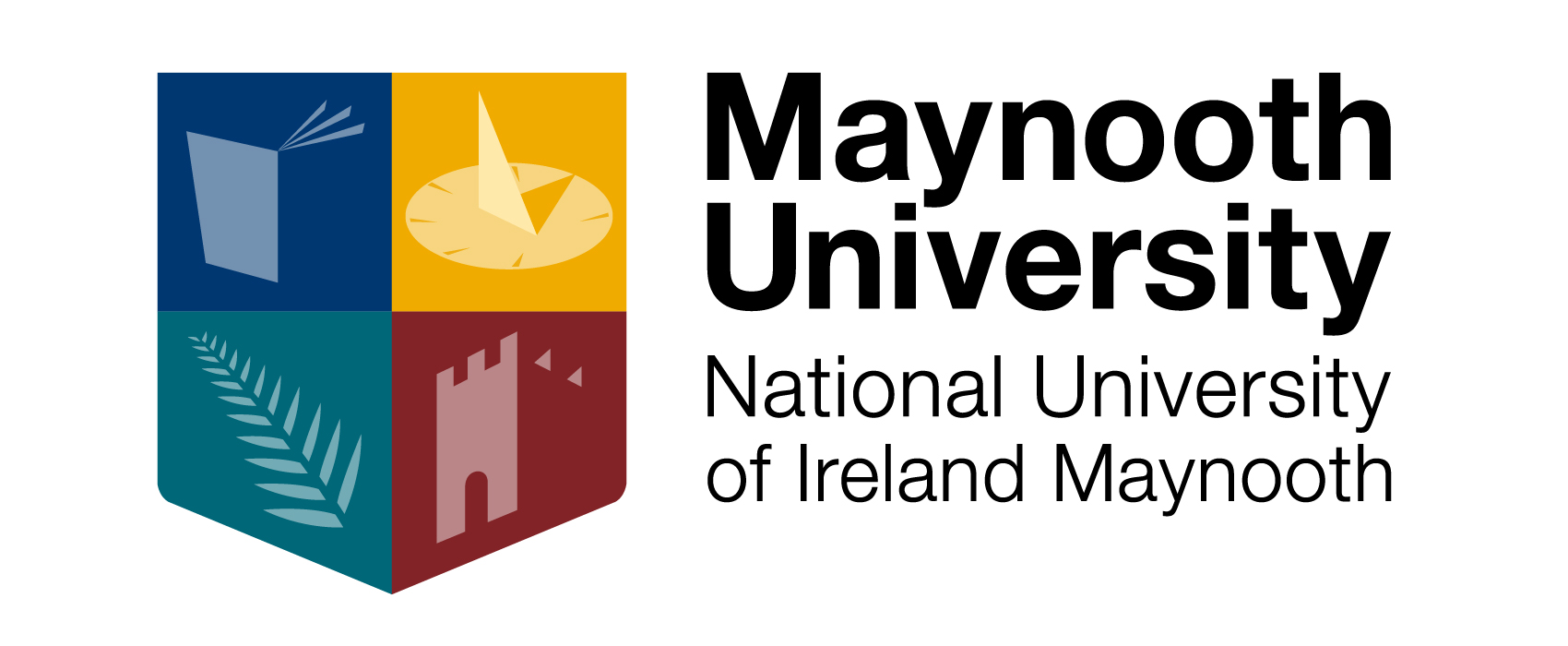}
\end{figure}

\vspace{5mm}
\small{Thesis presented for the degree of}\\
\vspace{2mm}
\Large{Doctor of Philosophy}\\
\vspace{3mm}
\small{to the}\\
\vspace{3mm}
\Large{National University of Ireland Maynooth}\\
\Large{Department of Theoretical Physics}\\
\vspace{6mm}
\large{November 2023}\\
\vspace{6mm}
\small{Department Head}\\
\vspace{1mm}
\large{Dr Joost Slingerland}\\
\vspace{6mm}
\small{Research advisor}\\
\vspace{1mm}
\large{Professor Brian Dolan, Professor Peter Coles}\\
\end{center}

\frontmatter

\clearpage
\thispagestyle{plain}
\par\vspace*{.35\textheight}{\centering To my family and friends\par}

\vspace{-1cm}
\tableofcontents

\vspace*{\fill}

\section*{\centering Declaration}

\vspace{6em}

This thesis has not been submitted in whole, or in part, to this or any other university for any other degree and is, except where otherwise stated, the original work of the author.

\vspace{20em}

\begin{flushright}
\makebox[17em]{\hrulefill}\\
Aonghus Hunter-McCabe, 30th November 2023
\end{flushright}

\vfill

\chapter*{Acknowledgements}
\vspace{-0.5cm}
I'd like to acknowledge the financial support of Maynooth University through the John and Pat Hume Scholarship program and the COVID-19 costed extension.

\bigskip  \begin{center} -=-=-=-=-=-=-=-=-=-= * =-=-=-=-=-=-=-=-=-=- \end{center}  \bigskip

\noindent First, I would like to thank my mother. No words can adequately capture the influence you’ve had on me or the unwavering support you’ve given me throughout my life, especially during my PhD. All I can say is thank you, Mam.

I am grateful to my supervisor, Brian, for taking me on as his PhD student and for guiding me through more physics than I ever thought I was capable of understanding. The breadth and depth of your knowledge sets a standard I suspect I'll spend a lifetime trying to meet. I’d also like to thank my co-supervisor, Peter, for stepping in after Brian’s retirement and supporting me through to the completion of my PhD. Your guidance has been invaluable.

To my family—Ciara, Kerrie, Dad, Lucy, Thomas, Tommo, Adi, and Antonio—your support, encouragement, and the fun you bring into my life every day kept me grounded throughout this experience and lifted my spirits during the most challenging times.

To the faculty, admin, postgrads and undergrads of the Theoretical Physics Department: I could list off a bunch of names, and undoubtedly forget some, but I hope that while I was there I was open and honest enough about how much this place and its people meant to me. All of you are so tremendously special and have influenced my thinking, teaching and learning in innumerable ways. Spending the best part of a decade in the department has been a transformative experience. The kindness, warmth and curiosity there is such a prominent feature because of the people, and I consider myself extremely lucky to have gotten the opportunity to work, laugh and argue with you all.

Finally to my partner, Bisola: we set off on parallel journeys all those years ago—you started med school, and I started my PhD. We're both doctors now (though one of us is a "real'' doctor). Our paths, however, didn't stay parallel; they became entwined. I followed you around to each county for your rotations, and as just much as I am a part of your med school experience, you are a part of this PhD. Your love, encouragement and determination inspired me and has helped me maintain a drive that I didn't know I had.

\chapter*{Abstract}

This thesis explores the application of differential geometric and general relativistic techniques to deepen our understanding of quantum mechanical systems. We focus on three systems, employing these mathematical frameworks to uncover subtle features within each. First, we examine Unruh radiation in the context of an accelerated two-state atom, determining transition frequencies for a variety of accelerated trajectories via first-order perturbation theory. For harmonic motion of the atom in a vacuum, we derive transition rates with potential experimental realizations. Next, we investigate the quantum Hall effect in a spherical geometry using the Dirac operator for non-interacting fermions in a background magnetic field generated by a Wu-Yang monopole. The Atiyah-Singer index theorem constrains the degeneracy of the ground state, and the fractional quantum Hall effect is studied using the composite fermion model, where Dirac strings associated with the monopole field supply the statistical gauge field vortices. A unique, gapped ground state emerges, yielding fractions of the form \( \nu = \frac{1}{2k+1} \) for large particle numbers. Finally, we examine the AdS/CMT correspondence through a bulk fermionic field in an RN-AdS\(_4\) background (with a U(1) gauge field), dual to a boundary fermionic operator. Spherical and planar event horizon geometries are discussed, with the temperature of the RN black hole identified with that of the dual system on the boundary. By numerically solving for the spectral functions of the dual theory, for a spherical event horizon at zero temperature, we identify a shift in the Fermi surface from that which arises in the planar case. Preliminary evidence of a phase transition emerges upon examining these spectral functions, again for the spherical horizon, at non-zero temperature.

\clearpage

\mainmatter
\chapter{Introduction}
\label{Introdution}

In Dirac's 1931 paper \cite{Dirac1931}, in which he outlined his eponymous quantisation condition, he stated:
\begin{quotation}
Non-euclidean geometry and non-commutative algebra, which were at one time considered to be purely fictions of the mind and pastimes for logical thinkers, have now been found to be very necessary for the description of general facts of the physical world.
\end{quotation}
Indeed it is commonplace in a modern undergraduate course in theoretical physics for students to have some exposure to differential geometry, especially in the context of studying general relativity. The utilisation of the tools and methods developed in differential geometry however extend well beyond general relativity to numerous areas in physics and serve as the basis for a thriving area of research. Numerous introductory texts have been written on the application of topology and differential geometry to physics, see \cite{Schutz1980, NashSen, Eguchi, Baez} for example.

However an area of physics which does not make itself easily amenable to the techniques of differential geometry is quantum mechanics and its extension into quantum field theory, though there have been significant advances made to place quantum field theories into the domain of a geometric footing, such are the efforts of string theorists. This particular effort of string theorists is an attempt to get at the heart of the largest discrepancy that exists in the theories of modern physics, the inability to unify the theory of general relativity and quantum mechanics.

There is a separate approach that one can take which does not seek to reconcile the fundamental theories of physics but instead employ the elegant mathematics of differential geometry and general relativity to look more deeply in to the complex world of quantum systems. Much like employing the method of images for classical electrostatics \cite{Jackson1999}, exploiting the geometry of the system can greatly simplify calculations. The mathematician Vladimir Arnold said of mathematics' role in physics \cite{ArnoldParis}
\begin{quotation}
Physics is an experimental science, a part of natural science. Mathematics is the part of physics where experiments are cheap.
\end{quotation}
The overarching theme of this thesis thus follows in this vein. We consider quantum systems in different geometric settings in an effort to make favourable the calculation of meaningful quantities involved. We do this for three separate systems, with a chapter dedicated to each. As each of the systems in question require their own brief overview of the necessary background, this main introduction will be solely an outline of the structure and contents of the following chapters.

In Chapter \ref{Chpt_1} we look at Unruh radiation \cite{Unruh1976} specifically in the context of an accelerated two state atom. Unruh Radiation, and the closely related phenomena of Hawking radiation \cite{Hawking1974}, was one of the first signiﬁcant marriages of quantum ﬁeld theory and general relativity. By setting a quantum field theory in a curved space-time, in the presence of a black hole, Hawking had discovered that black holes in a vacuum radiate thermal energy. Unruh extended this analysis to show that a linearly accelerated observer would too experience a thermal bath with temperature $T = a\hbar / 2\pi c k_{B}$, and $a$ the proper acceleration of the observer. There are significant experimental barriers to detecting Unruh radiation, i.e. to achieve $T \sim 1 $K the proper acceleration of the observer must be $a \sim 10^{20}$m$/$s$^{2}$. In an effort to overcome this the authors in \cite{Svidzinsky2018} consider a two state atom in its ground state being linearly accelerated towards a mirror in an effort to stimulate photon emission at lower accelerations. We follow along a similar scheme as is outlined in \cite{Svidzinsky2018} but look at a more general scenario where a variety of other trajectories to accelerate the two state atom is considered. The structure of this chapter then is as follows; we review the geometry of Rindler coordinates that give rise to linearly accelerated observers in flat space-time and present the standard derivation for Unruh radiation. We outline the scheme found in \cite{Svidzinsky2018} and finally proceed to our results in which we consider numerous other types of accelerated motion, with a specific emphasis on simple harmonic motion of the atom. The contents of this chapter are an edited and expanded version of the results found in our work \cite{DolanTwam}.

The focus of chapter \ref{Chpt_2} is the application of the Atiyah-Singer index theorem \cite{Atiyah} to the quantum Hall effect in a spherical geometry using the Dirac operator for non-interacting fermions in a background magnetic field. The magnetic field is supplied by a Wu-Yang magnetic monopole \cite{Wu+Yang} at the centre of the sphere. The use of a spherical geometry to analyse the quantum Hall effect is not new and was introduced most notably by Haldane in \cite{Haldane}. The use of the Atiyah-Singer index theorem for a spinor field on a sphere to approach the quantum Hall effect however we believe is a novel application. This chapter then proceeds as follows; we first introduce the background to the quantum Hall effect, both integer and fractional, and then briefly review the central idea of the Atiyah-Singer index theorem, in particular for the case of a U(1) spinor bundle on a sphere. We derive wave functions for higher Landau levels that are cross-section of a non-trivial U(1) bundle where the zero point energy vanishes and no perturbations can lower the energy. The Atiyah-Singer index theorem constrains the degeneracy of the ground state. The fractional quantum Hall effect is also studied in the composite Fermion model. Vortices of the statistical gauge ﬁeld are supplied by promoting Dirac strings associated with the monopole ﬁeld to physical vortices. A unique ground state is attained only if the vortices have an even number of flux units and act to counteract the background field, reducing the effective field seen by the composite fermions. There is a unique gapped ground state and, for large particle numbers, fractions $\nu= \frac{1}{2k+1}$ are recovered. This chapter is an expanded version of our paper \cite{DolanMe}.

Chapter \ref{chp3_ADS-CFT_for_Reissner_Nordstrom_BH} deals with the application of techniques devised from the anti-de Sitter/conformal field theory correspondence (AdS/CFT) first proposed by Maldacena in \cite{Maldacena}, to condensed matter systems. This area of research is sometimes referred to as the AdS/CMT correspondence. In the two and half decades since Maldacena's conjecture of the correspondence a flood of research has followed and the applications have expanded to a broad range of topics. A comprehensive introduction to the general topic of this Gauge/Gravity duality can be found in \cite{Ammon2015}. There is no analytic proof of the AdS/CFT correspondence and thus the broader topic of Gauge/Gravity duality does not admit a mathematically rigorous way in which to apply its techniques to condensed matter systems. There is however a collection of prescriptive approaches which have been devised. The layout of this chapter then is as follows; we start with a broad introduction to the general topic of holographic duality, drawing a thread between the concepts on which it relies. This will give us some intuition as to why the prescription we will utilise throughout the remainder of the chapter is justified. We will motivate the applications of this approach to condensed matter systems and discuss the relevance of phase transition temperatures of black holes in asymptotically AdS space and their possible impact on the dual boundary theory. We will review the prescription from \cite{Iqbal+Liu} and its application by \cite{McGreevyetal} in which they look for signatures of a non-Fermi liquid in the spectral function of a fermionic operator dual to a bulk fermionic field in an asymptotically AdS Reissner-Nordstr\"om (RN-AdS$_{4}$) background with a U(1) gauge field. They do this for a flat event horizon and zero black hole temperature (corresponding to zero temperature of the dual boundary system). The final section of this chapter then will concern our numerical results for the spectral functions for a spherical event horizon, at zero temperature, which appears to shift the location of the Fermi surface in the boundary theory. We also present results for the behaviour of the boundary theory at non-zero temperature, with a spherical event horizon, in particular at the phase transition temperature of the black hole. We find possible indications of a phase transition in the boundary theory of the type predicted for a (2+1) dimensional U(1) fermionic theory \cite{QED31}, also known as QED$_3$.

\chapter{Unruh Radiation and Shaking Photons from the Vacuum}
\label{Chpt_1}


\section{Introduction}


The advent of black hole thermodynamics was the first significant marriage of quantum field theory and general relativity. The microscopic origin of black hole radiation and black hole thermodynamics, investigated by Hawking \cite{Hawking1974} and Bekenstein in \cite{Bekenstein}, began by considering quantum field theories in curved space-time. This was just the beginning of the investigation into how geometric considerations impact quantum systems. The success and utility of these considerations have had consequences for many fields, in particular theories of quantum gravity, string theories and the birth of the study into gauge/gravity duality, a comprehensive introduction to which can be found in \cite{Ammon2015}. 

A direct impact of the work of Hawking and others had done in the thermodynamics of black holes, with which we will be primarily concerned, was the work of W.G.Unruh \cite{Unruh1976}. Unruh considered the case of an uniformly accelerated detector (or observer) in the vacuum of flat Minkowski space-time. The purpose of which was to investigate whether just acceleration in a quantum vacuum could give rise to thermal radiation, analogous to the aforementioned Hawking radiation but without the black hole. Given certain parameterised coordinates of flat Minkowski space-time so that we have the trajectory for a linearly accelerated observer in a quantum vacuum, we do indeed find that the detector witnesses what is aptly named Unruh radiation (or the Unruh effect) \cite{Unruh1976}. The accelerated detector (which in later sections we take to be a two level system - TLS), rather than seeing the vacuum, experiences a thermal photon bath with temperature $T=a\hbar/(2\pi c k_B)$, where $a$ is the proper acceleration of the detector.
One interpretation of this effect is that, in the case of our TLS, the virtual photons that normally dress the internal states are promoted to be real excitations due to the highly non-adiabatic nature of the acceleration. Virtual photons can have measurable signatures in atomic physics, e.g. in the Lamb shift and in Raman scattering \cite{Lamb1947, Raman1927}. However, experimentally observing the Unruh effect has proved challenging since to achieve $T\sim 1$ K requires an extreme acceleration $a\sim 10^{20} \,{\rm m/s^2}$, and to substantially excite the TLS  the latter would need a transition frequency $\omega_0/2\pi\sim 20\,{\rm GHz}$.

Along with this, experimental verification of Hawking radiation is itself a difficult task as the temperature is inversely proportional to the mass of the black hole \cite{Hawking1974}. The thermally relevant black holes are extremely small and have short lifetimes as the rate at which they radiate away their energy increases as they evaporate. The expectation is that primordial black holes would be the only candidates and, these being remnants of the earlier universe, it's unlikely we will ever find one to test this hypothesis. This does not spell out a hopeless scenario though. The importance of the Unruh effect and its analogous effect in black holes, has led to a number of proposals over the past three decades towards an experimental test of the existence of acceleration radiation. These proposals include detecting Unruh radiation  via electrons orbiting in storage rings \cite{Bell1983, Bell1987, Unruh1998},  in Penning traps \cite{Rogers1988}, in high atomic number nuclei \cite{Rad2012},  via shifts in accelerating hydrogen-like atoms \cite{Pasante1998}, via   decay processes of accelerating protons or neutrons \cite{Matsas1999}, when electrons experience ultra-intense laser acceleration \cite{Chen1999,Schutzhold2006}, by examining the Casimir-Polder coupling to an infinite plane from an accelerating two-level system \cite{Rizzuto2007}. Researchers have also investigated using cavities to enhance the effect \cite{Scully2003, Belyanin2006,Lopp2018}, and using the Berry phase or entanglement as probes of Unruh radiation \cite{Martin-Martinez2011,Hu2012,Tian2017}. With the advent of circuit quantum electrodynamics - cQED, researchers have investigated simulations of Unruh radiation via the Dynamic Casimir Effect - DCE, \cite{Johansson2009,Wison2011, delRey2012, Lahteenmaki2013}, or by using cQED to simulate relativistically moving systems \cite{Felicetti2015,Su2017}, using nuclear magnetic resonance - NMR \cite{Jin2016} or by studying the interaction between pairs of accelerated atoms \cite{Rizzuto2016}, or via the DCE \cite{Farias2019}. More recent work has probed  whether real motion can produce acceleration radiation and in \cite{Sanz2018,Wang2019}, the authors consider a mechanical method of modulating the electromagnetic fields in cQED DCE photon production.

\begin{figure}[tp] 
\begin{center}
\setlength{\unitlength}{1cm}
\begin{picture}(8.5,7)
\put(-2,-.2){\includegraphics[width=.85\columnwidth]{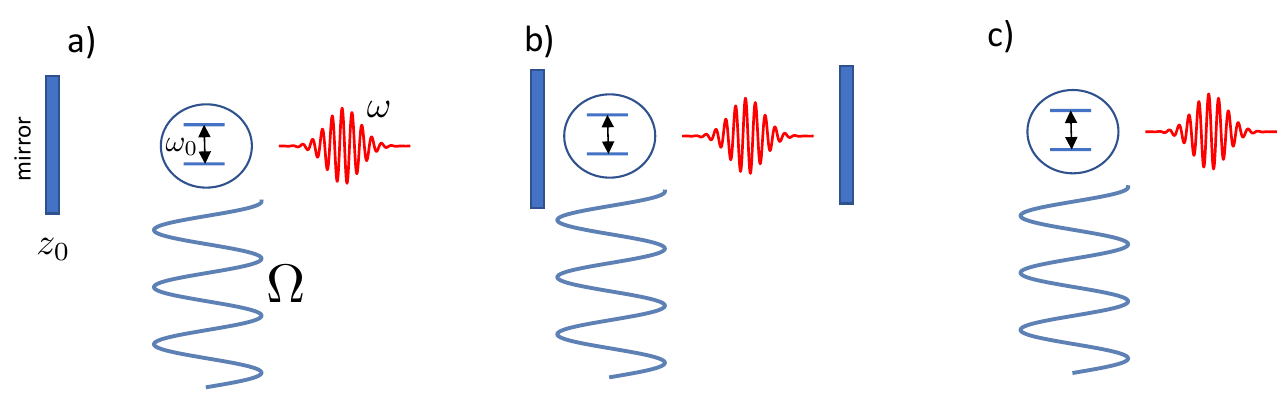}}
\end{picture}
\end{center}
\caption{We consider the generation of photons of frequency $\omega$, from a two level system (atom), with internal transition frequency $\omega_0$, mechanically oscillating at frequency $\Omega$ and amplitude $A$, and initially in the ground state with the electromagnetic field in the vacuum. a) oscillating in front of a mirror, b) oscillating inside a cavity, c) oscillating in free space.}
   \label{Fig1}
\end{figure}

In our work we instead consider a model where the centre of mass of a TLS moves in an accelerated manner that could be more simply achieved in a laboratory setting e.g. oscillatory motion.  In \cite{Svidzinsky2018} the authors discussed the possibility that a TLS, uniformly accelerating away from a mirror and initially in its ground state, could experience a transition to its excited state accompanied by the emission of a photon. This raises the question of what other kinds of acceleration might lead to such a process? In this chapter we show that, within the same scheme as \cite{Svidzinsky2018}, we can adjust the trajectory of the TLS for different forms of accelerated motion and still stimulate photon production. We focus particularly on simple harmonic oscillation and show that this can also result in photon emission accompanied by an excitation from the ground to the excited state of the TLS. Considering acceleration radiation from oscillatory motion has an advantage over continuous linearly accelerated motion in that the TLS stays in a compact region and thus should be more feasible for direct experimental implementation. With this in mind we derive closed compact expressions for the rate of photon production in the case of an oscillating TLS in the presence of a mirror, within a cavity, and just coupled to the vacuum, see figure \ref{Fig1}.


Before proceeding on with the discussion of our work we will; review the relationship between linear acceleration in a relativistic setting and the geometry of Minkowski space, derive an expression for Unruh radiation, and revise the scheme set out in \cite{Svidzinsky2018}.




\subsection{Rindler Coordinates}

The study of hyperbolic motion in Minkowski space-time i.e.~the case of rectilinear motion with \emph{constant} proper acceleration has been an important field arising from the study of the geometric properties of Minkowski space-time. Moreover it crucially played a role in the understanding of general relativistic phenomena such as the acceleration of a suspended particle near the event horizon of a black hole \cite{Unruh1976, Hawking1974}. Rindler coordinates are a natural system with which to study hyperbolic motion thus, in this section, we will review their connection and some of the relevant geometric properties of Rindler space, for a more complete treatment see \cite{Rindler, Steane, Wald, Socolovsky}.

\begin{figure}[h]
  \includegraphics[height=8cm, width=14cm]{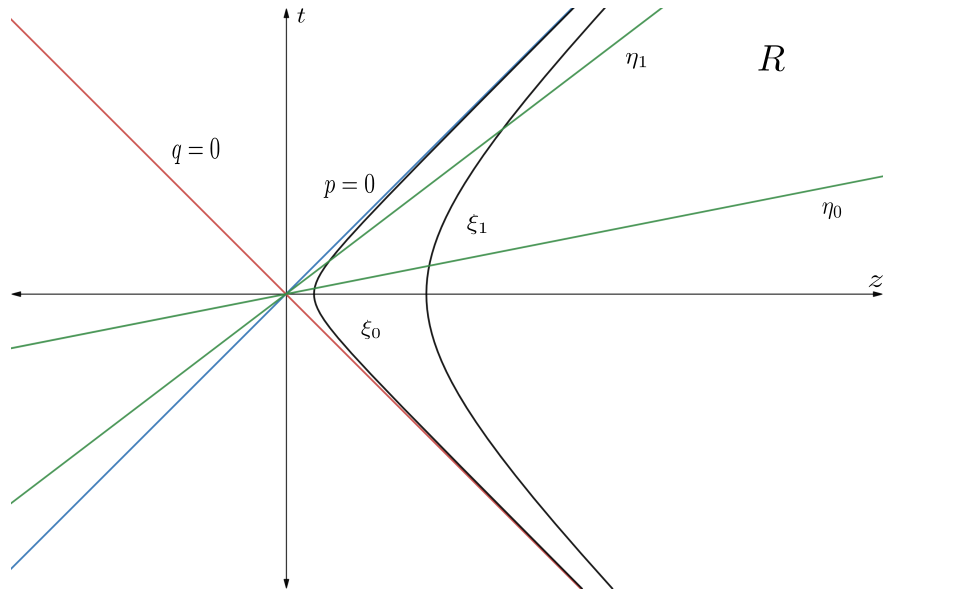}
  \caption{Trajectories in Rindler Coordinates. The labels for the hyperbolic curves $\xi_{0}, \xi_{1}$ are constant values for $\xi$ where $\eta$ is allowed to vary, similarly for constant values of $\eta$ we have the straight lines $\eta_{0},\eta_{1}$ with $\xi$ allowed to vary. Here $\xi_{0} < \xi_{1}$ and $\eta_{0} < \eta_{1}$. Note that we have the light-cone at the origin for $p=q=0$, and that the trajectories for constant $\xi$ are asymptotic to these lines. This illustrates the apparent horizon quality for linearly accelerated observers, i.e. for an observer undergoing constant acceleration away from the origin they can receive no information from the region beyond $z=t$. $R$ here is the right Rindler wedge bounded by $z = |t|$.}
\label{Fig2.1}
\end{figure}

To begin we look at the familiar global coordinates for Minkowski space-time, $x^{\mu} = (t,x,y,z)$, where we are using natural units and have set the speed of light $c=1$. The line element is given by
\begin{equation}
ds^2 = -dt^2 + dx^2 + dy^2 + dz^2
\end{equation}
and the metric is of course
\begin{equation}
g_{\mu \nu} = \begin{pmatrix} -1 & 0 & 0 & 0 \\ 0 & 1 & 0 & 0 \\ 0 & 0 & 1 & 0 \\ 0 & 0 & 0 & 1 \end{pmatrix}.
\end{equation}
Going forward we suppress the $x$ and $y$ coordinates and look only at the 1+1 dimensional Minkowski space-time. The line element is simply
\begin{equation}
ds^2 = -dt^2 + dz^2
\end{equation}
We can define light-cone coordinates $(p,q)$ as follows 
\begin{align}
p &= z - t \\
q &= z + t
\end{align}
where it's clear to see that in Minkowski coordinates, for constant $p$ or $q$, we get straight lines with a slope of $\pm 1$, i.e. the speed of light. To solidify this understanding we can look at the metric in these coordinates
\begin{equation}
ds^2 = dpdq = \frac{1}{2}(dp \otimes dq + dq \otimes dp)
\end{equation}
which gives the metric as 
\begin{equation}
g_{\mu \nu} = \begin{pmatrix} 0 & \frac{1}{2} \\ \frac{1}{2} & 0\end{pmatrix}.
\end{equation}
From this we can see that the vectors $\frac{\partial}{\partial p}$ and $\frac{\partial}{\partial q}$ have length $g(\partial_{p},\partial_{p}) = g(\partial_{q},\partial_{q}) = 0 $. Thus $\frac{\partial}{\partial p}$ and $\frac{\partial}{\partial q}$ are null vectors as expected. Clearly a grid of these coordinates, in $z-t$ space, show that at the intersection of each line we have the light-cone for an observer at that event (see figure \ref{Fig2.1} for an illustration of this fact for the light-cone of an observer at the origin).

Now when considering rectilinear motion with constant proper acceleration in Minkowski space-time we can follow along similar lines to \cite{Rindler} and first look at the change of the components of the 4-velocity. Let $U^{\mu} = (\gamma, \gamma \vec{u})$ where $\vec{u}$ is the 3-velocity and $\gamma$ is the usual Lorentz factor $\gamma = \frac{1}{\sqrt{1-u^2}}$. From this we can define a 4-acceleration by simply taking the time derivative, with respect to the proper time, of $U^{\mu}$
\begin{equation}
A^{\mu} = \frac{dU^{\mu}}{d\tau} = \gamma \frac{dU^{\mu}}{dt} = \gamma \left(\frac{d\gamma}{dt}, \frac{d(\gamma \vec{u})}{dt}\right) \label{rind.1.0}
\end{equation}
where the relationship between coordinate time and proper time is used, $d\tau = ds = \sqrt{-g_{\mu \nu}dx^{\mu}dx^{\nu}}= \sqrt{1-\left(\frac{dz}{dt}\right)^2}dt = \frac{dt}{\gamma}$. The components of the 4-acceleration can be split into their time and spatial coordinates
\begin{equation}
A^{0} = \gamma \frac{d\gamma}{dt}, \quad \vec{A} = \gamma \frac{d(\gamma \vec{u})}{dt}.
\end{equation}
For an instantaneously co-moving inertial reference frame, i.e. the frame in which the observer is instantaneously at rest \cite{Steane}, $\gamma|_{u=0}=1$, $\frac{d \gamma}{dt} = 0$ and the 4-acceleration becomes 
\begin{equation}
A^{\mu} = \left(0, \frac{d\vec{u}}{dt}\right) = (0,\vec{a}) \label{rind.1.1}
\end{equation}
where $\vec{a}$ is the proper acceleration of the observer. Hyperbolic motion, as previously mentioned, is defined when the proper acceleration is constant, interpreting this as acceleration between inertial frames. If we return to the case of 1+1 dimensional Minkowski space, with $\vec{u}=u=\frac{dz}{dt}$, and use the fact that the magnitude of the 4-acceleration is an invariant scalar under Lorentz transformations we can equate the magnitudes of \eqref{rind.1.0} and \eqref{rind.1.1}
\begin{equation}
A^{\mu}A_{\mu} = a^2 =\gamma^2 \left( -\left(\frac{d\gamma}{dt}\right)^2 + \left(\frac{d(\gamma u)}{dt}\right)^2 \right).
\end{equation}
Using the two relations $\frac{d\gamma}{dt} = \gamma^3 u \frac{du}{dt}$ and $\frac{d(\gamma \vec{u})}{dt} = \gamma^3 \frac{du}{dt} $ the above becomes
\begin{equation}
a = \frac{d(\gamma u)}{dt} 
\end{equation} 
and upon integrating this along with the initial condition that $u(t=0)=0$, we have
\begin{equation}
at = \frac{u(t)}{\sqrt{1-u(t)^2}}, \quad \Rightarrow \quad u(t) = \frac{at}{\sqrt{1+(at)^2}}. \label{rind.1}
\end{equation}
As $u(t) = \frac{dz}{dt}$ we can again integrate this expression with respect to $t$ to give
\begin{equation}
z(t) = a\int^{t}_{0} \frac{\tilde{t}d\tilde{t}}{\sqrt{1+(a\tilde{t})^2}}= \frac{1}{a}\sqrt{1+(at)^2} \label{rind.2}
\end{equation}
where we have $z(0) = 1/a$ and the trajectory of the observer undergoing constant proper acceleration is a hyperbola in Minkowski space-time
\begin{equation}
z^2 - t^2 = (1/a)^2. \label{rind.3}
\end{equation}
Note, the trajectory \eqref{rind.2} is confined to the region which is bounded by the light cone coordinates $p = 0$ for $t>0$ and $q=0$ for $t<0$, or more concretely $z \ge \vert t \vert$. This is known as the right Rindler wedge ($R$) of 1+1 dimensional Minkowski space-time, or Rindler space. It is also evident that the trajectories for observers with constant proper acceleration $a$ are asymptotic to these lines, the world-lines of photons, as displayed in figure \ref{Fig2.1} and the point $z(0)=1/a$ approaches the point $z=0$ only for $a \rightarrow \infty$. An accelerated observer that starts at $z=1/a$ cannot receive signals from the region beyond the line $z = t $, it is inaccessible to them and thus this line marks out an apparent horizon beyond which the observer can receive no information. We distinguish here between an apparent horizon, which is only present for the accelerated observer in Minkowski space, and an event horizon which is a global feature of a curved space-time. This is a motivating factor for us to look at quantum vacua in accelerated frames of reference in search of an analogous effect to Hawking radiation.

Before this though we note that \eqref{rind.3} admits parameterisations of our $z,t$ coordinates in terms of hyperbolic trigonometric functions. If we recall the relationship between the proper time and coordinate time $d\tau = \frac{dt}{\gamma}$ and using \eqref{rind.1} we can integrate both sides and find
\begin{equation}
\tau = \int^{t}_{0} d\tilde{t} \sqrt{1-u(\tilde{t})^2} = \int^{t}_{0} \frac{d\tilde{t}}{\sqrt{1+(a\tilde{t})^2}} = \frac{1}{a}\sinh^{-1}(at)
\end{equation} 
and thus have the coordinates for a uniformly accelerated observer in terms of their proper time
\begin{align}
t &= \frac{1}{a}\sinh(a\tau) \\
z &= \frac{1}{a}\cosh(a\tau) 
\end{align}
where we have gotten the equation for the $z$ coordinate from substituting our expression for $t$ in to \eqref{rind.3}. Dimensional analysis reveals the factors of $c$ that are required.
The equations then look like
\begin{align}
t &= \frac{c}{a}\sinh(\frac{a\tau}{c}) \\
z &= \frac{c^2}{a}\cosh(\frac{a\tau}{c}) 
\end{align}
these equations will be of particular relevance for our discussion around the accelerated TLS. Setting $c=1$ again, we can use the above equations to write down a natural coordinate system for Rindler space 
\begin{align}
t &= \frac{e^{\alpha \xi}}{\alpha}\sinh(\alpha\eta) \label{rind.coord.1}\\
z &= \frac{e^{\alpha \xi}}{\alpha}\cosh(\alpha\eta) \label{rind.coord.2}
\end{align}
where, for constant $\xi$, we have the trajectories for accelerated observers in Minkowski space-time, as can be seen from figure \ref{Fig2.1}.

The relationship between the proper acceleration and $\alpha$ is given by
\begin{equation}
a = \alpha e^{-\alpha \xi}
\end{equation}
and the proper time is related to $\eta$ via 
\begin{equation}
\tau = \frac{\alpha}{a} \eta = e^{\alpha \xi} \eta.
\end{equation}
In these coordinates the line element takes the form
\begin{equation}
ds^2 = e^{2\alpha\xi}(-d\eta^2 + d\xi^2) \label{rind.4.1}
\end{equation}
which displays this coordinate transformation as a conformal one of the space. This fact will be useful when looking at a massless scalar field in Rindler space.
Finally looking at one last coordinate transformation, $\rho = \frac{e^{\alpha \xi}}{\alpha}$  we recover the familiar Rindler space line element \cite{Rindler}
\begin{equation}
ds^2 = -(\alpha \rho)^2 d\eta^2 + d\rho^2. \label{rind.4}
\end{equation}
This sequence of coordinate transformations has an intimate relationship with the Schwarzschild metric. There is an apparent singularity in the inverse metric of equation \eqref{rind.4} at $\rho = 0$, this singularity is not essential however, but a consequence of our coordinates. If one were to work backwards through these transformations from the final equation above \eqref{rind.4} we would arrive at Minkowski space-time and find no such singularity and no event horizon. Minkowski space is thus the maximally extended space-time of Rindler space. Analogously the coordinate singularity in the Schwarzschild metric can too be transformed away given the correct choice of coordinates, those being Kruskal–Szekeres coordinates \cite{Kruskal1960}. These coordinates cover the entire space-time manifold of the maximally extended Schwarzschild solution apart from the essential singularity at the origin, and the event horizon is no longer  a boundary. Therefore given the insight from Hawking to analyse the behaviour of quantum fields near the event horizon of a Schwarzschild black hole it seems natural to then attempt to apply a similar scheme to the above for an accelerated observer. This will be the task for the next section.


\subsection{Unruh Radiation}

There are various ways to begin the calculations that lead to the Unruh effect. A good introduction to the thermodynamics of a black hole and a pretty neat and concise calculation of the Hawking temperature can be found in \cite{AZeeQFT}, albeit not particularly illuminating as to the origins of these effects. Interestingly the nature of the calculation is to approximate the Schwarzschild metric near the horizon, which results in the Rindler metric \eqref{rind.4} above, and then to introduce imaginary time. Approximating the Schwarzschild metric with the Rindler metric was also Unruh's initial motivation and what led to the discovery of the Unruh effect \cite{Unruh1976}. To provide a brief introduction to this effect we will begin by first looking at the solutions for a massless scalar field and then their vacua in different coordinate systems, following a similar scheme to that which can be found in \cite{DolanGT}. More complete treatments can be found in \cite{Unruh1976, Socolovsky}.

The Klein-Gordon equation is often the starting point when embarking on the study of quantum field theory. It follows from finding the extrema of the action
\beq
\mathbf{S} = \int d^{4}x \mathcal{L} 
\eeq
where the Lagrangian $\mathbf{L}= \int d^{3}x \mathcal{L}$. For a massless, classical, scalar field theory, $\mathcal{L}$ in a curved space-time, or non-inertial coordinate system in flat space-time as in \eqref{rind.4}, is
\beq
\mathcal{L} = -\half \sqrt{-g}\del_{\mu} \phi\del^{\mu} \phi = -\half \sqrt{-g} g^{\mu \nu}\del_{\mu} \phi\del_{\nu} \phi.
\eeq
The Euler-Lagrange equations (the equations of motion) are then
\beq
g^{\mu \nu} \del_{\nu} \del_{\mu} \phi = 0
\eeq 
where, for Minkowski space-time, the derivative operator just becomes the d'Alembertian operator, though we will restrict our analysis to just 1+1 dimensions. In the Rindler coordinates we have 
\beq
e^{-2\alpha \xi}\left\{(\partial^{2}_{\xi}- \partial^{2}_{\eta})\phi\right\} = 0 \label{no.1.1}
\eeq
and in Minkowski coordinates we have
\beq
(\partial^{2}_{z}- \partial^{2}_{t})\phi = 0.
\eeq 
Provided the term in the braces of equation \eqref{no.1.1} vanishes, our equations of motion are invariant under this coordinate transformation
\beq
(\partial^{2}_{z}- \partial^{2}_{t})\phi = (\partial^{2}_{\xi}- \partial^{2}_{\eta})\phi = 0. \label{K.G._eq}
\eeq
Note again though that our coordinates $\eta$, $\xi$ are only defined in the right Rindler wedge. Now $\phi$ can be decomposed, in the Rindler coordinates, into a complete set of orthonormal solutions $u_i$, with coefficients $a_i$
\beq
\phi = \sum_i a_i u_i(\eta,\xi) + a^{*}_i u^{*}_i(\eta,\xi).
\eeq
A separate, yet equivalent set of orthonormal solutions can also be found for the decomposition of $\phi$ in the $z$-$t$ frame. We label these functions $u'_j$ and have coefficients $a'_j$ and $\phi$ can then be written
\beq
\phi = \sum_i a'_{i} u'_i(t,z) + a'^{*}_i u'^{*}_i(t,z).
\eeq
Obviously each element of these sets of functions satisfy \eqref{K.G._eq}. Looking at normalised out going modes as solutions to this equation, for the Rindler observer, we have
\beq
u_i = \frac{1}{2\sqrt{\pi\omega_{i}}}e^{i\omega_{i}(\xi-\eta)}.
\eeq 
We can find the coefficients of these functions via Fourier transforming along the world-line of the Rindler observer. By this we mean; for a fixed $\xi = \xi_0$ and $\eta$ being the proper time for our Rindler observer, $d\eta$ is tangent to this trajectory, integrating along it thus gives us our desired Fourier coefficients
\beq
a_{i} = e^{-i \omega_{i} \xi_{0}}\sqrt{\frac{\omega_{i}}{\pi}}\int_{-\infty}^{\infty} e^{i \omega_{i} \eta} \phi(\eta,\xi_{0})d\eta.
\eeq
with $\omega \ge 0$. The complex conjugates of the above functions and coefficients are straightforward. Now we can can relate the coefficients of these functions in these alternate coordinate systems (with what amounts to essentially a change of basis) by
\beq
a_{i} = \sum_{j} \alpha_{ij} a'_{j} + \beta_{ij}^{*} a'^{*}_{j}
\eeq
where $\alpha_{ij}$ and $\beta_{ij}$ are the Bogoliubov transformation coefficients \cite{Bogoliubov}. These coefficients are found to be
\begin{align}
\alpha_{ij} =& e^{-i\omega_{i}\xi_{0}}\sqrt{\frac{\omega_{i}}{\pi}}\int_{-\infty}^{\infty} e^{i\omega_{i}\eta} u'_{j}(t(\eta),z(\xi_{0}))d\eta \\
\beta_{ij} =& e^{i\omega_{i}\xi_{0}}\sqrt{\frac{\omega_{i}}{\pi}}\int_{-\infty}^{\infty} e^{-i\omega_{i}\eta} u'_{j}(t(\eta),z(\xi_{0}))d\eta. \label{Unruh_1}
\end{align}
Quantizing our scalar field, the details of which we won't go through here but can be found in \cite{AZeeQFT}, $\phi$ becomes an operator on the Hilbert space of quantum states and our Fourier coefficients become the familiar creation and annihilation operators $\hat{a_i}^{\dagger}$, $\hat{a_i}$ respectively. The Hilbert space in our two coordinate systems have their own vacua, such that
\beq
\hat{a_i} \ket{0_{R}} = 0
\eeq
for Rindler space and
\beq
\hat{a'_i} \ket{0_{M}} = 0
\eeq
for Minkowski space. Note that, as we have already stated that these coefficients, now operators, are related via a basis change, we can act on the Minkowski vacuum with the creation operator in the Rindler coordinates and vice versa. Moreover we can define a number operator, $\hat{N}_{i}=\hat{a_i}^{\dagger}\hat{a_i}$ in the Rindler coordinates and it is then a straightforward calculation to see that
\beq
\bra{0_{M}}\hat{N}_{i} \ket{0_{M}} = \sum_{j} \vert \beta_{ij} \vert^{2}
\eeq
or in words, the particle number is \emph{observer dependent}, provided $\beta_{ij}\neq 0$. Let us go about calculating this quantity from \eqref{Unruh_1}. Subbing in for $u'_{j}$ we have
\begin{align}
\beta_{ij} =& e^{i\omega_{i}\xi_{0}}\sqrt{\frac{\omega_{i}}{\pi}}\int_{-\infty}^{\infty} e^{-i\omega_{i}\eta} \left( \frac{1}{2\sqrt{\pi\omega'_{j}}}e^{i\omega'_{j}(z-t)}\right)d\eta \\
=& \frac{e^{i\omega_{i}\xi_{0}}}{2\pi}\sqrt{\frac{\omega_{i}}{\omega'_{j}}}\int_{-\infty}^{\infty} e^{-i\omega_{i}\eta} e^{(i\omega'_{j}/ \alpha) e^{\alpha(\xi_{0}-\eta)}}d\eta
\end{align}
where we have used the fact that $z-t=e^{\alpha(\xi_{0} - \eta)}/\alpha$ from \eqref{rind.coord.1}. Introducing a change in integration variable $iy = \frac{\omega'_{j}}{\alpha}e^{\alpha(\xi_{0} - \eta)}$, which gives $d\eta = \frac{-dy}{\alpha y}$, our integral becomes
\beq
\beta_{ij} = \frac{e^{-\frac{\pi\omega_{i}}{2\alpha}}}{2\pi\alpha}\sqrt{\frac{\omega_{i}}{\omega'_{j}}}e^{i\frac{\omega_{i}}{\alpha}\ln{(\frac{\alpha}{\omega'_{j}})}}\int_{0}^{-i\infty} y^{\frac{i\omega_{i}}{\alpha}-1} e^{-y} dy. \label{Unruh_2}
\eeq
with the branch $\ln{i}=i\pi/2$. This is almost a $\Gamma$-function, except the integrals are along the imaginary 
axis rather than the real axis. 
Provided $\frac{\omega_{i}}{\alpha}>0$ (so $\omega_{i} \rightarrow \omega_{i} +  i \epsilon$ with
$\epsilon >0$) we can rotate the contour to get\footnote{Imagine doing an integral around a slice of pie with
opening angle $90^\circ$ and vertex at the origin, in the SE quadrant for the upper sign and the NE quadrant for the lower sign. There is a branch point at the origin
but this is harmless provided $\frac{\omega_{i}}{\alpha}>0$ as usual for $\Gamma$-functions.
The exponential kills the integral around the quarter-circle at infinity.}
\begin{align}
\beta_{ij} &= \frac{e^{-\frac{\pi\omega_{i}}{2\alpha}}}{2\pi\alpha}\sqrt{\frac{\omega_{i}}{\omega'_{j}}}e^{i\frac{\omega_{i}}{\alpha}\ln{(\frac{\alpha}{\omega'_{j}})}}\int_{0}^{\infty} y^{\frac{i\omega_{i}}{\alpha}-1} e^{-y} dy \\
&= \frac{e^{-\frac{\pi\omega_{i}}{2\alpha}}}{2\pi\alpha}\sqrt{\frac{\omega_{i}}{\omega'_{j}}}e^{i\frac{\omega_{i}}{\alpha}\ln{(\frac{\alpha}{\omega'_{j}})}} \Gamma(\frac{i\omega_{i}}{\alpha}). 
\end{align}
Now the quantity $\vert \beta_{ij} \vert^{2}$ is
\beq
\vert \beta_{ij} \vert^{2} = \frac{\omega_{i} e^{-\frac{\pi\omega_{i}}{\alpha}}}{4\pi^{2}\alpha^{2}\omega'_{j}}\frac{\pi}{\frac{\omega_{i}}{\alpha}\sinh{(\frac{\pi\omega_{i}}{\alpha})}} = \frac{1}{2\pi\alpha\omega'_{j}}\frac{1}{e^{\frac{2\pi\omega_{i}}{\alpha}}-1} \label{Unruh_2}
\eeq
where we have used the fact that $\vert\Gamma(iy)\vert^{2} = \frac{\pi}{y\sinh{\pi y}}$. It is clear that we have a Bose-Einstein distribution from equation \eqref{Unruh_2}. The term in the exponential for a Bose-Einstein distribution is normally $\frac{\hbar \omega}{k_{B}T}$ and thus, with factors of c restored throughout the above integration and setting $\xi_{0} = 0$, we can relate the temperature to the proper acceleration of the observer
\beq
T = \frac{\hbar a}{2\pi k_{B} c}
\eeq
which is the Unruh temperature.

Now that we are on sound footing with the prospect of acceleration radiation and its origins in quantum field theory we can move to the work from \cite{Svidzinsky2018} and the setup of an accelerated two state atom with a mirror.
\newpage
\section{Accelerated Two State Atoms}


\subsection{Deriving Probabilities} We will begin by showing how to derive the probability for the two-level system, moving in one dimension along a pre-set space-time trajectory, in the presence of a mirror, to become excited and emit a photon using first order perturbation theory.  Using the derived expression we first confirm that if the two level system is at rest the probability to emit a photon is zero. In the subsequent subsections we consider various other types of motions. Following \cite{Svidzinsky2018}, we consider a two level system, or atom, coupled to the  electromagnetic field with a coupling strength $g$, and atomic transition frequency  $\omega_0$, so the excitation energy is $\Delta E=\hbar \omega_0>0$. 
For simplicity we consider motion in
one-dimension given by position $z(\tau)$. Later in the chapter, from \S \ref{mirror} on, our analysis will be non-relativistic with $\tau = t$, i.e., ordinary Newtonian time rather than proper time.~This is a suitable choice for our purposes as we achieve significant transition frequencies in the non-relativistic regime with oscillatory motion.\footnote{The calculation can be done relativistically but this is not necessary. If we assume a mechanical oscillation frequency $\Omega/2\pi \sim\, 10\, {\rm GHz}$, and a maximum amplitude of oscillation as $A\sim 10 {\rm nm}$, then the maximum velocity achieved of the atom is $v_{max}\sim 600\, {\rm m/s}\ll c$. As we show below we predict that these parameter values will yield significant acceleration radiation.}

\noindent In the interaction picture the Hamiltonian for the interaction of the atom with the electric field
 $\phi_\omega$ is \cite{Svidzinsky2018}
\begin{multline}H_I(\omega,t) = \hbar g\bigl\{a_\omega \phi_\omega(t(\tau),z(\tau))
+ a^\dagger_\omega \phi^*_\omega(t(\tau),z(\tau)) \bigr\}
\times \bigl(\sigma_- e^{-i\omega_0 \tau} + \sigma_+ e^{i\omega_0 \tau}  \bigr)\label{Hamomega}
\end{multline}
which is the dipole interaction of the atom with the field. The $\sigma_+$ operator raises the internal atomic state and $\sigma_-$ lowers it, $a_\omega^\dagger$, $a_\omega$ are photon creation and annihilation operators and
$\phi_\omega$ are field modes that depend on the boundary conditions. The modes associated with the internal transition frequency $\omega_{0}$ are parameterised in $\tau$, i.e.~capturing the state transitions and their associated frequencies in the rest frame of the atom. {\color{black} We note that (\ref{Hamomega}), describes the interaction at a specific frequency $\omega$, and the full Hamiltonian is obtained by $H_I(\tau)\equiv \int\,H_I(\omega, \tau)\,d\omega$, and thus our treatment encapsulates the potential excitation of radiation at any wavelength. In what follows we compute the probability to excite the atom and emit a photon of frequency $\omega$, where the latter is not taken as a fixed quantity.}
{\color{black} This interaction has been used numerous times in the literature to model the coupling between a two-level atom and a quantum field but here we do not make the rotating-wave approximation and the position of the atom is allowed to vary in time. The rotating-wave approximation has the effect of suppressing the terms in the interaction Hamiltonian that allow, for example, the creation of a photon and a jump to the excited state of the atom, see \cite{quantstat} for details. These states are significant for accelerated two-state systems, as we demonstrate in the rest of this chapter. Further, we permit the photon field mode frequency $\omega$ to be arbitrary and thus the atom can couple to vacuum modes of any frequency. This is unlike the so-called ``single mode approximation''  (SMA), where authors consider the atom to couple preferentially to a small number of modes concentrated at a single frequency e.g. for an atom within an accelerating cavity \cite{Alsing2003, FSchuller2005, Alsing2006, Datta2009, Wang2010, Dragan2011, Hwang2012, Brown2012, Lopp2018}. Research has shown that making the SMA can lead to difficulties in superluminal propagation effects at strong couplings \cite{Sanchez2018}, and entanglement generation \cite{Bruschi2010}, however some works have  explored beyond the SMA including an NMR analogue simulation of Unruh radiation \cite{Jin2016} [2 modes].    }
If the two level system is moving on the entire real line and its position is a function of $t$, we would use right and left moving field modes
\bea \phi_\omega=e^{-i (\omega t(\tau) \mp  k z(\tau))}, \label{eq:infinite-phi}  \eea
where $k=\omega/c>0$, is the $z$-component of the photon's wave-vector.
In the presence of a mirror fixed at $z=z_0$, we would instead use
\beq \phi_\omega=e^{-i(\omega t(\tau) - k z(\tau) + k z_0)}- e^{-i (\omega t(\tau) + k z(\tau) - k z_0)},\label{eq:mirror-phi}\eeq
thus ensuring that the photon field, and hence the transition amplitude,  vanishes when $z(\tau)-z_{0}$ is an integer multiple of the wavelength and the atom is at a node of the field.  In a cavity of length $L$, we would again use 
(\ref{eq:mirror-phi}), but the frequency and wave-vector would be restricted by the condition that  $k=\omega/c=2 \pi   n/L$ for some positive integer $n$.
We shall work with (\ref{eq:mirror-phi}) on the half-line for the moment,
and later adapt  the results to the case of a cavity or the entire real line as in (\ref{eq:infinite-phi}). 
In first order perturbation theory the probability of exciting the atom (via the raising operator $\sigma_+ e^{i\omega_0 \tau}$ in the interaction), and at the same time creating  a photon {\color{black}  of frequency $\omega$ }
via the $\hat a_{\omega}^\dagger \phi_\omega^*$,
term in the interaction potential, is given by Fermi's golden rule 
\begin{align} 
P=& \frac{1}{\hbar^2} \left| \int_{-\infty}^\infty \bra{1_\omega,a} H_{I} \ket{0,b} d\tau \right| ^2 \label{prob.1} \\
=& g^2 \left| \int_{-\infty}^\infty\bigl[ e^{i k (z(\tau)- z_0)} -c.c. \bigr] e^{i \omega t(\tau) + i \omega_0 \tau} d \tau \right|^2 \label{eq:phi-mirror} 
\end{align} 

where $a$ and $b$ here refer to the excited and ground states of the TLS respectively, as in \cite{Svidzinsky2018}. If the atom is at rest, $z$ is constant, $t=\tau$, and  the integral gives $\delta(\omega + \omega_0)$, which is zero since $\omega>0$ and $\omega_0>0$.  Thus $P=0$, and the probability to become excited and emit a photon vanishes. We now use the expression \eqref{eq:phi-mirror}, in the following to consider various types of space-time motions $z(\tau)$, to discover how they can give rise to non-vanishing probabilities $P$.

\subsection{Linear Acceleration Towards a Mirror} \label{Lin_Acc}
We will briefly review the trajectories laid out by \cite{Svidzinsky2018} to elucidate our motivations and to lay the path we will follow in subsequent sections. For simplicity we set $c=1$ so $k=\frac{2 \pi }{\lambda}=\omega$
where $\omega$ is the angular frequency and $\lambda$ the wavelength of the emitted photon. Recall $\omega_0$ is the angular frequency of the transition between the two internal states of the atom.
The trajectory of the atom undergoing constant acceleration $a$ in its own instantaneous rest frame (proper acceleration) we found previously to be 
\begin{equation} 
t(\tau) = \frac{1}{a}\sinh(a \tau), \qquad z(\tau) = \frac{1}{a}\cosh(a\tau) \label{eqno.1}. 
\end{equation} 
The probability from \eqref{eq:phi-mirror} above with this trajectory is
\begin{equation}
P= g^2 \left|
\int_{-\infty}^\infty\bigl[
e^{i \omega (z(\tau)-z_0)} -c.c. \bigr] e^{i \omega t(\tau) + i \omega_0 \tau}
d \tau \right|^2,
\end{equation}
where, upon substituting in the following formula obtained from equation \eqref{eqno.1},
$$\pm z(\tau) + t(\tau) =  \pm\frac{1}{a}e^{\pm a\tau},$$
the integral above becomes
\begin{eqnarray*}
\int_{-\infty}^\infty 
e^{i \frac{\omega}{a} e^{a\tau}} e^{-i\omega z_0}  e^{i \omega_0 \tau}  d \tau 
-&&\kern -27pt\int_{-\infty}^\infty 
e^{-i \frac{\omega}{a} e^{-a\tau}} e^{i\omega z_0}  e^{i \omega_0 \tau}  d \tau \\
&=&
\int_{-\infty}^\infty 
e^{i \frac{\omega}{a} e^{a\tau}} e^{-i\omega z_0}  e^{i \omega_0 \tau}  d \tau 
-\int_{-\infty}^\infty 
e^{-i \frac{\omega}{a} e^{a\tau}} e^{i\omega z_0}  e^{- i \omega_0 \tau}  d \tau \\
&=&
\int_{-\infty}^\infty 
e^{i \frac{\omega}{a} e^{a\tau}} e^{-i\omega z_0}  e^{i \omega_0 \tau}  d \tau -c.c.
\end{eqnarray*}
where we've changed integration variable $\tau \rightarrow -\tau$ in the second
integral.
So 
\begin{equation}
P= g^2 \left|
\int_{-\infty}^\infty\bigl[e^{i \frac{\omega}{a} e^{a\tau}} e^{-i\omega z_0}  e^{i \omega_0 \tau}
 -c.c. \bigr] d \tau
\right|^2 .
\end{equation}
Now changing integration variables from $\tau$ to $x=\frac{\omega }{a} e^{a\tau}$, so $d \tau = \frac{1}{a x} d x $, we have
\begin{equation}
\int_{-\infty}^\infty e^{i \frac{\omega}{a} e^{a\tau}} e^{-i\omega z_0}  e^{ i \omega_0 \tau} d\tau
=\frac{1}{a} e^{-i \bigl(\omega z_0 - \frac{\omega_0}{a}\ln\bigl(\frac{a }{\omega}\bigr)\bigr) }
\int_0^\infty e^{i x}  x^{i\frac{\omega_0}{ a}-1 }d x.
\end{equation}
We notice that the integrals
\begin{equation}
I_\pm =\int_0^\infty e^{\pm i x}  x^{\pm i\frac{\omega_0}{a}-1} d x
\end{equation}
are identical to equation \eqref{Unruh_2} in our section on Unruh radiation, up to some superficial sign changes that don't effect the result. Following along the same calculation we find that
\begin{equation}
I_\pm\left(\frac{\omega_0 }{a}\right) =e^{-\frac{\pi \omega_0 }{2a}}
\lim_{\epsilon\rightarrow 0^+} \Gamma\left(\pm \frac{i \omega_0 }{a} + \frac{\epsilon}{a}\right) =  e^{-\frac{\pi \omega_0 }{2a}}\Gamma\left(\pm\frac{i \omega_0 }{a}\right),
\end{equation}
which we can write as
\begin{equation}
\Gamma\left(\pm\frac{i \omega_0 }{a}\right) = e^{\pm i\psi}\left|\Gamma\left(\pm\frac{i \omega_0 }{a}\right)\right|.
\end{equation}
The phase $\psi$ will depend on $\frac{\omega_0 }{a}$, and though we don't have its analytic form it doesn't matter. What we do know is
\begin{equation}
\left|\Gamma\left(\pm \frac{i \omega_0 }{a}\right)\right|^2 =\frac {a \pi }{\omega_0\sinh\left(\frac { \pi\omega_0 }{a}\right)},
\end{equation}
which is all we need to write the probability as
\begin{eqnarray}
P&=&\frac{\pi g^2 e^{-\frac{\pi \omega_0 }{a}} }{a \omega_0 \sinh\left(\frac{\pi \omega_0 }{a}\right)}
\left|e^{-i \Bigl(\omega z_0 - \frac{\omega_0}{a}\ln\bigl((\frac{a }{\omega}\bigr)\Bigr) }e^{i\psi}
- e^{i \Bigl(\omega z_0 - \frac{\omega_0}{a}\ln\bigl(\frac{a }{\omega}\bigr)\Bigr) }e^{-i\psi} 
\right|^2   \\
&=&  
\frac{4 \pi g^2  \sin^2(\omega z_0 - \varphi) }{a \omega_0 e^{\frac{\pi \omega_0 }{a}} \sinh\left(\frac{\pi \omega_0 }{a}\right)}=  \frac{8 \pi g^2}{a \omega_0} \frac{\sin^2(\omega z_0 - \varphi) }{e^{\frac{2\pi \omega_0 }{a}}-1} \label{mirror_prob}
\end{eqnarray}
where 
\begin{equation}
\varphi = \frac{\omega_0}{a} \ln \left(\frac{a}{\omega } \right) + \psi.
\end{equation}
This is the result found in \cite{Svidzinsky2018}, with the mathematical details worked out, so we're confident that the expression in equation \eqref{eq:phi-mirror} for the probability is correct and we can play around with it for any trajectory. 
\newpage

\subsection{Probabilities from other initial states:}

From the expression for the probability above,equation \eqref{prob.1}, we see that there are four possible probabilities where the system changes its internal state and a photon is created i.e. four combinations of the inner product of the interaction Hamiltonian with the bra-ket's in different states. These are ; the case mentioned above $ \bra{1_\omega,a} H_{I} \ket{0,b}$, which, due to the absolute value in \eqref{prob.1}, is identical to the case of $\bra{0,b} H_{I} \ket{1_\omega,a}$ with just $\tau \rightarrow -\tau$ (i.e. the first process in reverse). There is also the case of spontaneous emission, where we have $\bra{1_\omega,b} H_{I} \ket{0,a}$, which will naturally have the same probability as $\bra{0,a} H_{I} \ket{1_\omega,b}$, again with $\tau \rightarrow -\tau$. This is the case of a TLS moving along an accelerated trajectory in its excited state, transitioning to its ground state and a photon emerging. It seems reasonable then to see this as spontaneous emission of the accelerated atom in the presence of a mirror. We expect this to have a much higher likelihood of occurring due to the atom already being in its excited state but its still a useful exercise.

\subsubsection{Spontaneous emission with a mirror}
The calculation is similar to that in the previous section but with certain subtleties that alter the final probability. To begin with we consider the probability of the event
\begin{equation}
P = \frac{1}{\hbar^2} \left|\int_{-\infty}^\infty d \tau\bra{1_\omega,b}\hat H_{I}\ket{0,a} \right|^2
\end{equation}
which yields
\begin{equation}
P = g^2 \left|\int_{-\infty}^\infty d \tau [e^{i \omega (z(\tau) - z_0)} - c.c.] e^{i \omega t(\tau) - i \omega_0 \tau} \right|^2
\end{equation}
following similar calculations as \S \ref{Lin_Acc} we have 
\begin{equation}
\pm z(\tau) + t(\tau) =  \pm\frac{1}{a}e^{\pm a\tau},
\end{equation}
so the integral becomes
\begin{equation}
\int_{-\infty}^\infty e^{i \frac{\omega}{a} e^{-a\tau}} e^{-i\omega z_0}  e^{i \omega_0 \tau}  d \tau -c.c.
\end{equation}
where again we've changed $\tau \rightarrow -\tau$ in the first integral. A change of integration variables from $\tau$ to $x=\frac{\omega }{a} e^{-a\tau}$, so $d \tau = \frac{-d x}{a x}  $ gives us a slightly different expression from \S \ref{Lin_Acc} but is the right choice to arrive at the gamma function. This change of variables gives
\begin{equation}
\int_{-\infty}^\infty e^{i \frac{\omega}{a} e^{-a\tau}} e^{-i\omega z_0}  e^{ i \omega_0 \tau} d\tau
=\frac{1}{a} e^{-i \bigl(\omega z_0 + \frac{\omega_0}{a}\ln\bigl(\frac{a }{\omega}\bigr)\bigr) }
\int_0^\infty e^{i x}  x^{-i\frac{\omega_0}{ a}-1 }d x.
\end{equation}
which, upon considering the integrals
\begin{equation}
I_\pm =\int_0^\infty e^{\pm i x}  x^{\mp i\frac{\omega_0}{a}-1} d x.
\end{equation}
we can follow the previous calculation almost identically and arrive at the probability
\begin{equation}
P = \frac{8 \pi g^2}{a \omega_0} \frac{e^{\frac{2\pi \omega_0 }{a}}\sin^2(\omega z_0 + \varphi)}{e^{\frac{2\pi \omega_0 }{a}}-1}
\end{equation}
where again $\varphi = \frac{\omega_0}{a} \ln \left(\frac{a}{\omega } \right) + \psi$. It is straight forward to see then that this has a significantly higher probability of occurring as the form is almost identical to \eqref{mirror_prob} but now with an exponential term in the numerator to contribute. Physically this seems reasonable as a TLS in its excited state has a chance to drop to its ground state and radiate away energy regardless but under constant acceleration there is further energy added to entice photon emission.
While this result is somewhat expected from a physical picture of the system it assures us we are on solid footing and can proceed to consider not just different initial states but different trajectories entirely. In the following we look at the case of an atom undergoing simple harmonic motion in the presence of a mirror, in a cavity and in a vacuum.

\newpage
\section{Oscillating 2-state atom}\label{mirror}


We now consider an atom forced to oscillate in the presence of a stationary mirror, the latter located at $z_0$, oscillating with a motional angular frequency $\Omega$,  around $z=0$
with amplitude $0<A<z_0$ 
(so we do not hit the mirror),  and set  
\begin{equation}
t=\tau, \qquad z(t) = A \sin(\Omega t).
\end{equation}
Using this in \eqref{Hamomega}, the probability of creating a photon of frequency
$\omega$ from the vacuum, and at the same time exciting the atom from its ground state to its excited state is
$$P(\omega)= g^2 \left|
\int_{-\infty}^\infty\bigl[
e^{i k (A \sin(\Omega t) - z_0)} -c.c. \bigr] e^{i (\omega +  \omega_0)t} d t
\right|^2. $$
Simplify the notation by absorbing $\Omega$ into $t$ and defining dimensionless variables 
$\tau = \Omega t$, $\tilde \omega = (\omega +\omega_0)/\Omega$,
$\tilde A = k A$ and $\tilde z_0 = k z_0$, we obtain
\begin{eqnarray} P(\omega) &=& \frac{g^2}{\Omega^2} \left|
\int_{-\infty}^\infty\bigl[
e^{i (\tilde A \sin \tau - \tilde z_0)} -c.c. \bigr] e^{i \tilde \omega \tau} d \tau\right|^2 \nonumber\\ 
&=& \frac{g ^2}{\Omega^2} \left|
e^{-i\tilde z_0}\int_{-\infty}^\infty e^{i (\tilde A \sin \tau+ \tilde \omega \tau)} d \tau \right.\nonumber\\ 
& &\qquad\qquad\left.-e^{i\tilde z_0}\int_{-\infty}^\infty  e^{i (\tilde A \sin \tau - \tilde \omega \tau)} d  \tau\right|^2.
\label{totalSHO-probability} 
\end{eqnarray}

The integrals appearing in (\ref{totalSHO-probability}) 
 are related to
Anger functions \cite{Abramowitz1970} (section 12.3.1),
$$ {\bf J}_\nu(x)=\frac{1}{2\pi}\int_{-\pi}^\pi e ^{i(x \sin\theta - \nu \theta)}d \theta
= \frac{1}{\pi}\int_0^\pi \cos(x\sin\theta - \nu\theta)d \theta.$$
But Anger functions are not quite what we want since what we have in (\ref{totalSHO-probability})
, is
\begin{eqnarray}
\int_{-\infty}^\infty  e ^{i(x \sin\theta - \nu \theta)} d \theta
&=&\!\! \sum_{s=-\infty}^\infty\, \int_{-\pi}^\pi e^{i\bigl(x \sin\theta - \nu(\theta+2s\pi)\bigr)}d \theta \\
&=&\!\! \pi {\bf J}_\nu(x) \sum_{s=-\infty}^\infty  e^{-2is\nu\pi}.
\label{eq:totalamplitude}
\end{eqnarray}
Letting $ \nu = n + \{\nu\}$, 
where $n$ is an integer, and $0 \le  \{\nu\}<1$, is the non-integral part of $\nu$,
then
$ \sum_{s=-\infty}^\infty  e^{-2is\nu\pi}= \sum_{s=-\infty}^\infty  e^{-2is\{\nu\}\pi} = 1 + 2\sum_{s=0}^\infty  \cos(2s\{\nu\}\pi) = \pi \delta(\{\nu\})$
and (\ref{eq:totalamplitude}), vanishes unless $\nu$ is an integer, in which case it
diverges. This is not unexpected since Fermi's Golden Rule says that if the transition rate is constant, integrating over all $t$ will necessarily give an infinite answer for any transition with non-zero probability.  For our periodic case its more informative to estimate a transition rate over a motional cycle rather than the accumulated  probability over all time.

\sloppy Before estimating this let us consider the case when $\nu$ is a non-zero rational number, $\nu = p/q$ with $p$ and $q$ mutually prime and positive, in the integral
$$ \int_{-\infty}^\infty e^{i(x\sin \theta -\nu \theta)} d \theta
= q \int_{-\infty}^\infty e^{i(x\sin (q \psi)  - p \psi)} d \psi 
$$
where $\psi = \theta/q$. The integrand is periodic in $\psi$
with period $2\pi$ so again the integral will diverge (unless the integral over one period vanishes).  Integrating over just one period in $\psi$, we can define
\begin{eqnarray}
{\cal J}(x;p,q) &=& 
\frac{1}{2 \pi} \int_{-\pi}^\pi e^{i(x \sin (q \psi)  - p \psi )} d \psi\\\nonumber
&=& \frac{1}{\pi} \int_0^\pi \cos\big\{x \sin (q \psi)  - p \psi \bigr\} d \psi,
\label{eq:J-def}
\end{eqnarray}
which, in terms of Anger functions, becomes
\begin{eqnarray}
{\cal J}(x;p,q) &=& 
\frac{1}{2 \pi q} \int_{-q\pi}^{q \pi} e^{i(x \sin \theta  - \frac p q  \theta )} d \theta \\\nonumber
&=&\frac{1}{q}  {\bf J}_{\frac p q }(x)\sum_{s=0}^{q-1} e^{i \pi (2 s - q + 1)\frac{p}{q}}.\nonumber
\end{eqnarray}
We note however that $$\sum_{s=0}^{q-1} \exp({i \pi (2 s - q + 1)p/q}) = 
\exp({-i\pi\frac{p(q-1)}{q}}) \sum_{s=0}^{q-1} \exp({2i \pi p s/q})=0,$$ 
since $\sum_{s=0}^{q-1} \exp(2i \pi p s/q)$,
vanishes for any two integers $p$ and $q$, provided $p/q$ is not an 
integer.
 From this observation we conclude  that the transition rate is zero unless $\tilde \omega$ is a
positive integer, that is 
\begin{equation}
\omega + \omega_0 = n\Omega,\label{eq:Omega-restriction}
\end{equation}
where $\Omega$ is the mechanical frequency, $\omega_0$ is the atomic transition frequency, $\omega$ is the frequency of the photon field and $n$ is an integer.
Since $\tilde \omega$ is  positive, $n>0$, and to obtain a closed expression for the transition probability over a mechanical cycle we can use the integral
representation of the Bessel function,  \cite{Abramowitz1970} (section 9.1.21),
$J_n (\tilde A) = \frac{1}{2 \pi}\int_{-\pi}^{\pi} 
\exp[{i(\tilde A \sin \tau - n \tau)}] d \tau = (-1)^n J_{-n}(\tilde A)$,
to write  $\int_{-\pi}^{\pi} \exp [{i(\tilde A \sin \tau + \tilde z_0 -  n  \tau)} ]d \tau= 2 \pi e^{i \tilde z_0}  J_n(\tilde A)$, and $\int_{-\pi}^{\pi} \exp[{i(\tilde A \sin \tau -
 \tilde z_0 +  n \tau)} ]d  \tau
= 2 \pi e^{-i (\tilde z_0 - n \pi)}  J_n(\tilde A)$.
By dividing by the period of mechanical oscillation, $2\pi/\Omega$, we get a transition rate (in ${\rm Hz}$), as
\begin{eqnarray}
\overline P_{n} & = &\frac{\Omega }{2 \pi}
\frac{g^2}{ \Omega^2} \left|
 \Bigl[ 2 \pi \left(e^{-i (\tilde z_0 - n \pi)} -  e^{i \tilde z_0}\right) \Bigr]
J_n(\tilde A)\right|^2\nonumber\\
&=&
\frac{8 \pi g^2}{\Omega} \sin^2\left(\tilde z_0 -\frac{\pi n}{2}\right) J_n^2(\tilde A)\\
&=&
\frac{8 \pi g^2}{\Omega} \sin^2\left(k z_0 -\frac{\pi n}{2}\right) J_n^2(k A),\label{eq:mirror-rate}
\end{eqnarray}
where $\omega = n \Omega - \omega_0 >0$.


\subsection{2-dimensional motion}

The formalism can be applied to a 2-level atom following any closed trajectory $\vec x(t)$ in the two dimensional $y-z$ plane, with a flat mirror located at $z_0$ extending in the $y$-direction.   For an electromagnetic wave with wave-vector $(k_y,k_z)$ we simply replace $k (z(t) - z_0)$ in (\ref{eq:phi-mirror}) with
$\vec{k} \cdot \vec{x}(t) - k_z z_0$. One cannot obtain analytic answers for a general trajectory but some simple cases are immediate:
\vspace{.25cm}\\

\subsubsection{ 2-level atom oscillating parallel to a mirror}  for oscillation in the $y$-direction replace $k (z(t) - z_0)$ in (\ref{eq:phi-mirror}) with
$ k_y A \sin(\Omega t)  - k_z z_0 $  with $A$ again a constant amplitude.
The analysis is identical and (\ref{eq:mirror-rate}) still holds, but with $k z_0$ replaced with $k_z z_0$ and $k A$ replaced with $k_y A$.
\vspace{.25cm}\\

\subsubsection{Rotating 2-level atom with a mirror} For an atom rotating around the fixed point $(0,0)$ in a circle with radius $R$
and constant angular velocity $\Omega$, $\vec x(t) = R(\cos(\Omega t),\sin(\Omega t))$ and
$\vec k \cdot \vec x(t)=k_y R \cos(\Omega t) + k_z R \sin(\Omega t)$.
If we parameterise the direction of the electromagnetic wave by a phase $\delta$ with
 $k_y=k\sin\delta, \quad k_z = k\cos\delta$
then
  $k_y R \cos(\Omega t) + k_z R \sin(\Omega t) - k_z z_0  = k R \sin(\Omega t + \delta) - k z_0 \cos \delta $
and the phase $\delta$ 
in the sine function can be
absorbed into $t$ in (\ref{eq:phi-mirror}), which does not affect the result. We just replace the
amplitude $A$ in (\ref{eq:mirror-rate}) with the rotation radius $R$ and replace 
$k z_0$ with $k_z z_0 \cos \delta$.


\subsection{Oscillating 2-level atom in a cavity}\label{cavity}

 In a one-dimensional cavity containing no photons the transition rate to excite the atom and at the same time emit a photon of frequency $\omega$ is still given by (\ref{eq:mirror-rate}), except that the allowed values of $\omega$ are discrete, (reinstating $c$ here)
 $\omega = c k = \pi m c/L$,
with $m$ a positive integer and Eqn (\ref{eq:Omega-restriction}) imposes the condition
\[ \frac{\pi m c}{L} + \omega_0=n \Omega\]
on $\Omega$. The rate is enhanced however if the cavity already contains $N$ photons of 
frequency $\omega$, since then $\langle N+1|a^\dagger_\omega|N\rangle = \sqrt{N+1}$,
giving the transition rate to excite/de-excite the atom given $N$ photons in the cavity as,
\begin{equation}
\overline P_{n,m,N,\pm} = 
\frac{8 \pi \chi_\pm\, g^2}{\Omega} \sin^2\left(\frac{\pi m z_0}{L} -\frac{\pi n}{2}\right) J_n^2\left(\frac{\pi m A}{L}\right),\label{eq:cavity-rate}
\end{equation}
where $\chi_+=N+1$, to excite and emit a photon, while $\chi_-=N$ to de-excite the atom and absorb a photon, and
$n \Omega = \omega_0 - \omega = \omega_0 - \pi m c/L$. {\color{black} We note that in order to observe this phenomena one could consider one of the mirrors to be slightly imperfect and then one could perform spectroscopy on the photons leaking from the optical cavity. The signature of the photons created via (\ref{eq:cavity-rate}), will prove challenging to detect over any background of existing photon occupation within the cavity.}
\vspace{.25cm}\\

\begin{figure}[tp] 
\begin{center}
\setlength{\unitlength}{1cm}
\begin{picture}(8.5,7.5)
\put(-2,-.3){\includegraphics[width=.65\columnwidth]{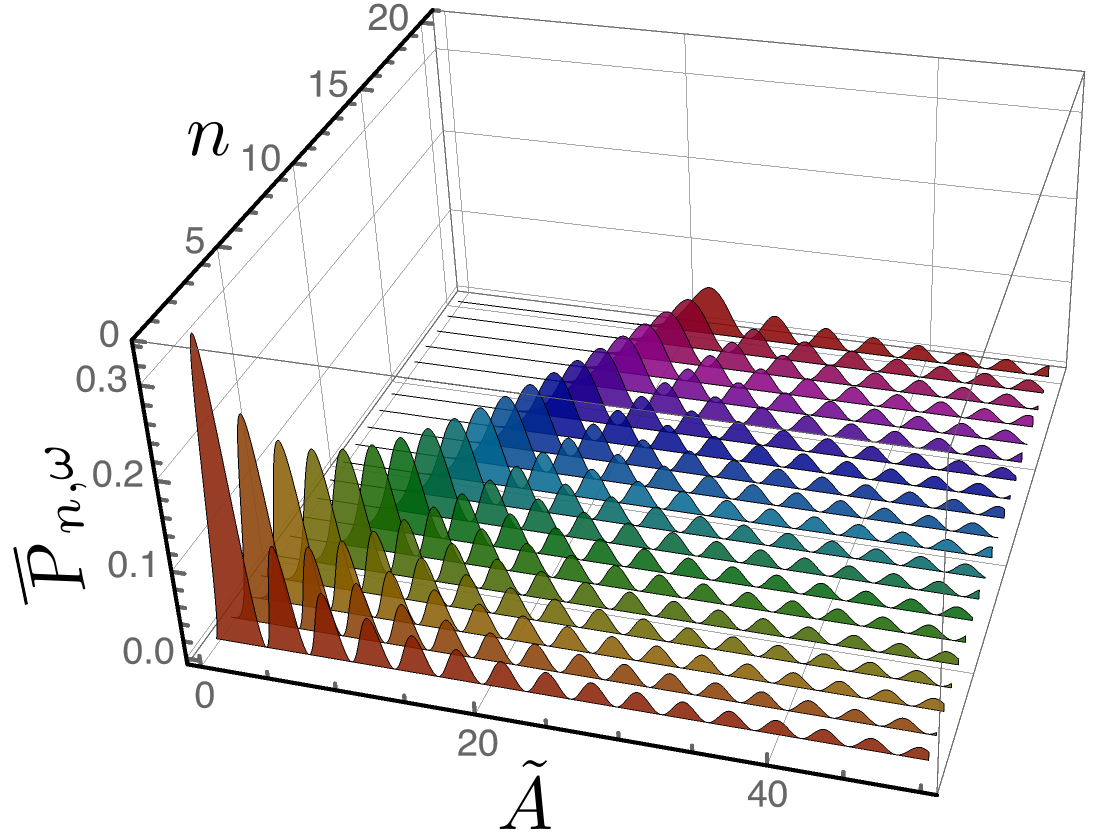}}
\end{picture}
\end{center}
\caption{Transition rate for SHO forced atom in free space to emit a photon with angular frequency $\omega = n\Omega - \omega_0$ and wavelength $\lambda={2\pi c}/{\omega}$, equation  (\ref{eq:P_n-no-mirror}), as a function of dimensionless oscillation amplitude $\tilde A =({2\pi}/{\lambda}) A$ and sideband index $n$. The prefactor 
${2 \pi g^2}/{\Omega}$ is omitted.  We see that transition rate is negligible until $\tilde A$ is of order $n$.}
   \label{Fig2}
\end{figure}

\newpage
\subsection{Atom performing SHO in free space}\label{free} 

Repeating the analysis of the previous section when there's no mirror, the transition rate to emit a   photon is,  using (\ref{eq:infinite-phi}),
\begin{equation}
\overline P_{n,\omega}  = \frac{\Omega }{2 \pi}
\frac{g^2}{\Omega^2} \left|
\int_{-\pi}^\pi \left(e^{i \tilde A \sin \tau} - c.c \right) e^{i \tilde \omega \tau} d \tau
\right|^2 
  = \frac{8 \pi g^2 }{\Omega} J_n^2( k A) ,
\label{eq:P_n-no-mirror}
\end{equation}
with $n>0$. Even with no mirror an atom can be excited and emit a photon at the same time by shaking the atom hard enough to supply the required energy.  
The transition rate (\ref{eq:P_n-no-mirror}), with  $\omega = n\Omega - \omega_0$ is 
\begin{equation}
\overline P_{n}
  = \frac{8 \pi g^2 }{\Omega} J_n^2\left(\frac { (n \Omega -\omega_0)  A }{c}\right),
\label{SHOfreespace}  
\end{equation}
and for a given $n$ the rate will be largest at the first peak of $J_n$. From Fig \ref{Fig2}, we observe that for a given $n$, the transition rate is negligible unless $\tilde{A}\sim n$. The maximum rate occurs when $n=1$ and $\tilde{A}\sim 1.8$, and here $\overline P_1\approx 2.1g^2/\Omega$, but to achieve this one must have a mechanical oscillation amplitude of $A\sim 1.8 c/(\Omega-\omega_0)$, which will be extremely large when compared to the highest realistically achievable values of $\Omega-\omega_0$, in an experiment. 

Another possible avenue one could take would be to consider the case when the amplitude of oscillation $A$ is small enough so that the argument of the Bessel function in (\ref{SHOfreespace}), is small and we can use the expansion $J_1^2(x)\sim x^2/4,\; 0<x\ll 1$. With this approximation the rate is maximised when $\omega_0=\Omega/2$, and taking this for small amplitude oscillations we obtain $\overline P_{1}\approx\pi (A\alpha)^2\Omega^3/(32 c^2)$, with $\alpha = g \omega_{0}$. 
There is a cQED setup for a two level superconducting artificial atom that is placed on a vibrating cantilever/membrane as in the recent experimental work by \cite{Viennot2018}. By using a diamond based nanomechanical resonator one can achieve motional frequencies in the GHz regimes \cite{Gaidarzhy2007,Rodriguez-Madrid2012}.  Unfortunately to achieve this rate one would need to consider ultra-strong coupling $g$ as is done here \cite{Forn-Diaz2016,  Magazzu2018, PuertasMartinez2019, FriskKockum2019, Forn-Diaz2019}. Our analysis is perturbative so any transition rate calculated for large $g$ should be taken with caution. A non-perturbative numerical calculation of the transition probabilities may be possible given the simplicity of the system, but we leave that for future work.

{\color{black} We finally compare our results with recent related works. In \cite{Lo2018}, and \cite{Ferrari2019}, the authors respectively consider acceleration radiation from a mechanically oscillating two level system in free space  and the modification of spontaneous emission in a two level system adjacent to an oscillating mirror. However their studies are restricted to the case where $\Omega<\omega_0$, and only multi-photon off-resonant processes play a role in these cases. Such off-resonant processes cannot be captured via our first order perturbative analysis and thus we cannot make any clear comparison with their results. However the work of \cite{Souza2018}, which looks at the radiation emitted from a two-level atom oscillating in free-space, does consider the regime $\Omega>\omega_0$, and include resonant processes and find the  emission of photons with the frequency $\omega_1=\Omega-\omega_0$, but does not find the higher modes $\omega_n=n\Omega-\omega_0$, which we predict to also exist, though with greatly reduced probabilities. To make a more quantitative comparison we study the small photon frequency case when $\omega_1\equiv\Omega- \omega_0$, is small and $\omega_1 A/c\ll 1$.  Making the approximation $J_1(x)^2\sim x^2/4$, for $x\ll 1$, and comparing our rate (\ref{SHOfreespace}), with Eqn (5) \cite{Souza2018}, in this regime, (in the notation of \cite{Souza2018}, this is when $\omega_1\equiv\omega_{\rm cm}- \omega_0$ is small), we find that both rates scale as $\overline P_{1}\sim \Gamma_{\rm MIE}\sim A^2/\Omega$. However in our case we find $\overline P_{1}\sim \omega_1^2$, while in \cite{Souza2018}, $\Gamma_{\rm MIE}\sim \omega_1^3$, a difference which may be due to the differences between the Hamiltonians. In our study we assume the Hamiltonian \eqref{Hamomega},  a  model for coupling between two levels systems and vacuum fields used by many works and which also was used in \cite{Svidzinsky2018}, to derive the Unruh temperature for the case of uniform acceleration, while \cite{Souza2018}, Eqn (1), includes both the normal dipole coupling but also a term linear in the velocity of the two level system. }

\chapter{The Quantum Hall Effect on a Sphere and the Atiyah-Singer Index Theorem }
\label{Chpt_2}



\section{Introduction}

The Hall effect and its offspring, the quantum Hall effects, both integer and fractional, have been a fascinating area of study since their discovery. These effects are observed when a thin, effectively two-dimensional, conducting material with a current running through it is placed in a  constant magnetic field that is perpendicular to the plane of the material. In the classical picture a somewhat accurate description can be drawn from a kinetic theory of an electron gas scattering off the lattice of heavy immobile atoms in the material, as in the Drude model \cite{Drude}. Obviously this model is not the full story and breaks down in its descriptive power for materials at low temperatures and strong magnetic fields. This discrepancy between observation and theory is mostly resolved via the understanding of the quantum mechanical origins of the phenomena, though open questions still linger. 
The integer quantum hall effect is, in part, a consequence of Landau quantization of the cyclotron orbits of the electrons in the material, first described in \cite{Ando1975}. However at lower temperatures ($T < 5$K) and stronger magnetic fields, fractional quantum hall states were discovered by \cite{Tsui1982} and this has led to an explosion of research in the area. Laughlin constructed trial ground state wave functions on the plane in \cite{Laughlin} and provided an explanation for the fractional states with filling factor $\nu = \frac{1}{2k+1}$ with $k$ an integer.

As the interest and research in to quantum hall systems grew, novel approaches and more sophisticated mathematical techniques began to arise. In particular, Haldane \cite{Haldane} considered a model of particles moving on the surface of a magnetic sphere ---
a sphere with a magnetic monopole at the centre. Haldane's use of the spherical geometry in this instance was to consider particles moving on the surface of the sphere to create a rotationally invariant version of Laughlin's wave function. It is common in such analyses to ignore the electron spin, since in strong magnetic fields it is assumed that electron spins are split and only the lower energy state is relevant to the problem so the spin can be ignored. In constructing the wave functions the particles are essentially considered to be spinless, but obey Fermionic statistics so that the many particle wave-function is anti-symmetric.
In particular the spin connection for spin-$\frac{1}{2}$ particles moving in a curved space plays almost no role in Haldane's construction.
Neither does the topology of the sphere play any real role, the sphere is merely a mathematical device that simplifies the analysis. In an effort to explain the hierarchy of fractional filling factors of Landau-levels this was an effective exploitation of the symmetry inherent to the geometry of the sphere. However the mathematics of spinors on compact manifolds is very rich in both geometry and topology.  In particular the Atiyah-Singer index theorem \cite{Atiyah} tells us that the Dirac operator on a sphere with a magnetic monopole at the centre has zero modes, an energy gap and constrains the number of positive and negative chirality zero modes.
This is true both relativistically and non-relativistically, the mathematics is essentially the same in both cases (the latter is basically the square of the former). 

There is a number of advantages in focusing on these zero modes: they are topological invariants, for any fixed monopole charge they must always be there even if the magnetic field is distorted (provided the total magnetic charge does not change), and are therefore topologically stable, and they ensure that the zero point energy vanishes, there is an absolute minimum for the energy which no perturbation can lower.
The number of zero modes is constrained by the index theorem, for a magnetic monopole of charge $m$ the difference between the number of positive chirality zero modes $n_+$ and the number of negative chirality zero modes $n_-$ is\footnote{There is a choice of sign on the right hand side which
depends on the definition of chirality, in our conventions it will be $-m$.}
\begin{equation}
n_+ - n_- =-\frac{1}{2\pi}\int_{S^2} F =-m \label{index.1}
\end{equation}
where $m$ is an integer (the first Chern class of a $U(1)$ bundle).
In particular if $m$ is positive $n_+$ can vanish in which case
\[n_-=m, \]
while if $m$ is negative $n_-$ can vanish and then
\[ n_+ = m.\]
When the zero modes have only one chirality
the number of linearly independent zero modes is exactly $|m|$ this reflects the degeneracy of the ground state.

The main focus of this chapter will be to investigate the quantum Hall effect (QHE) on a sphere from the point of view of the Atiyah-singer index theorem and show how the zero modes relate to Haldane's version of the Laughlin ground state wave function. 
While the role of topology has long been appreciated in the quantum Hall effect to our knowledge the Atiyah-Singer index theorem has not been exploited to any great extent, except for the case of relativistic 4-component fermions in graphene, \cite{Park+Yi,Ezawa}, and in non-commutative geometry in the higher dimensional QHE \cite{Hasebe}.  In this work we are dealing with ordinary 2-component non-relativistic electrons and this has not been investigated before to our knowledge. 
The fact that the filling factor is related to the Chern class of a $U(1)$ bundle over a torus, which is a Brillouin zone in $k$-space, was pointed out in \cite{NTW},
but while Chern classes are part of the index theorem for the Dirac operator, the theorem itself is much more than just Chern classes, in the context studied here it is about zero modes of the Dirac equation.  Another topological aspect of the quantum Hall effect is its relation to Chern-Simons theory but this is not relevant to the index theorem, Chern-Simons theories are only defined in odd dimensions and the index always vanishes in odd dimensions. The Chern-Simons action is relevant in an effective action approach, in this setup the coefficient of the Chern-Simons actions is related to the Hall conductivity.

The integer QHE is studied first, with a uniform magnetic flux through the surface of the sphere.  The exact ground state for $N$ non-interacting Fermions
is calculated and reproduces Haldane's result, equation (\ref{eq:Haldane}), for filling factor $\nu = 1$.

The fractional quantum Hall effect is then studied in the context of Jain's composite Fermion picture \cite{Jain}.
Magnetic vortices, represented by Dirac monopoles for which the Dirac string is replaced by a physical vortex of a statistical gauge field of strength $v$ and is not a gauge artefact, are attached to the electrons.  The resulting composite particles move in the total magnetic field generated by the monopole plus the vortices.  For the wave function (a cross-section of a $U(1)$ bundle) to be free from singularities the vortices necessarily have strength $|v|=2k$, where $k$ is an integer, and act so as to reduce the strength of the uniform background field. Again zero modes can be constructed, equation (\ref{eq:fractional-ground-state}), and there is a unique ground state with an energy gap where for large $N$ the filling factor is $\nu = \frac{1}{2k +1}$. This ground state can be related to Laughlin's ground state wave-function for the fractional QHE through a singular gauge transformation that removes the vortices.

The layout of this chapter then is as follows. We begin with a brief introduction to the background of the quantum Hall effect and the Atiyah-Singer index theorem in \S \ref{bground}. \S \ref{sec:Dirac} reviews the Dirac operator on the surface of a sphere with a magnetic monopole at the centre. In \S \ref{sec:ZeroModes} zero modes are constructed and shown to give a stable ground state with an energy gap for filling factor $\nu=1$. For completeness wave-functions for energy eigenstates in the higher Landau levels are exhibited in terms of Jacobi polynomials in \S \ref{sec:HigherLL}.  Vortices are introduced and ground state wave functions for the fractional quantum Hall effect are presented in \S \ref{sec:Vortices}. The results are summarized and conclusions presented in \S \ref{sec:Conclusions}.


\section{Background to The Hall Effect and Index Theorem}\label{bground}
\subsection{The Quantum Hall Effect}
Introductory material and explanations for these effects are numerous \cite{Jain, Tong, Chakraborty, Prange+Girvin}, and as the purpose of this thesis is to look at the impact of the geometry on the system we will briefly recap here the central points of this vast area as it pertains to our research.

To begin we should consider the classical case and the observation first made by Edwin Hall, from whom the name derives \cite{Hall}. Hall noticed that when a current is put through a two-dimensional conducting plate and a magnetic field passes through the plate perpendicularly, there is a build up of charge on the edges of the plate. This creates a voltage difference, known as the Hall voltage, which runs in the transverse direction to the original direction of the current. A typical introductory explanation as to why this occurs, not known to Hall at the time, can be derived simply from looking at the Lorentz force for a charged particle confined to move in two-dimensions in a constant, perpendicular, magnetic field. The Lorentz force for an electron is
\begin{equation} 
\vec{F} = -e \vec{v}\times \vec{B} \label{QH.1}
\end{equation}
where $q$ is the charge of the particle, $\vec{v} = v_{x} \hat{\imath} + v_{y} \hat{\jmath}$ is its velocity and $\vec{B} = B \hat{k}$ is the constant magnetic field, with Cartesian coordinates $(x(t),y(t),z(t))$ and basis vectors $\hat{\imath},\hat{\jmath},\hat{k}$. From Newton's second law the equations of motion from \eqref{QH.1} are a set of coupled O.D.Es given by
\begin{align}
M\ddot{x} &= -eB \dot{y} \\
M\ddot{y} &= eB \dot{x},
\end{align}
where $M$ is the mass of the particle. The trajectory of the particle subject to this force is a circle and its position as a function of time is given by
\begin{align}
x &= x_{0} + r\sin(\omega_{B} t +\phi) \\
y &= y_{0} - r\cos(\omega_{B} t +\phi).
\end{align}
Here $\omega_{B} = \frac{eB}{M}$ and is known as the cyclotron frequency, $r$ is the radius of the circle and $(x_{0},y_{0})$ is the center of the circle.

This description however doesn't qualitatively describe the flow of electrical current and charges in the two-dimensional plate we are considering. To do so we must consider a model of charge transport, the simplest of which is the Drude Model \cite{Drude}. Within this model we consider the particles of a charged electron gas scattering off the stationary heavy atoms of the conducting material. This requires adding to the Lorentz force above an electic field $\vec{E} = E \hat{\imath}$ in the direction of the current and we must also introduce a term known as the scattering term. The addition of these two terms gives 
\begin{equation} 
\vec{F} = -e \vec{E} -e \vec{v}\times \vec{B} -\frac{1}{\tau} M\vec{v} \label{QH.2}
\end{equation}
where the last term is the scattering term and $\tau$ is called scattering time. This is still a classical model and assumes that we can treat this problem in terms of kinetic theory of gases. The time $\tau$ is mean-free time and is simply the average time between collisions in this kinetic model between electrons and the static molecules making up the metal. Now we need not solve these equations as they are but instead consider the circumstance when the system has settled to an equilibrium state, i.e. when $\vec{F} = 0$. With this condition and writing \eqref{QH.2} in terms of the current density $\vec{J} = -e n_{e}\vec{v}$ we have
\begin{equation}
\frac{n_{e} e^2 \tau}{M} \vec{E} = \omega_{b}\tau(J_{x}\hat{\jmath}-J_{y}\hat{\imath}) + \vec{J}
\end{equation}
which can be written as a matrix equation given by 
\begin{equation}
\vec{E} = \frac{M}{\tau n_{e} e^2}\begin{pmatrix} 1 & \omega_{b}\tau \\ -\omega_{b}\tau & 1 \\ \end{pmatrix} \vec{J}
\end{equation}
where here we see that we have the standard relationship between the electric field and the current density by identifying the matrix in front as the resistivity tensor.
\begin{equation} 
\mathbf{\rho} = \frac{M}{\tau n_{e} e^2}\begin{pmatrix} 1 & -\omega_{b}\tau \\ \omega_{b}\tau & 1 \\ \end{pmatrix}
\end{equation}
Note that in the absence of a magnetic field we have just direct current and thus $\rho_{DC} = \frac{M}{\tau n_{e} e^2}$. For a scalar resistivity the conductivity is just the reciprocal of the real number but for a matrix we have the inverse. Thus the conductivity tensor is 
\begin{equation} 
\mathbf{\sigma} = \frac{\sigma_{DC}}{1+ (\omega_{B} \tau)^2}\begin{pmatrix} 1 & \omega_{b}\tau \\ -\omega_{b}\tau & 1 \\ \end{pmatrix}
\end{equation}
with $\sigma_{DC} = \frac{1}{\rho_{DC}}$. The transverse terms of these tensors are the interesting feature. The relationship between the transverse resistivity and the magnetic field give us our first insight 
\begin{equation}
\rho_{xy} = \rho_{DC}\omega_{B}\tau = \frac{B}{n_{e}e} \label{Drude.1}
\end{equation}
a linear relation. 
This means that when treating the transverse components of the resistivity $\rho_{xy}$ as functions of the magnetic field we have a physical prediction that should be experimentally detectable; in a strong magnetic field, for a two dimensional plate with current running in one direction, we should detect a change in the resistance in the transverse direction and thus a voltage. The Hall voltage. This is in good alignment with experimental results for weak magnetic fields $(B < 1 \ T)$ but there is more to the story. At lower temperatures and stronger magnetic fields, plateaus occur!
\begin{figure}[h!]
\centerline{\includegraphics[width=0.7\linewidth]{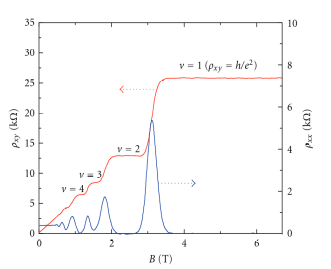}}
\caption{Image can be found here \cite{Bao}. The data is taken at 0.3 K from a GaAs-based heterostructure, displaying the integer quantum hall effect. The red line tracks the transverse resistivity $\rho_{xy}$, $\nu$ is the filling factor and the blue line shows the longitudinal resistivity $\rho_{xx}$. We can see for weak magnetic fields the transverse resistivity is in good agreement with the classical explanation, i.e. the linear relation from \eqref{Drude.1}.}
\label{IQHE}
\end{figure}

As can be seen from the above image in figure \ref{IQHE}, there is a good fit for the classical linear relation for weak magnetic fields. The plateaus however are an indication of something more intricate occurring (a persistent trend in the story of the quantum hall effect) and where classical mechanics fails, a quantum approach must be tried.

To look at the quantum mechanical version of the Hall effect we begin where we normally do with quantum mechanics and look at the classical Lagrangian and Hamiltonian. It is not immediately clear what the Lagrangian for this system is. We need to consider the kinetic energy of the particle, knowing that it's moving in a background electromagnetic field, and thus must have interaction terms. There is the standard intuition for the Lagrangian $L = T - V$, where $T$ and $V$ are the kinetic and potential energies respectively. This gives us a hint for where to begin. We can start by writing the electric and magnetic fields in terms of their potentials
\begin{align}
\vec{E} &= - \nabla \phi - \frac{\partial \vec{A}}{\partial t} \\
\vec{B} &= \nabla \times \vec{A}
\end{align}
and the Lorentz force becomes 
\begin{align}
\vec{F} &= e\nabla \phi + e\frac{\partial \vec{A}}{\partial t} - e \vec{v}\times (\nabla \times \vec{A}) \label{QH.3} \\ 
&= e\left( \nabla \phi + \frac{\partial \vec{A}}{\partial t} - \nabla(\vec{v}\cdot \vec{A}) + (\vec{v}\cdot\nabla)\vec{A} \right) \label{QH.4}\\ 
&= e\left(\nabla(\phi -\vec{v}\cdot\vec{A}) + \frac{d\vec{A}}{dt}\right)
\end{align}
where we have used a vector calculus identity between lines \eqref{QH.3} and \eqref{QH.4} and the chain rule between \eqref{QH.4} and the last line as $(\vec{v}\cdot\nabla) = \frac{dx}{dt}\frac{\partial}{\partial x} + \frac{dy}{dt}\frac{\partial}{\partial y} + \frac{dz}{dt}\frac{\partial}{\partial z}$. Rearranging the above 
\begin{equation}
\frac{d}{dt}\left(M\dot{\vec{r}} -e \vec{A}\right) = e\nabla(\phi- \vec{v}\cdot\vec{A})
\end{equation}
and comparing with the Euler-Lagrange equations $\frac{d}{dt}\left(\frac{\partial L}{\partial \dot{x}^{\mu}}\right) = \frac{\partial L}{\partial x^{\mu}}$ we can see our Lagrangian should be
\begin{equation}
L = \frac{1}{2} M |\dot{\vec{r}}|^{2} -e \dot{\vec{r}}\cdot\vec{A} + e\phi
\end{equation}
with the canonical momentum given by $\vec{\pi} = \frac{\partial \mathcal{L}}{\partial \dot{\vec{r}}} = M\dot{\vec{r}} - e\vec{A}$. The Hamiltonian then is 
\begin{align}
{H} &= \dot{\vec{r}}\cdot\vec{\pi} - L \\
&=M |\dot{\vec{r}}|^{2} - e \dot{\vec{r}}\cdot\vec{A} - \frac{1}{2} M |\dot{\vec{r}}|^{2} + e \dot{\vec{r}}\cdot\vec{A} - e\phi\\
&= \frac{1}{2} M |\dot{\vec{r}}|^{2} -e \phi
\end{align}
Which in terms of the canonical momentum is
\begin{equation}
{H} = \frac{1}{2M} |\vec{\pi} + e\vec{A}|^{2} -e\phi. \label{Hamiltonian.1}
\end{equation}

Upon quantising our system and promoting these quantities to operators the move to the quantum mechanical picture is straightforward. 

So far, we have established the Hamiltonian for a particle moving in a background electromagnetic field. We need to restrict this scenario further to a particle confined to move in two dimensions and subject to a perpendicular magnetic field. With this setup, we can choose an appropriate gauge potential that simplifies the analysis.

When introducing the quantum hall effect it is common to work with the Landau gauge with $\vec{A} = xB\hat{\jmath}$. This allows one to simplify the Hamiltonian equation to a shifted harmonic oscillator, and the wavefunctions are proportional to Hermite polynomials. For details of this approach and a nice introduction see \cite{Tong}. 

For our purposes we will instead work in what is called the symmetric gauge. The reasons for this are as follows; it is used to to describe the fractional quantum hall effect and used in the Laughlin wavefunction. Also we use complex coordinates on the sphere in the main body of this chapter and so we will have a natural extension from the QHE on the plane to the QHE on a sphere. The symmetric gauge in Cartesian coordinates is given by $\vec{A} = \frac{B}{2}\left( x\hat{\jmath} - y\hat{\imath}\right)$. However we want to employ the techniques of differential geometry and extend our analysis more freely to our investigation into the QHE on the sphere. With this in mind we will express our gauge potential using the language of differential forms.

Using complex coordinates  $z=\frac{1}{\sqrt{2}} (x+i y)$ on the flat 2-dimensional plane our gauge potential for a perpendicular magnetic field becomes

\begin{equation}
A= \frac{i B}{2} (z d \bar z - \bar z dz) 
\end{equation}

which gives

\begin{equation}
F=d A = i B\, (dz \wedge d\bar z) = B\, (d x \wedge d y).
\end{equation} 

As was previously mentioned it is common in the analyses of these systems to ignore spin as it is assumed that the strong magnetic field splits the electron spin states and only the lower energy state is relevant. We will instead include spin, in keeping with our later analysis of fermions on the sphere, and thus work with 2-component spinors on the plane. To obtain the Hamiltonian we will need to look at the spin-$\frac{1}{2}$ covariant derivatives for the Dirac operator. In our complex coordinates these are given by
\begin{align}
D_z &= \partial_z + \frac{1}{ 2}\omega_{1 \bar 1,z}\gamma^{1 \bar 1} + \frac{i e A_z}{\hbar} \\
&= \partial_z -\frac{ieB \bar{z}}{2\hbar} \\
D_{\bar z} &= \partial_{\bar z} + \frac{1}{ 2}\omega_{\bar 1 1, \bar z} \gamma^{\bar 1 1} + \frac{i e A_{\bar z}}{\hbar} \\
&= \partial_{\bar z} + \frac{ieBz}{2\hbar}
\end{align}
where the $\omega_{ij}$'s are spin connection one-forms and we choose the two dimensional gamma matrices to be $\gamma^{1} = \sigma^{1}$, $\gamma^{2} = \sigma^{2}$. The connection one-forms are zero in this instance because we are in flat space and only the gauge potential contributes. From the form of the covariant derivatives we can see how they are the operators we would arrive at for the Hamiltonian in \eqref{Hamiltonian.1}. However, as we are dealing with spinors we have these derivatives as components of the Dirac operator given by
\begin{equation} 
i\Dslash   = i(\sigma_+ D_z + \sigma_- D_{\bar z} \bigr) 
= i \begin{pmatrix}
0 & \partial_{z} - \frac{b \bar z}{2}  \\ 
\partial_{\bar{z}} +\frac{b z}{2} & 0
\end{pmatrix}.\label{eq:Dirac-Plane}
\end{equation}
where $b=\frac{ e B}{\hbar} = \frac{1}{l_B^2}$ and $l_B$ is known as the magnetic length.
The square of this operator gives
\begin{equation}
(i\Dslash)^2 = \left( -\partial_z \partial_{\bar z} -\frac{b}{2}L_3
 + \frac{b^2}{4} \bar z z\right) - \frac{b}{2} \sigma^3, \label{Dirac_Plane.1}
 \end{equation} 
and thus the Hamiltonian is given by 
\begin{equation}
H = -\frac{\hbar^2}{2 M}{\Dslash^2}
\end{equation}
with $M$ the mass of the fermion, and $L_{3} = z\partial_{z} - \bar{z}\partial_{\bar{z}}$, the orbital angular momentum operator in the direction perpendicular to the plane. This Hamiltonian is almost identical to what we would arrive at if we ignored the spin of the fermions. The contribution of spin is accounted for in our Hamiltonian by the last term in equation \eqref{Dirac_Plane.1}. 
Eigenspinors for the squared Dirac operator 
\[ \psi = \begin{pmatrix} 
\phi_+ \\ \phi_- 
\end{pmatrix} \]
satisfy
\[ \left(
\partial_z \partial_{\bar z} +\frac{b}{2}(z\partial_z - \bar z \partial_{\bar z})
\pm \frac{b}{2} - \frac{b^2}{4} \bar z z\right)\phi_\pm = -\lambda^2 \phi_\pm \]
and are given in \cite{Jellaletal} in terms of associated Laguerre polynomials $L_n^{(\alpha)}(y)$. We can construct eigenspinors using either of the forms
\begin{align*} 
\phi_\pm = z^{p_\pm} f_{\pm}(r)  \qquad &\qquad   \qquad  \qquad \hbox{(1)} \\
\phi_\pm = \bar{z}^{\bar p_\pm} g_{\pm}(r) \qquad &  \qquad  \qquad  \qquad  \hbox{(2)}  
\end{align*}
with $r= \bar z z $. We will look just at the first case of these two equations as much of the analysis is identical. With $b>0$ and $f'_\pm = \frac{ d f_\pm}{d r}$ we have
\begin{align}
\partial_z \phi_\pm & = p_\pm  z^{p_\pm -1} f_{\pm} + z^{p_\pm} \bar z f'_\pm
=z^{p_\pm -1} ( p_\pm f_\pm + r f'_\pm)\\
\partial_{\bar z} \partial_z \phi_\pm & 
=z^{p_\pm} \bigl\{r f''_\pm +(p_\pm+1)f'_\pm\bigr\}
\end{align}
and
\begin{equation} (z \partial_z - \bar z \partial_{\bar z}) \phi_\pm = p_\pm \phi_\pm
\end{equation}
giving
\begin{equation} 
r f''_\pm +(p_\pm +1) f'_\pm + \left( \frac{b}{2}(p_\pm \pm 1) - \frac{b^2 r}{4}\right) f_\pm  = -\lambda^2 f_\pm
\end{equation}
We get Laguerre's equation by writing $f_\pm = e^{-b \bar z z/2} h_\pm $ leading to
\begin{equation}
r h''_\pm +(p_\pm + 1 - b r) h'_\pm  -\frac{b}{2}(1\mp 1)h_\pm = -\lambda^2 h _\pm.
\end{equation}
In terms of the dimensionless variable $\tilde{r}= b \bar z z$
 \begin{align*} 
\tilde{r} \ddot h_+ + (p_+ + 1 - \tilde{r}) \dot h_+   & = -\widetilde \lambda^2 h _+ \\ 
\tilde{r} \ddot h_- + (p_- + 1 - \tilde{r}) \dot h_-   & = -(\widetilde \lambda^2-1 )h _- 
\end{align*}
where $\dot h_\pm = \frac{d h_\pm}{d \tilde{r}}$ and $\widetilde \lambda^2 = \frac{\lambda^2}{b}$.
These are associated Laguerre equations: the eigenvalues are 
$\widetilde \lambda^2 = n =0,1,2,\ldots$ and $h_{\pm} = L_{n_\pm}^{(p_\pm)}(y)$ are associated Laguerre polynomials of order $n_\pm$ ($n_+ = n_-+1 =n$). Obviously $p_\pm=0,1,2,\ldots$, the eigenvalues of the $L_3$ operator, are non-negative integers for the solutions to be regular at the origin.

Thus our eigenspinors have unnormalised components 
\begin{equation}
 \phi_+ \propto z^{p_+} e^{-\tilde{r}/2} L_n^{(p_+)}(\tilde{r}) \ , \qquad \phi_- \propto z^{p_-} e^{-\tilde{r}/2} L_{n-1}^{(p_-)}(\tilde{r})
 \end{equation}
which is in line with the wavefunctions mentioned previously when considering the Landau gauge; Hermite polynomials are a special case of associated Laguerre polynomials \cite{abramowitz+stegun}. The eigenvalues $\lambda^2$ can be related to the energy levels of the system as they are proportional to the eigenvalues of the Hamiltonian, $E_{n} = \frac{\hbar^2 \lambda^2}{2M} = \frac{eB\hbar}{2M} n$ and are known as Landau levels. The energy gap between the $n$ and $n+1$ states is given by $\Delta E = \frac{eB \hbar}{M}$. From this we can see why we need the samples of 2-dimensional conducting materials to be maintained at low temperatures for the QHE to be detectable. We need the average amount of thermal energy in the system to be less than that of the separation between the $n$ and $n+1$ energy levels i.e. $k_{B}T \ll \frac{eB \hbar}{M}$.

Returning to the components of our spinors, for $n=0$, $\phi_-=0$ and the zero-mode chiral spinor component is just $\phi_+= z^p e^{-\tilde{r}/2}$. For $n \ge 1$, we determine the components based on their relationship to the Dirac operator, not its square, with \(\lambda = \pm \sqrt{b n}\),

\begin{equation}
\left(\partial_{\bar z} + \frac{b}{2} z\right) \phi_+ = \mp i \lambda \phi_- = \mp i \sqrt{b n}\, \phi_-
\end{equation} 
we can take $\phi_+ = z^{p_+} e^{-\tilde{r}/2} L_n^{(p_+)}(\tilde{r})$ and 
\begin{equation}
\phi_- = \pm \frac{i}{\sqrt{ b n}}  \left( b z^{p_+ +1} \frac{d L_n^{(p_+)}}{d \tilde{r}}\right)e^{-\tilde{r}/2}. 
\end{equation}
Using the identity
\begin{equation}
\frac{d L_n^{(\alpha)}} {d \tilde{r}}=- L_{n-1}^{(\alpha+1)},
\end{equation} 
for $n\ge 1$, so
\[ \phi_- =  \mp i\sqrt{ \frac b n} z^{p_+ +1} L_{n-1}^{(p_++1)} e^{-\tilde{r}/2}
\]
and $p_ - = p_+ +1$.
Thus, for $n\ge 1$ our unnormalised eigenspinor, in terms of $z\bar{z}/ l_{B}^{2}$ is 
\begin{equation}
\psi_{n,p} = z^p e^{-z\bar{z}/2l_{B}^{2}}\begin{pmatrix}
L_n^{(p)}(\frac{z\bar{z}}{l_{B}^{2}}) \\ \frac{\mp i}{\sqrt{n}}\frac{z}{l_{B}} L_{n-1}^{(p+1)}(\frac{z\bar{z}}{l_{B}^{2}})
\end{pmatrix} \label{eq:Lzp-spinor}\end{equation}
with eigenvalue $\lambda = \sqrt{b n}$  and $p=0,1,2,\ldots$.
Note that there is no upper bound to $p$, which here characterise the degeneracy of the Landau levels in the infinite plane.

To get a better understanding of the degeneracy of these states in a finite area consider the $n=0$ state; $\phi_-=0$ and the zero-mode chiral spinors are $\phi_+= z^p e^{-z\bar{z}/2l_{B}^2}$. This component is the unnormalised solution one arrives at for the lowest Landau level for a scalar-wavefunction in the symmetric gauge \cite{Tong}. We can get an idea for the profile of the wavefunction by differentiating $|\phi_{+}|^2$ with respect to $z\bar z$. What we find is that it takes its maximum value at $z\bar{z} = pl_{B}^2$ which, in terms of the $x$ and $y$ coordinates in the plane gives
\begin{equation}
R =\sqrt{x^2 + y^ 2}= \sqrt{2p}l_{B}.
\end{equation}
Now if we consider the area $A = \pi R^2 = 2\pi pl_{B}^2 = p\frac{h}{eB}$. We find that the degeneracy of the states per Landau level, for a finite area $A$ is given by the angular momentum eigenvalue $p = \frac{eBA}{h} = \frac{\Phi}{\Phi_{0}}$, where $\Phi_{0} = h/e$ is known as the elementary magnetic flux quantum. So the number of available quantum states, $N_{B} = p$, is the ratio between that magnetic flux through the area, $\Phi= BA$, and $\Phi_{0}$. This degeneracy leads us to a quantity we have already reffered to that is central to the discussion of the QHE. The filling factor $\nu$. The filling factor for a 2-dimensional electron gas is defined as the ratio between the number of electrons in the gas and the number of available quantum states given by the degeneracy of each Landau level; $\nu = \frac{N_{e}}{N_{B}}$. We can also write the filling factor in terms of the density of electrons and density of available states per unit area, like so $\nu = \frac{n_{e}}{n_{B}}$.

Now recall the form of our transverse resistivity above in equation \eqref{Drude.1}, $\rho_{xy} = \frac{B}{n_{e}e}$. We can write this in terms of the filling fraction like so $\rho_{xy} =  \frac{B}{e n_{B}\nu}$. As $n_{B}=\frac{N_{B}}{A}=\frac{eB}{h}$ we have the relationship between the transverse resistivity and the filling factor given by
\begin{equation}
\rho_{xy} =  \frac{h}{e^2}\frac{1}{\nu}.
\end{equation}
This is why the resistivity in graph \ref{IQHE} presented are in units of $\frac{h}{e^2}$, quite natural units to use. The filling factor here, along with the energy gap $\Delta E$, gives a heuristic explanation as to why we see quantised plateaus at integer filling factors. As the magnetic field increases, the gap between Landau levels increases as \(\Delta E \propto B\). Electrons occupy states up to the Fermi energy. For \(\nu = 2\), which corresponds to \(\rho_{xy} = \frac{h}{2e^2}\), the lowest two Landau levels are fully occupied, while higher Landau levels remain energetically out of reach for the electrons. With further increases in magnetic field strength, the filling factor \(\nu\) decreases, forcing electrons into fewer occupied Landau levels. At sufficiently high magnetic fields, only the lowest Landau level is filled, which we see as \(\nu = 1\) in figure \(\ref{IQHE}\), corresponding to \(\rho_{xy} = \frac{h}{e^2}\).

Note that in this explanation we are not expanding on the role that disorder plays in these systems. The above figure \ref{IQHE} is for a sample of a significantly disordered system. A proposed explanation from \cite{Laughlin} as to why these plateaus exist, along with the existence of the Landau levels is also due to Anderson localisation \cite{Anderson}. Anderson localisation is a form quantum interference in strongly disordered metals that can cause the system to become insulating. In the context of the Landau levels discussed above, the disorder of the systems effects the degeneracy of the unperturbed system by causing the states to broaden around the Landau levels, i.e nearest the unperturbed Landau level the states remain extended but moving further away the states become localised. For a neat summary of this effect see \cite{GirvYang}. A natural question then is to consider what happens when our samples of conducting material are particularly clean, i.e. of low disorder.

As many scientists know, more detailed experiments often raise more questions. In 1982 the fractional quantum Hall (FQHE) effect was discovered by the authors of \cite{Tsui1982} for filling fraction $\nu = 1/3$. More fractional states were since discovered and yet a complete analytic solution for the fractional quantum hall effect still evades us. 
\begin{figure}[h!]
\centerline{\includegraphics[width=0.7\linewidth]{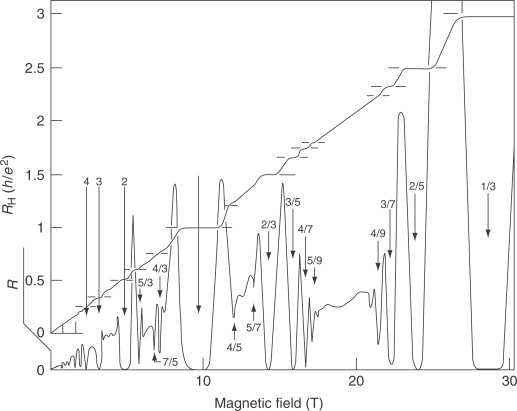}}
\caption{Image from the press release of the 1998 Nobel prize, awarded to Tsui, Stormer and Laughlin for the discovery and explanation of the FQHE.}
\end{figure}
We have qualitative descriptions from the trial wavefunction first written down by Laughlin. As the disorder in the quantum hall system is decreased we find that fractional states become visible and the integer quantum hall plateaux become less prominent. Interestingly we find better agreement with the classical hall Drude model discussed previously. These fractional hall states are those with filling factor  $\nu = p/q$ with $p,q \in \mathbb{Z^{+}}$. The reason this occurs, whereby we end up with more pronounced plateaux and more intervals instead of fewer and a blurring of the hall effect is proposed to be due to the Coulomb interaction between the electrons. This working understanding is what inspired Laughlin to write down the ansatz in complex coordinates for a wave function for the FQHE that incorporates this Coulomb interaction and thus gives rise to fractional states of odd denominator \cite{Laughlin}. 
The Laughlin trial ground-state wavefunction is given as
\begin{equation}
\psi(z_{1},...,z_{N}) = \prod^{N}_{i<j}(z_{i} - z_{j})^{2k+1}e^{-\frac{1}{2l_{b}^2}\sum_{i=1}^{N} \bar{z}_{i}z_{i}} \label{Laugh.1}
\end{equation}
for $N$ particles, with coordinates in the complex plane $z_{i}$ and $k \in \mathbf{Z}^{+}$. Note that the power of the polynomials in $z_{i}$ is odd. This is so that the wavefunction is anti-symmetric and thus obeys fermi-dirac statistics. Now recall from the IQHE, we had that the ground state for the scalar wavefunction, which would be the $N = 1$ case for \eqref{Laugh.1}, had a "radius" associated with its wavefunction given by its angular momentum $R=\sqrt{2p}l_{B}$. If we focus our attention on a single particle, with coordinate $z_{1}$ for example, we can see that from the form of the wavefunction that its highest power, and thus maximum angular momentum, will be $(2k+1)(N-1)$. Following along a similar calculation from above, with the area $A = \pi R^2 = 2\pi l_{B}^2(2k+1)(N-1)$ our degeneracy is given by $N_{B} = (2k+1)(N-1)$ and the filling factor 
\begin{equation}
\nu = \frac{N}{N_{B}} = \frac{N}{(2k+1)(N-1)}
\end{equation} 
which for $N \rightarrow \infty$ gives our fractional filling factor with odd denominator
\begin{equation}
\nu = \frac{1}{(2k+1)}
\end{equation} 
Haldane extended this analysis to explain the hierarchy of fractional states that appear in \cite{Haldane} and used the symmetry of the sphere to show that states followed a continued fraction pattern of states of odd denominator. He showed that the filling factors were given by the continued fraction
\begin{equation}
 \nu = [2k+1,p_{1},...,p_{n}] 
\end{equation}
where $p_{i}=2,4,6,...$. We use the continued fraction notation here as in \cite{Haldane}. The wave function Haldane uses, in our complex coordinates, is given by
\begin{equation} \psi(z_{1},...,z_{N})= \left(\prod_{i=1}^{N} \frac{1}{(1+ z_i \bar z_i)^{\frac{|m| -1}{2}}}\right)
\prod_{i<j}^{N} (z_i - z_j)^{2 k+1}.\label{eq:Laughlin.1}\end{equation}
with $m$ here the magnetic monopole charge. We arrive at this expression in \S \ref{sec:Vortices} via considering flux attachments bound to each of the $N$ particles in question, these are known as composite-fermions. This approach was first presented by J.K. Jain in \cite{Jain}. Jain’s composite fermion picture introduces vortex fluxes of a statistical gauge
field attached to the charge carriers. In our analysis of the FQHE on a sphere in \S \ref{sec:Vortices} the vortices have the effect of opposing the background magnetic field supplied by the magnetic monopole encompassed by the sphere, similar to Jain's original approach. The mathematical details of these vortices can be found in appendix (1.1).

\subsection{Magnetic Monopoles and The Index Theorem}

The theory of magnetic monopoles is rich and has deep historical ties to the development of electromagnetism. Pierre Curie considered seriously their existence in 1887 \cite{Curie} and noted that an ``isolated sphere in space, charged with free magnetism'' would have ``magnetic fields all oriented along the radii and all directed outward'' \cite{Curie}. Dirac used the existence of a magnetic monopole to explain the quantisation of electric charge \cite{Dirac1931}, though they have eluded experimental detection. Dirac posited the existence of monopoles via a thought experiment; by first imagining a solenoid which is made infinitesimally thin and where one end is sent infinitely far away. Focusing on the end not at infinity, the magnetic field produced by the solenoid extends radially from the center where the solenoid ends. There is a singular point where the infinitesimally thin solenoid would puncture a sphere enclosing the monopole. This solenoid is known as a Dirac string.
\begin{figure}[h!]
\centerline{\includegraphics[width=0.7\linewidth]{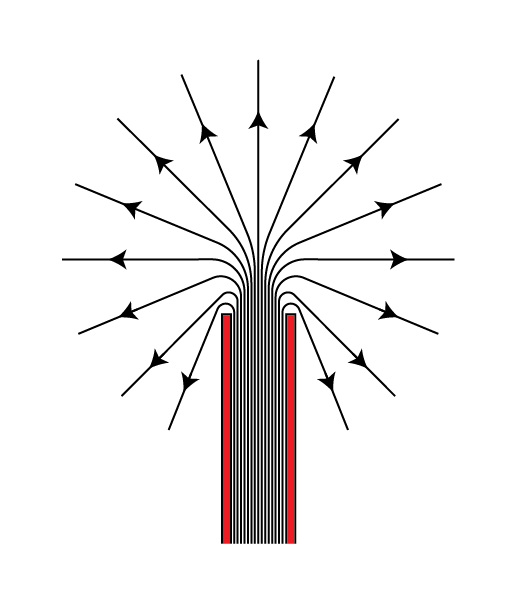}}
\caption{Exaggerated illustration of the Dirac string. Image taken from open-source code sharing site Zenodo \cite{zenodo}}
\end{figure}
\newpage
Considering this along with the quantization of angular momentum Dirac arrived at the quantization condition
\begin{equation}
\frac{em}{2\pi\hbar} = N \label{dirac.quant}
\end{equation}
where $e,m$ are the electric and magnetic charges respectively and $N \in \mathbb{Z}$. The full details of the calculation can be found in \cite{Dirac1931} and a neat derivation can be found in \cite{DolanHallog}.
This expression also contains within it the expression we had previously for the quantum of magnetic flux $\Phi_{0} = h/e $ and $m = N\Phi_{0}$. While this physical intuition is helpful, the Dirac string itself is not physical but rather a gauge artefact; a magnetic monopole can indeed be constructed without Dirac strings. To achieve this, we adopt an alternative formulation of the magnetic monopole. Given our focus on spinor fields on a sphere, we follow the approach of Wu and Yang in \cite{Wu+Yang}, which avoids Dirac strings and eliminates singular points on the sphere.

The 2-sphere is a manifold without boundary that cannot be covered with a single coordinate chart. With this in mind we consider potentials defined for the northern and southern hemispheres. Expressing the potentials in terms of differential forms, in spherical coordinates this looks like
 \begin{equation}
A^{(\pm)}=\frac{m}{2} (\pm 1-\cos\theta) d \phi
\end{equation}
where the gauge transformation between the two potentials is given by \newline
$A^{(+)} = A^{(-)} + m d \phi$. The electromagnetic tensor, or Maxwell 2-form, is then
\begin{equation}
F=dA=\frac{m}{2}\sin\theta d\theta \wedge d\phi.
\end{equation}
Integrating this 2-form over the unit sphere we have
\begin{equation}
\frac{1}{2\pi} \int_{S^2} F = m \label{mono.1}
\end{equation}
as we would expect for a monopole. For a $U(1)$ spinor bundle the potential $A^{\pm}$ plays a crucial role as part of the connection on the sphere. In this context, the Maxwell 2-form is associated with the curvature 2-form $\Omega = idA = iF$ for the $U(1)$ bundle. This relationship is key in the physical interpretation of the Atiyah-Singer index theorem for our system.

The Atiyah-Singer index theorem, proved by Michael Atiyah and Isadore Singer in 1963 \cite{Atiyah}, is a generalisation of earlier results in differential geometry and algebraic topology. The well known Gauss-Bonnet theorem for example is a special case. The central idea of the Atiyah-Singer index theorem, for a manifold without boundary, is about drawing a relationship between a topological invariant of a smooth compact manifold $M$ to an invariant of a differential operator $D$, known as a Fredholm operator, which acts on sections of a smooth vector bundle $V$ on $M$. The invariants on either side of the relation are known as their index. The index of the operator $D$, sometimes called the analytic index, is given by
\begin{equation}
\text{ind}(D) := \text{dim(ker}(D)) - \text{dim(ker}(D^{*}))
\end{equation}
For our Dirac operator on a sphere, which is a map of sections of the $U(1)$ spinor bundle (the spinor fields) to itself, the above index is given by the difference between the dimension of the space of spinor fields that get mapped to zero by the Dirac operator and the dimension of the space that gets mapped to zero by its adjoint. From a physical perspective what we have is that the analytic index is given by the difference between the number of positive and negative chirality zero modes of our system
\begin{equation}
\text{ind}(D) = n_{+} - n_{-}.
\end{equation}

The topological side is significantly more involved, the details of which can be found in \cite{Eguchi, NashSen}, but it suffices to say that for a 2-dimensional compact manifold without boundary with a spin structure the topological index is given by
\begin{equation}
 \int_{M} c_{1}(V) = \frac{i}{2\pi}\int_{M} \text{Tr} \ \Omega
\end{equation}
where $c_{1}(V)$ is the first Chern form of a vector bundle $V$ on $M$ and $\Omega$ is the curvature 2-form. For a sphere equipped with a spin structure, with a monopole charge at the center and a principal $U(1)$ bundle we have
\begin{equation}
\frac{i}{2\pi}\int_{S_{2}} \text{Tr} \ \Omega = -\frac{1}{2\pi} \int_{S^2} F = - m.
\end{equation}
where we have used the fact that $\Omega = iF$. Thus our topological index is equal to $-m$. The Atiyah-Singer index theorem tells us that the topological index and analytic index are equal and thus we arrive at the result stated in the introduction \eqref{index.1} 
\begin{equation}
n_+ - n_- =-m.
\end{equation}
The number of positive and negative chirality zero modes is constrained by the strength of the monopole charge.

The Atiyah-Singer index theorem has far reaching consequences in both mathematics and physics, well beyond what has been presented here. For more of the mathematical details on this topic see \cite{Shanahan} and for texts geared towards a theoretical physics perspective see \cite{Eguchi, NashSen}. For our purposes though this result is sufficient and we can to proceed to the main body of this chapter.
\newpage

\section{The Dirac operator on a sphere \label{sec:Dirac}}


\subsection{The Hamiltonian}


The full spectrum and eigenfunctions of the Dirac operator 
on a sphere in the absence of a magnetic monopole were studied in 
\cite{Abrikosov}.  On a magnetic sphere the spectrum can be derived from group theory \cite{Dolan}.  The eigenstates can be expressed simply in terms of Jacobi polynomials which were found to describe spinless particles on a magnetic sphere by Wu and Yang \cite{Wu+Yang}.

First consider a single non-relativistic spin-$\frac{1}{2}$ charged particle of mass $M$ confined to move on the surface of a sphere with a magnetic monopole at the centre of the sphere. The Hamiltonian is
\[ H = -\frac{\hbar^2}{2 M}{\Dslash^2}\]
where $i \Dslash$ is the (Hermitian) Dirac operator in the presence of the monopole and the sphere has unit radius.
This is bounded below and if there are zero modes of the Dirac operator they must be ground states with vanishing zero point energy.

The gauge potential for a monopole at the centre of the sphere is taken to be
\begin{equation*}A^{(\pm)}=\frac{m}{2} (\pm 1-\cos\theta) d \phi \qquad \Rightarrow \qquad
F=dA=\frac{m}{2}\sin\theta d\theta \wedge d\phi\end{equation*}
(the upper (lower) sign is for the upper (lower) hemisphere).
The monopole charge is
\[ \frac{1}{2\pi} \int_{S^2} F = m\]
with $m$ an integer, as mentioned previously.

We shall use a complex co-ordinate on $S^2$, where for the northern hemisphere we have
\begin{equation*}
z=\tan\left(\frac{\theta}{2}\right) e^{i\phi},
\end{equation*}
in terms of which the metric is given by 
\begin{equation}
ds^2 =  \frac{4 d z d\bar{z}}{(1 + z \bar{z})^2}
\end{equation}
and the potential is
\begin{equation}
  A^{(+)}(z) = \frac{i m}{2} \frac{(zd\bar{z} - \bar{z}dz)}{(1+ z \bar z)},
  \qquad
A^{(-)}(z) = \frac{i m}{2} \frac{1}{(1+ z \bar z)}
  \left( \frac{d z}{z} - \frac{d \bar z}{\bar z} \right)\label{eq:NS-gauge-transformation}
\end{equation}
and
\begin{equation}
  F = i m\frac{dz \wedge d\bar{z}}{(1 + z \bar z)^2}.
  \end{equation}

Complex connection one-forms are obtained from the metric by first defining orthonormal one-forms like so
\begin{equation} 
d s^2 = \frac{1}{2} \left( e^1 \otimes e^{2} + e^{2} \otimes e^1 \right) = \frac{4 d z d\bar{z}}{(1 + z \bar{z})^2} 
\end{equation}
which gives
\begin{equation} 
e^1 = \frac{\sqrt{2}dz}{1 + z\bar{z}}, \qquad e^{2} = \frac{\sqrt{2}d\bar{z}}{1 + z\bar{z}}.
\end{equation}
Indices are raised and lowered with
\begin{equation*}\delta_{1 2} \qquad \hbox{and} \qquad \delta^{1 2}.\end{equation*}
We can then use Cartan's first structure equation 
\begin{equation}
de^{a} = -\omega^{a}_{b} \wedge e^{b}
\end{equation}
to determine the connection one-forms
\begin{align*}
d e^1 + \omega^{1 2} \wedge e_{2} &= - \frac{\sqrt{2} z d \bar z \wedge d z}{(1 + z \bar z)^2} + \omega^{12}\wedge e^1=0\\
\Rightarrow \qquad \omega^{1 2} &= \frac{z d \bar z }{ (1+ z \bar z)} + f dz,\\
\omega^{2  1} &= \frac{\bar z d z }{ (1+ z \bar z)} + g d \bar z,
\end{align*}
where $f(z,\bar z)$ and $g(z,\bar z)$ are functions of our complex variable. It is consistent to take $f=-\bar{z}/(1+z\bar{z})$ and $g= -z/(1+z\bar{z})$ which effectively leaves us with one connection one-form
\begin{equation}\omega^{1 2}= \frac{z d \bar z - \bar z d z}{ (1+ z \bar z)}\end{equation}
with $\omega^{12} = -\omega^{21} = -\omega_{12}$.

\subsection{The Dirac operator}

Choosing $\gamma^1 = \sigma^1$, $\gamma^2=\sigma^2$ the Dirac operator on the unit sphere is
\begin{equation} 
-i\Dslash = -i(1+z \bar z) \bigl(\sigma_+ D_z + \sigma_- D_{\bar z} \bigr)\label{eq:Dirac0},
\end{equation}
with spin-$\frac{1}{2}$ co-variant derivatives, defined on the northern hemisphere (for electric charge $e=-1$, $\hbar =1$), given by
\begin{align}D_z &= \partial_z - \frac{i}{ 2}\omega_{1 2,z}\gamma^{1 2} + i A_z \nonumber \\
& = \partial_z + \frac{1}{ 2}\frac{\bar z \sigma^3}{ (1+ z \bar z)} + \frac{m}{ 2} \frac{\bar z }{ (1+ z \bar z)} 
= \partial_z + \frac{(m+\sigma^3)}{ 2} \frac{\bar z }{ (1+ z \bar z)},\label{eq:Dz}\\
D_{\bar z} &= \partial_{\bar z} - \frac{i}{ 2}\omega_{2 1, \bar z} \gamma^{2 1} + i A_{\bar z}\nonumber \\
                 & = \partial_{\bar z} - \frac{1}{ 2} \frac{z \sigma^3}{ (1+ z \bar z)} -\frac{m}{2}\frac{z}{(1+ z \bar z)} 
= \partial_{\bar z} - \frac{(m+\sigma^3)}{ 2} \frac{z }{ (1+ z \bar z)},\label{eq:Dbarz}
\end{align}
where 
\begin{equation} 
\gamma^{1 2} = \frac{1}{2} (\gamma^1 \gamma^{2} - \gamma^{2} \gamma^1) = i\sigma^3
\end{equation}
and it is clear $\sigma^3$ arises in the co-variant derivatives as part of the spin connection on the sphere.
More explicitly
\begin{equation}
\Dslash = \begin{pmatrix}0 & (1 + z \bar z)\partial_{z} + \frac{(m-1)}{ 2} \bar z  \\ 
(1 + z \bar z)\partial_{\bar z} - \frac{(m+1)}{ 2} z  & 0
\end{pmatrix},\label{eq:D-slash}
\end{equation}
which is anti-hermitian. The curvature associated with the co-variant derivatives is
\[ [D_z,D_{\bar z}]= -\frac{(m+\sigma^3)}{(1+ z \bar z)^2}.\] 
The spin connection can be viewed as effectively increasing the magnetic charge by one for positive chirality spinors and decreasing it by one for negative chirality spinors.


\subsection{Angular Momentum}

The energy eigenstates can be classified by additional quantum numbers, in particular angular momentum will be a good quantum number but the definition involves some subtleties.
There are two aspects to the discussion of angular momentum: the presence of the
magnetic field and the orthonormal frame necessary to define spinors.
The magnetic field can be accommodated by defining
\[ L_a =  \epsilon_{a b}{}^c x^b ( p_c +  A_c)
= -i \epsilon_{a b}{}^c x^b ( \partial_c + i   A_c),\]
but the algebra does not close, rather
\begin{equation}  
[L_a , L_b]=i \epsilon_{a b c}  \bigl( L_c  + e   x_c({\bf r}.{\bm B})\bigr).
\label{eq:JJB}
\end{equation}
In particular for a monopole field
\[ [L_a,L_b] = i \epsilon_{a b}{}^c \left(L_c - \frac{ m x_c}{2 r}\right),\]
but this can be countered by defining \cite{Fierz}
\[ J_a = L_a - \frac{ m x_a}{2 r},\]
giving a closed algebra
\begin{equation}
  [J_a,J_b] = i \epsilon_{a b}{}^c J_c.\label{eq:SU2-algebra}\end{equation} 
In terms of $z$,
\begin{align}
J_+ & = z^2  \partial_z + \partial_{\bar z} + \frac{m z}{2},\nonumber\\
J_- & = -{\bar z}^2  \partial_{\bar z} - \partial_z + \frac{m \bar{z}}{2},
\label{eq:Jm-def}\\
J_3 & = z \partial_z - \bar{z} \partial_{\bar z} + \frac{m}{2}.\nonumber
\end{align}
But this is not sufficient for spinors, Lie derivatives will also drag the orthonormal frame.
In the absence of any magnetic field the Lie derivative of a spinor $\psi$ with respect to a vector field $\vec L$ can be defined as \cite{SpinorLie}
\begin{equation}
 L^i D_i \psi + \frac 1 4 (d L)_{i j}\gamma^{i j}\psi \label{eq:Lie-psi}\end{equation}
where $\gamma^{i j} = \frac 1 2 (\gamma^i \gamma^j - \gamma^j \gamma^i)$, which in our context we have $\gamma^{i}=\sigma^{i}$, and $d L$ is the exterior derivative of the 1-form that is metric dual to the vector $\vec L$.
In terms of $z$,
\begin{align*} 
d L_+ &  = \frac{4 z}{(1 + z \bar z)^3} d z \wedge d \bar z\\ 
d L_- &  = - \frac{4 \bar z}{(1 + z \bar z)^3} d z \wedge d \bar z\\
d L_ 3 & =2 \frac{(1- z \bar z)}{(1+ z \bar z)^3} d z \wedge d \bar z 
\end{align*}
The prescriptions (\ref{eq:Jm-def}) and (\ref{eq:Lie-psi}) can be combined to give the Lie derivative of
a spinor in the presence of a magnetic monopole at the centre of the unit sphere 
in the following way
\begin{align}
{\bm J}_+ & = z^2 D_z + D_{\bar z} +\frac{(m+ \sigma^3) z}{1+ z \bar z} = z^2  \partial_z + \partial_{\bar z} + \frac{(m+ \sigma^3) z}{2},\nonumber\\
{\bm J}_- & = -{\bar z}^2  D_{\bar z} - D_z + \frac{(m+ \sigma^3)\bar{z}}{1+ z \bar z}=-{\bar z}^2  \partial_{\bar z} - \partial_z + \frac{(m+ \sigma^3)\bar{z}}{2},
\label{eq:J-def}\\
{\bm J}_3 & = z D_z - \bar{z} D_{\bar z} + \frac{(m + \sigma^3)}{2}\left(\frac{1- z \bar z}{1 + z \bar z}\right)
= z \partial_z - \bar{z} \partial_{\bar z} + \frac{(m + \sigma^3)}{2}\nonumber
\end{align}
on the northern hemisphere.
These satisfy
\begin{align} [{\bm J}_+,{\bm J}_-]&=2 {\bm J}_3, \qquad [{\bm J}_3,{\bm J}_\pm]= \pm {\bm J}_\pm\label{eq:SU2}\\
[{\bm J}_3,D_z] & =-D_z, \qquad [{\bm J}_3,D_{\bar z}]=D_{\bar z}, \nonumber\\
[{\bm J}_3,\Dslash] & =0,\qquad [{\bm J}^2,\Dslash]=0.\nonumber\end{align}
The square of the Dirac operator is related to the quadratic Casimir ${\bm J}^2$,
\[ -\Dslash^2 = {\bm J}^2 - \frac 1 4 (m^2 -1).\]
The eigenvalues of the square of the Dirac operator on a coset space can be calculated from group theory by expressing them in terms of quadratic Casimirs, details of which can be found here \cite{Dolan}. For the sphere
$S^2 \approx SU(2)/U(1)$, with a monopole at the centre, 
\[ \lambda^2 = n(n+|m|) \]
with degeneracy 
\begin{equation}
2 n + |m|
\end{equation}
where $n$ is a non-negative integer. Thus, with ${\bm J}^2 = J(J+1){\bm 1}$,
\[ J+\frac 1 2 = n + \frac {|m|}{2}.\]
There are zero-modes when $m\ne 0$ but when there is no background field
$n$ cannot vanish, in accordance with the Lichnerowicz theorem \cite{Lichnerowitz}.


\section{Zero modes \label{sec:ZeroModes}}


Spinors can be decomposed in terms of chiral eigenstates
\begin{equation*}\Psi = \begin{pmatrix}\chi_{+} \\ \chi_{-}\end{pmatrix}.\end{equation*}
Positive and negative chirality zero modes satisfy
\begin{align} D_{\bar z} \chi_+ &=0\\
D_{z} \chi_- &=0\end{align}
respectively.  From (\ref{eq:Dz}) and (\ref{eq:Dbarz}) it is immediate that,
on the northern hemisphere,
\begin{align}
\chi_+ & = {z^p}(1+ \bar z z)^{\frac{m+1}{2}}\label{eq:chi+p}\\
  \chi_- & = {{\bar z}^{\bar p}}(1+ \bar z z)^{-\frac{m-1}{2}} \label{eq:chi-barp}
\end{align}
satisfy these equations for any powers $p$ and ${\bar p}$.
However $p$ and $\bar p$ must be non-negative integers for $\chi_\pm$ to be well behaved at the north pole.
We also want $\chi_\pm$ to be finite at the south pole, where
$|z|\rightarrow \infty$, so we must also require that 
 $\bar p - m +1 \le 0$ and $p + m +1 \le 0$.\footnote{At the upper limit of these bounds the magnitude of $\chi_\pm$ is finite but
the phase is undefined, this is a gauge artifact. A well defined phase is obtained at the south pole by performing the gauge transformation $\chi_\pm \rightarrow e^{i(m  \pm 1)\phi}\chi_\pm$ (the $\pm1$ arises from the spin connection).\label{footnote:phase}}
Thus, since $p$ and $\bar p$ are non-negative, positive chirality zero modes require  $0\le p \le  -m-1$ and negative chirality zero modes require $0 \le \bar p \le m -1$.
We see that for a positive chirality zero mode to exist it must be the case that $m\le - 1$ while a negative chirality zero mode
requires $m\ge 1$.  For no value of $m$ are there both positive and negative zero modes. The index theorem then tells us that
\[ n_+ - n_- = -m \qquad \Rightarrow \qquad n_\pm = \mp m.\]
Thus for $m\ge 1$, $\bar p = 0,\ldots m-1$ exhausts the possibilities and for $m\le -1$, $p = 0,\ldots |m|-1$ exhausts the
possibilities. Thus, the index theorem, together with the explicit construction of $\chi_{\pm}$, accounts for all possible modes.

The index theorem tells us that, provided the geometry and background magnetic field do not allow for simultaneous positive and negative zero modes, the number of zero modes here is $|m|$. This differs from Haldane's result that the degeneracy is $|m|+1$ and the difference is called the shift \cite{Wen+Zee}. In the multi-particle wave-function (discussed below) 
the shift is the difference between the number of flux quanta and the number of particles and it is non-zero in Haldane's analysis precisely because the electron spin and its coupling to the curvature of the sphere is ignored.  When electron spin and the spin connection on the sphere are treated properly the shift is zero and this is clearly shown here, it can be traced to 
the $(m+1)$ and $(m-1)$ terms in (\ref{eq:D-slash}), electrons with opposite spin couple to the spin connection with the opposite sign.  

The most general (un-normalised) zero modes are linear combinations of (\ref{eq:chi+p}) and (\ref{eq:chi-barp}) with constant co-efficients,
\begin{align}
\chi_+ & = \sum_{p =0}^{|m|-1} \frac{ a_{p} z^{p}}{(1 + \bar z z)^{\frac{|m|-1}{2}}}, \qquad \hbox{for}\ m\le -1,\label{eq:single-particle-chi+}\\
  \chi_- & = \sum_{\bar p =0}^{m-1} \frac{ a_{\bar p} {\bar z}^{\bar p}}{(1 + \bar z z)^{\frac{m-1}{2}}}, \qquad \hbox{for}\ m\ge 1.\label{eq:single-particle-chi-}
\end{align}

We shall analyze the $m < 0$ case (for positive $m$ simply complex conjugate the ground state wave functions).
The single particle ground state (\ref{eq:single-particle-chi+}) has degeneracy $|m|$,
which is a consequence of the index theorem.

The quantum Hall effect is a many particle phenomenon. Suppose we have $N$ identical particles on the sphere
and denote their co-ordinates by $z_i$, $i=1,\ldots,N$.
Ignoring interactions between the particles, the total ground state has zero energy and again consists of zero modes, but now for the zero mode associated with particle $i$ the co-efficients $a_p$ or $a_{\bar p}$ can be polynomials of the other $N-1$ co-ordinates.
The most general multi-particle ground state is
\begin{equation}
 \chi_+(z_1,\cdots,z_N)  = \left(\prod_{i=1}^N\frac{1}{(1 + \bar z_i z_i)^{\frac{|m|-1}{2}}}\right)\sum_{p_1, \ldots, p_N =0}^{|m|-1} 
                          a_{ p_1 \ldots p_N} {z_1}^{p_1} \cdots {z_N}^{p_N}.
\end{equation}
Since the particles are fermions the wave-function should be anti-symmetric, so
$a_{ p_1 \ldots p_N} $ should be anti-symmetric in its indices.  This requires $N\le |m|$ and leads to a degeneracy $\frac{|m|!}{N!(|m|-N!)}$. If $N>|m|$ then all the particles cannot fit into the ground state and some must go into the second Landau level. If $N< |m|$ the
ground state is degenerate and cannot be expected to be stable under perturbations. There is a unique ground state, stable under small perturbations, if and only if $N=|m|$ in which case 
\begin{equation}
  \chi_+(z_1,\cdots,z_N)  = \left(\prod_{i=1}\frac{1}{(1 + \bar z_i z_i)^{\frac{|m|-1}{2}}}\right)\prod_{i<j} (z_i - z_j).\label{eq:Haldane}
\end{equation}
which is equation \eqref{eq:Laughlin.1}, for $k=0$. Thus there is a unique stable ground state if and only if the filling factor
\[\nu = \frac{N}{|m|}=1. \]
These are ground-state wave functions, the spherical versions of the Laughlin wave-functions on the plane for the integer quantum Hall effect.
For a sphere of radius $R$, and reinstating $\hbar$, the energy gap is
\[ \Delta E = \frac{(|m|+1)\hbar^2}{2 M R^2}=\frac{(N+1)\hbar^2}{2 M R^2}.\]
In the planar limit, $R\rightarrow \infty$, $N\rightarrow\infty$, keeping the
particle density $\rho = \frac{N}{4\pi R^2}$ finite, the energy gap
is 
\begin{equation}
  \Delta E = \frac{2 \pi  \rho \hbar^2}{M}= \frac{e B \hbar}{M},\label{eq:DeltaE}
\end{equation}
where $\frac{e B }{h} = \frac{|m|}{4 \pi R^2}$.


\section{Higher Landau levels \label{sec:HigherLL}}


When $N>|m|$, some particles must go into higher Landau levels. The energy
eigenfunctions in the higher Landau levels can be described by Jacobi polynomials,
$P^{(\alpha,\beta)}_n(\cos\theta)$. For spinless particles, the appearance of Jacobi polynomials was first found by the authors of \cite{Wu+Yang}. For fermions, Jacobi polynomials also appear in \cite{Villalba}, but not in the context of Landau levels. While our approach differs in using complex coordinates on the sphere and a different initial ansatz our results are consistent with theirs.
To outline the calculation we begin with a generic eigenspinor
\begin{equation}
\psi = \begin{pmatrix} \phi_1 \\ \phi_2 \end{pmatrix}
\end{equation}
operating on this with the Dirac operator from equation \eqref{eq:D-slash} results in a set of coupled partial differential equations
\begin{align}
(1 + z \bar z)\partial_z \phi_2 + \frac{(m-1)}{ 2} \bar z \phi_2 &= -i\lambda \phi_1 \label{d1}\\
(1 + z \bar z)\partial_{\bar z} \phi_1 - \frac{(m+1)}{ 2} z \phi_1 &= -i\lambda \phi_2. \label{d2}
\end{align}
Squaring the Dirac operator will decouple these
\begin{align*}
&{\Dslash}^2 \psi = -{\lambda}^2 \psi \\
&{(1+z \bar z) \begin{pmatrix} 0 & D_z \\ D_{\bar z} & 0 \end{pmatrix}} {(1+z \bar z) \begin{pmatrix} D_z \phi_2 \\ D_{\bar z} \phi_1 \end{pmatrix}} = -{\lambda}^2 \begin{pmatrix} \phi_1 \\ \phi_2 \end{pmatrix} \\
&(1+z \bar z)\begin{pmatrix} \bar z D_{\bar z} \phi_1 + (1+z \bar z) D_z D_{\bar z} \phi_1 \\ z D_z \phi_2 + (1+z \bar z) D_{\bar z} D_z \phi_2 \end{pmatrix} = -{\lambda}^2 \begin{pmatrix} \phi_1 \\ \phi_2 \end{pmatrix} \\
\end{align*}
giving two second-order partial differential equations
\begin{align}
&(1+z\bar z)^2 {\partial_{z} \partial_{\bar z} \phi_1} + a (1+z \bar z)(\bar z\partial_{\bar z} - z\partial_z) \phi_1 \nonumber \\
&+({\lambda}^2 -a(1+z\bar z) -a(a-1)z\bar z)\phi_1 = 0 \\
&(1+z\bar z)^2 {\partial_{z} \partial_{\bar z} \phi_2} + (1-a) (1+z \bar z)(z\partial_z - \bar z\partial_{\bar z}) \phi_2 \nonumber \\
&+({\lambda}^2 -(1-a)(1+z\bar z) -a(a-1)z\bar z)\phi_2 = 0
\end{align}
Where $a= \frac{(1+m)}{2}$ and $(1-a) = \frac {1-m}{2}$. Focusing on the first equation above (as the analysis for the second equation is identical) we can simplify it by proposing the ansatz
\begin{equation}\phi_1 = \frac {z^p \bar z^{\bar p}}{(1+z\bar z)^q} f(z\bar z).\end{equation}
and introducing a change of variables $z\bar{z} = \frac{u}{1-u}$. Substituting these in we arrive at 
\begin{align}
&u(1-u)\ddot{f} +\Big[(p+{\bar p}+1)-2u(q+1)\Big]\dot{f} \nonumber \\&+ \frac{1}{1-u}\Big[\frac{p{\bar p}}{u} + u((q-a)(q+a)-{\bar p}(q-a)-p(q+a))\Big]f \nonumber\\ &+({\lambda}^2+a({\bar p}-p-1)-q(p+{\bar p}+1))f=0
\end{align}
where derivatives with respect to $u$ are denoted with dots ($\dot{f}$). Introducing the constraints
\begin{align}
&p{\bar p} =0 \\
&(q-a)(q+a)-{\bar p}(q-a)-p(q+a) = 0
\end{align}
with $p$ and $\bar{p}$ elements of $\mathbb{Z}$. This gives four possibilities
\begin{align}
\bar p &= 0, \quad q= p + a \quad \text{ or } \quad q= -a\\
p &= 0, \quad q= {\bar p} - a \quad \text{ or }\quad q= a,
\end{align}
leading to four hypergeometric differential equations
\begin{align}
&u(1-u)\ddot{f} +\big(p+1-u(1-m)\big)\dot{f}+({\lambda}^2)f=0\\
&u(1-u)\ddot{f} +\big(p+1-u(2p+m+2+1)\big)\dot{f}+({\lambda}^2-(m+1+p)(p+1))f=0\\
&u(1-u)\ddot{f} +\big(\bar{p}+1-u(m+2+1)\big)\dot{f}+({\lambda}^2-m-1)f=0\\
&u(1-u)\ddot{f} +\big(\bar{p}+1-u(2{\bar p}-m+1)\big)\dot{f}+({\lambda}^2-{\bar p}({\bar p}-m))f=0.
\end{align}
The solutions for these equations, which have general parameters $a,b,c$, i.e.
\begin{equation*}u(1-u)\ddot{f} +\big(d-u(b+c+1)\big)\dot{f}-(bc)f=0, \\\end{equation*}are of the form 
\begin{equation}F(b,c;d;u) = \sum_{n=o}^{\infty} \frac{{(b)_n}{(c)_n}}{(d)_n}{\frac{u^n}{n!}}.\label{Hyp.1} \end{equation}
Our solutions for each case are then
\begin{align}
&F_{1}(b,c;d;u) = F(\frac{-m}{2}+\delta,\frac{-m}{2}-\delta;p+1,u)\\
&F_{2}(b,c;d;u) = F(p+\frac{m}{2}+1+\delta,p+\frac{m}{2}+1-\delta;p+1,u)\\
&F_{3}(b,c;d;u) = F(\frac{m}{2}+1+\delta, \frac{m}{2}+1-\delta;\bar{p}+1,u)\\
&F_{4}(b,c;d;u) = F(\bar{p}-\frac{m}{2}+\delta,\bar{p}-\frac{m}{2}-\delta;\bar{p}+1,u).
\end{align}
with $\delta = \sqrt{\frac{m^2}{4}+\lambda^2}$. From here we follow along similar lines to \cite{Villalba} and truncate the hypergeometric functions above by setting the parameter $c = -n$ with $n \in \mathbb{Z}^{+}$. This brings us to Jacobi Polynomials of the form \begin{equation}F(\alpha+1+\beta+n, -n; \ \alpha+1; u)= \frac{n!}{(\alpha+1)_n}P^{(\alpha,\beta)}_{n}(1-2u) \end{equation}
which for our four solutions above we have
\begin{align}
&F_1(n-m,-n;1+p;u), \ &\lambda_{n} = \pm \sqrt{(n)(n-m)} \\
&F_2(2p+2+n+m,-n;1+p;u),\ &\lambda_{n} = \pm \sqrt{(n+1+p)(n+1+p+m)}\\
&F_3(n+m+2,-n;1+\bar{p};u), \ &\lambda_{n} = \pm \sqrt{(n+1)(n+1+m)}\\
&F_4(n+2\bar{p}-m,-n;1+\bar{p};u), \ &\lambda_{n} = \pm \sqrt{(n+\bar{p})(n + \bar{p}-m)}
\end{align}
along with their eigenvalues. We can shift the integer $n$ in the above equations
and re-express $F_1 - F_4$, and their associated eigenvalues, as 
\begin{align}
&F_1(n-m,-n;1+p;u),&\qquad \lambda = n(n-m)\\
&F_2(n+p+1+m,p+1-n;1+p;u),&\qquad \lambda = n(n+m)\\
&F_3(n+m+1,-n+1;1+\bar{p};u),&\qquad \lambda = n(n+m)\\
&F_4(n+\bar{p}-m,\bar{p}-n;1+\bar{p};u),&\qquad \lambda = n(n-m).
\end{align}
Similar solutions can be found for $\phi_{2}$ by starting from an identical ansatz 
\begin{equation}\phi_2 = \frac {z^s \bar z^{\bar s}}{(1+z\bar z)^r} f(z\bar z).\end{equation}
Similar to the approach used in \cite{Villalba} we find that there are redundancies given in the above set of solutions such that all we need for both spinor components are
\begin{itemize}
\item For $m>0$, with $p=-\bar p$ and $s=-\bar s$:
\begin{align}
\phi_1 & \propto e^{-i \bar{p} \phi} u^{\frac {\bar p} {2}} (1-u)^{\frac{m+1 - {\bar p}}{2}} P_{n-1}^{(\bar p,m+1-\bar p)}(1-2 u), \quad -n+1 \le \bar p \le n+m ,\label{eq:phi_1+}\\
\phi_2 & \propto e^{-i \bar{s} \phi} u^{\frac {\bar s} {2}} (1-u)^{\frac{m-1 - {\bar s}}{2}} P_n^{(\bar s,m-1-\bar s)}(1-2 u), \quad -n \le \bar s \le n+m-1.\label{eq:phi_2+}
\end{align}
\item For $m<0$ with $\bar p = -p$ and $\bar s = -s$:
\begin{align}
\phi_1 & \propto e^{i  p \phi} u^{\frac {p} {2}} (1-u)^{\frac{|m|-1-p}{2}} P_{n}^{(p,|m|-1-p)}(1-2 u), \quad -n \le  p \le n+|m|-1,\label{eq:-phi_1-}\\
\phi_2 & \propto e^{i  s \phi} u^{\frac {s} {2}} (1-u)^{\frac{|m|+1-s}{2}} P_{n-1}^{(s,|m|+1-s)}(1-2 u), \quad -n+1 \le  s \le n+|m|.\label{eq:-phi_2-}
\end{align}
\end{itemize}
The eigenspinors of $-i\Dslash$, with  $\lambda_n = \pm \sqrt{n(n+|m|)}$, $n\ge 1$, and the relative normalization of the components worked out, are perhaps best exhibited using polar co-ordinates $(\theta, \phi)$ (on the northern hemisphere)
\beq
\psi_{\lambda_n,\alpha} = {\cal N}_{n,\alpha,\beta}
 \begin{pmatrix}
 z^\alpha \left(\cos \frac {\theta}{2} \right)^{|m|-1} P_n^{(\alpha,\beta)}(\cos\theta)\\ 
\mp i \bigl(\sqrt{\frac{n+|m|}{n}}\,\bigr) z^{\alpha+1} \left(\cos \frac{\theta}{2}\right)^{|m|+1}  P_{n-1}^{(\alpha+1,\beta+1)}(\cos\theta)
\end{pmatrix}\label{eq:sphere-eigenspinors}
\eeq
where $\alpha =-n,\ldots n+|m|$ labels the $2n+|m|$ independent degenerate states, $\beta$ is fixed by $\alpha+\beta = |m|-1$ and
\[{\cal N}_{n,\alpha,\beta}^2 =\frac{(2 n + |m|)}{8 \pi}\frac{\Gamma(n+1)\Gamma(n+|m|+1)}{\Gamma(n+\alpha+1)\Gamma(n+\beta+1)}\]
  a normalisation constant.\footnote{The eigenspinors are associated with irreducible representations of $SU(2)$ and are also expressible as Wigner $d$-functions.
  The full degenerate set for a given $n$ constitute a single column of the matrix in the $2 n+|m|$ dimensional irreducible representation of SU(2).}

The higher Landau levels now have both chiralities at the same energy level, but these can be separated by adding a Zeeman splitting
term $\mu |m| ({\bf 1} - \sigma_3)$ to the Hamiltonian.

The second Landau level corresponds to $n=1$ and has degeneracy $|m|+2$.
By the same argument as before the anti-symmetrised ground state multi-particle wave function is degenerate unless
\[ N= |m| + (|m|+2) =2(|m|+1) ,\] in which case the filling factor is
\[\nu = \frac{N}{|m|} = \frac{2 N}{N-2}\
  \rightarrow \ 2 \quad \mbox{as} \quad N \rightarrow \infty. \]
The resulting wave-function is non-degenerate, it is stable under perturbations and the energy gap between the  second and third Landau levels is
\[ \Delta E = \frac{\hbar^2}{2 M R^2}\bigl( 2(|m|+2) - (|m| +1)\bigr)
  = \frac{(|m|+3)\hbar^2}{2 M R^2}  = \frac{(N+4)\hbar^2}{4 M R^2}.\]

Repeating the argument for larger, but finite, $n$ we recover the integer quantum Hall effect in the limit of large $N$. There is a unique stable ground state
when the $n$-th Landau level is fully filled
\[ N=\sum_{k=0}^n (2 k + |m|)= (n+1)(n+|m|)\]
so
\[ \nu = \frac{N(n+1)}{N-n(n+1)} \ \rightarrow\ n+1  \quad \mbox{as} \quad N \rightarrow \infty. \]
The energy gap between level $n$ and level $n+1$
is
\[\Delta E  = \frac{\hbar^2}{2 M R^2}\bigl( (n+1)(|m|+n+1) - n(|m| +n)\bigr)
  =  \frac{(|m| + 2 n +1)\hbar^2}{2 M R^2}
=  \frac{\bigl(N -n(n+1)\bigr)\hbar^2}{2 (n+1)M R^2}.\]
In the planar limit
\[ \Delta E \ \rightarrow\  \frac{|m| \hbar^2 }{2 M R^2} = \frac{ e B \hbar}{M},\]
as expected.


\section{Fractional filling fractions \label{sec:Vortices}}


Fractional filling fractions in the quantum Hall effect can often be understood in terms of flux attachment \cite{Jain}. A statistical gauge field is introduced
and the effective degrees of freedom are composite objects consisting of
electrons bound to statistical gauge field vortices.
These vortices then reduce the effective field seen by the composite fermions.

The gauge potential describing a uniform flux through the sphere arising from a monopole with charge $m$ at the centre of the sphere together
with $N-1$ vortices each of strength $v$ piercing the sphere at points $z_j$ is described in the appendix, (\ref{eq:A+}) and (\ref{eq:A-}). The gauge potential is 
\begin{align} 
A^{(+)}  & = \frac{v}{2 i} \sum_{j=1}^{N-1}\left( \frac{d z }{z-z_j}
-  \frac{d \bar z}{\bar z-\bar z_j}\right)
+ \frac{i m}{2}\left(\frac{z d \bar z - \bar z d z}{1+z \bar z}\right),\label{eq:A+a}\\
A^{(-)} 
&= \frac{v}{2 i} \sum_{j=1}^{N-1}\left( \frac{d z }{z-z_j}
-  \frac{d \bar z}{\bar z-\bar z_j}\right)
+ \frac{i}{2} \left( \frac{m}{(1+ z \bar z)}  +  (N-1) v \right) 
  \left( \frac{d z}{z} - \frac{d \bar z}{\bar z} \right).
\end{align}
Using the identities 
\begin{equation} \partial_z\left(\frac{1}{\bar z -\bar z_i}\right)= \partial_{\bar z}\left(\frac{1}{z - z_i}\right)= 2\pi \delta(z - z_i) \label{iden.2} \end{equation}
the field strength is
\[ F = i\left(2 \pi v \sum_{j=1}^{N-1} \delta(z-z_j) + 
\frac{m}{2 (1+z \bar z)^2}\right) dz \wedge d \bar z.\]

The spectrum of the Dirac operator can be determined when there are magnetic vortices threading through the surface of the sphere in addition to a monopole at the centre. 
Omitting the self-energy of a composite fermion with its own vortex, and assuming all the vortices have the same strength $v_i=v$, the Dirac operator (\ref{eq:Dirac0}) on the northern hemisphere then involves covariant derivatives
\begin{align}
D_z & = \partial_z
+ \frac v 2 \sum_{j=1}^{N-1}\frac{1}{z - z_j}
+ \frac{( m +\sigma_3)}{2} \frac{\bar z}{1+z \bar z }
,\label{eq:Dzv}\\
    D_{\bar z} 
& = \partial_{\bar z}
- \frac v 2 \sum_{j=1}^{N-1}\frac{1}{\bar z - \bar z_j}
- \frac{( m +\sigma_3)}{2} \frac{ z}{1+z \bar z }.\label{eq:Dbarzv}
\end{align}
Again using the identities from equation \eqref{iden.2} the commutator is
\[ [D_z,D_{\bar z}] = -2\pi  v \sum_{j=1}^{N-1} \delta(z-z_j) -\frac{(m + \sigma_3)}{(1+ z \bar z)^2}. \]
The index over the whole sphere, including the points associated with the vortices, is
\begin{equation}
 n_+ - n_- = -\frac{1}{2\pi}\int_{S^2} F  = -\bigl[ m +(N-1)v \bigr].\label{eq:index-m}\end{equation}
If the points representing the vortices are excluded the index over the sphere minus $N-1$ points is
\begin{equation}
 n_+ - n_- = -\frac{1}{2\pi}\int_{S^2 - (N-1)\ \hbox{\small points}} F = - m,\label{eq:index-mtilde}\end{equation}

Zero modes of (\ref{eq:Dzv}) are 
\begin{equation}\chi_-=  \frac{ \bar z^{\bar p} }{(1+z \bar z)^{\frac {m - 1} 2}}
  \prod_{j=1}^{N-1}\frac{(z - z_j)^{-\frac v 2}}{(\bar z - \bar z_j)^{\bar l}},
\label{eq:chilbar}\end{equation}
with $-\frac{v}{2}\ge \bar l$ for regularity at $z_j$.\footnote{When $\bar l \le 0$ this is immediate,
when $\bar l>0$ we invoke 
\[ D_ z \chi_- = 
- 2 \pi \bar l \left[\sum_{j=1}^{N-1}(\bar z - \bar z_j)\delta(\bar z - \bar z_j) \right]\chi_- =0.\]}
Similarly zero modes of (\ref{eq:Dbarzv})
are
\begin{equation}\chi_+= z^p (1+z \bar z)^{\frac {m + 1} 2}
  \prod_{j=1}^{N-1}\frac{(\bar z - \bar z_j)^{\frac v 2}}{(z - z_j)^l}.\label{eq:chil}
\end{equation}

However (\ref{eq:chilbar}) are not all linearly independent,
one could take linear combinations with $\bar z_i$ dependent co-efficients to construct a numerator that has
powers of $\bar z-\bar z_i$ which change $\bar l$, $\bar l$ and $\bar p$ are not independent in
(\ref{eq:chilbar}). Similarly $l$ and $p$
are not independent in (\ref{eq:chil}).  We seek a criterion for constraining
$l$ and $\bar l$ and we shall explore this by looking at the transformation properties under rotations. Of course $SU(2)$ is no longer a symmetry when there are vortices present, but we can still ask how the wave functions (\ref{eq:chilbar}) and (\ref{eq:chil}) change under rotations.

In the presence of vortices the spinor Lie derivatives introduced before (\ref{eq:J-def}) are modified to
\begin{align}
 {\bm J}_+   & = z^2 \partial_z + \partial_{\bar z} + 
\frac v 2 \sum_{j=1}^{N-1} \left( \frac{z^2}{z - z_j}  +  \frac{1}{\bar z - \bar z_j}\right)
        +\frac{(m+ \sigma_3)z}{2}
        \nonumber \\
 {\bm J}_-   & = -{\bar z}^2 \partial_{\bar z} - \partial_z + \frac v 2 \sum_{j=1}^{N-1}\left( \frac{\bar z^2}{\bar z - \bar z_j}  +  \frac{1}{z - z_j}\right)
 + \frac{(m + \sigma_3) \bar z}{2} \label{eq:J-def-vortices} \\  
{\bm J}_3  &= z \partial_z - \bar z \partial_{\bar z}+
 \frac v 2 \sum_{j=1}^{N-1}\left( \frac{z}{z - z_j}  +  \frac{\bar z}{\bar z - \bar z_j}\right)       
+ \frac{(m +\sigma_3)}{2}
        \nonumber
\end{align} 
on the northern hemisphere.
These generate $SU(2)$ at any point on the sphere away from the vortices, but not at the vortices themselves ---
at the vortices there will be delta function singularities that prevent the algebra from closing. 
The algebra is well defined and closes on the
sphere with $N-1$ points removed.

For ${\bm J}_3$
\begin{equation}
  [{\bm J}_3,D_z] = -D_z - 2 \pi v \bar z \sum_{j=1}^{N-1} \delta(z-z_j), 
\qquad [{\bm J}_3,D_{\bar z}]  = D_{\bar z} - 2 \pi v z \sum_{j=1}^{N-1}\delta(z-z_j).\end{equation}
This implies that ${\bm J}_3$ commutes with the Dirac operator on the sphere with $N-1$ points removed.
A short calculation gives
\begin{equation}
{\bm J}_3 \chi_+ =\left(p +\left(\frac v 2 - l\right) \sum_{j=1}^{N-1} \frac{z}{z-z_j} + 2 \pi l \bar z \sum_{j=1}^{N-1}(z-z_j)\delta (z-z_j)  +\frac{m +1}{2}\right)\chi_+,\label{eq:L_3chi_+}
\end{equation}
if $v$ and $l$ are both positive.
Choosing $l=\frac{v}{2}$ results in 
\[
\chi_+ = z^p (1+ z \bar z)^{\frac{m+1}{2}}
\prod_{j=1}^{N-1}\left( \frac{\bar z - \bar z_j}{z - z_j}\right)^{\frac v 2}\]
with
\[
{\bm J}_3 \chi_+ = \left(p + \pi v \bar z \sum_{j=1}^{N-1} (z-z_{j})\delta(z-z_j) + \frac{m + 1}{2}\right) \chi_+
\]
and $\chi_+$ is an eigenfunction of ${\bf J}_3$ on the punctured sphere with the $N-1$ points removed.

We restrict $p$ to be a non-negative integer, so as to render $\chi_+$ well behaved\footnote{By well behaved we mean that it is finite and, apart from the overall factor of $1/(1+ z \bar z)^{\frac{(|m|-1)}{2}}$, it is a product of a function analytic in $z$
and a function analytic in $\bar z$.} at $z=0$,
and take $m <0$ with $p \le |m|-1$ so that $\chi_+$ is well behaved as $|z|\rightarrow \infty$. 
Then
\begin{equation} \chi_+ = \frac{z^p}{(1+ z \bar z)^{\frac{|m|-1}{2}}}
\prod_{j=1}^{N-1}\left( \frac{\bar z - \bar z_j}{z - z_j}\right)^{\frac v 2}.\label{eq:chi+singular}\end{equation}
Singularities at the points $z_j$ are evident here as the phase of (\ref{eq:chi+singular}) is undefined there when $v\ne 0$.
So the points with the vortices have to be excised from the sphere and the index on the punctured sphere is given by (\ref{eq:index-mtilde}).  There are no normalisable negative chirality zero modes when $m$ is negative, as can be checked explicitly,
so $n_+ = |m|$.
For $\chi_-$ the analysis is similar, except $m >0$ and (\ref{eq:chi+singular}) is complex conjugated.  

Excising points and using (\ref{eq:chi+singular}) for the zero modes may seem natural
from a mathematical point of view but physically it is not satisfactory.
In the flux attachment picture each of the vortices is attached to a particle
and we wish to include all the particles in the dynamics, we do not want to 
remove these points.  We can avoid excising points yet still satisfy the index theorem by choosing $l=-\frac{v}{2}$.  Now
\begin{equation} \chi_+ = z^p ( 1 + z \bar z)^{\frac{m+1}{2}}
\prod_{j=1}^{N-1}(\bar z - \bar z_j)^{\frac v 2}(z - z_j)^{\frac v 2}\label{eq:chi+m}\end{equation}
is well behaved for
$p\ge 0$, $v=2 k \ge  0$, where $k$ is an integer,  and
\begin{equation} p + m  + 1 + 2(N-1) k \le 0 \qquad \Rightarrow \qquad
0 \le p \le -(m  + 1 + 2(N-1) k)=-m'-1,\end{equation}
where $m' = m + 2 k (N-1)$.
Thus $m' \le -1$ for positive chirality zero modes (there are no normalisable negative chirality zero modes for negative $m'$). 
The index is now (\ref{eq:index-m}) and $n_+ = |m'|$. With $0\le p \le |m'|-1$ equation (\ref{eq:chi+m})
then is a complete set for the zero modes,
though they are not eigenstates of ${\bm J}_3$.
$m'$ is the effective magnetic charge the
composite fermions see, since $m'$ and $m$ are both negative $|m'|<|m|$ and the composite fermions move in a weaker field than that generated by the monopole $m$, a consequence of the vortices is that the composite Fermions effectively move in a weakened background field.

The net result is that, if we do not wish to excise the vortices from the sphere, then the number of zero modes for negative $m'$ is $|m'|$ and
\begin{equation}
 \chi_+ = \frac{z^p}{( 1 + z \bar z)^{\frac{|m|-1}{2}}} \prod_{j=1}^{N-1}|z - z_j|^{2 k},\label{eq:chi+m-k}\end{equation}
with $v=2 k>0$ and $m = m' - 2(N-1) k <0$.  The vortices necessarily have even charge and act to oppose 
the background monopole field, thus reducing the effective magnetic field that
the composite fermions see.\footnote{Again a similar analysis for $\chi_-$ changes the sign of $m$, and $m'$ with $v=2 k$ and complex conjugates (\ref{eq:chi+m-k}).}  

A general zero mode is a linear combination,
\begin{equation}
 \chi_+ = \frac{1}{( 1 + z \bar z)^{\frac{|m|-1}{2}}} 
\prod_{j=1}^{N-1}|z - z_j|^{2 k}\sum_{p=0}^{|m'|-1}  a_p z^p ,
\label{eq:chi_+}
\end{equation} 
In the flux attachment picture each of the vortices is attached to a particle.
With $N$ particles the antisymmetrised many-particle wave function is
\[ \chi_+(z_1,\ldots,z_N) = \left(\prod_{i=1}^N \frac{1}{(1+ z_i \bar z_i)^{\frac{|m| -1}{2}}}\right)
\left(\prod_{i<j}^N |z_i - z_j|^{2 k}\right)
\sum_{p_1,\ldots,p_N=0}^{|m'|-1}
a_{p_1 \ldots p_N} z_1^{p_1} \cdots z_N^{p_N},\]
where $a_{p_1 \ldots p_N}$ is anti-symmetric.  The ground state is unique if and only
if $a_{p_1 \ldots a_N}$ is unique (up to an overall constant) and this requires
$|m'|=N$ with $a_{p_1 \ldots a_N} \propto  \epsilon_{p_1 \ldots a_N}$. The unique (un-normalised) ground state is
\begin{equation} 
\chi_+(z_1,\ldots,z_N) = \left(\prod_{i=1}^{N} \frac{1}{(1+ z_i \bar z_i)^{\frac{|m| -1}{2}}}\right)
\prod_{i<j}^{N} |z_i - z_j|^{2 k}(z_i - z_j).\label{eq:fractional-ground-state}
\end{equation}
This is the ground state for a system of non-interacting composite fermions each consisting of an electron bound to a vortex 
of strength $2k$ and 
subject to a background field consisting of a magnetic monopole of charge $m$.  
Wave functions of this form on a plane, and hence with a different geometrical factor, were considered by \cite{G-J} and studied numerically in \cite{FFMS}.

There is an energy gap as before  
and the filling factor is
\[ \nu = \frac{N}{|m|} = \frac{N}{N + 2 k(N-1)}\quad \rightarrow \quad \frac{1}{2 k+1} \qquad \hbox{as} \quad N\rightarrow \infty.\]
The system therefore describes the Laughlin series of the fractional quantum Hall effect.

The vortices can be removed by a singular gauge transformation,
\[\chi_+ \rightarrow e^{-i\Phi}\chi_+, \qquad A \rightarrow A + i e^{i \Phi} d e^{-i\Phi}\]
where the phase $\Phi$ is
\[ \Phi =
  \frac {v} {2 i}  \sum_{i<j}^N\ln
  \left(\frac{\bar z_i - \bar z_j}{z_i - z_j}\right).\]
With $v = 2 k$, the ground state $\chi_+$ (\ref{eq:fractional-ground-state}) gauge transforms to
\begin{equation} \chi_+ \ \rightarrow \ \widetilde \chi_+= \left(\prod_{i=1}^{N} \frac{1}{(1+ z_i \bar z_i)^{\frac{|m| -1}{2}}}\right)
\prod_{i<j}^{N} (z_i - z_j)^{2 k+1}.\label{eq:Laughlin}\end{equation}
This is Haldane's ground state for the quantum effect on a sphere, \cite{Haldane}, 
apart from the geometrical factor $\prod_{i=1}^{N} (1+ z_i \bar z_i)^{-\left(\frac{|m| -1}{2}\right)}$ it
is the Laughlin ground state \cite{Laughlin}.
This is a zero mode for $N$ electrons in a background gauge field for a monopole of charge $m <0$ with
potential
\[  \widetilde A^{(+)} = \frac{i m}{2}\left(\frac{z d \bar z - \bar z d z}{1+z \bar z}\right),
\qquad \widetilde A^{(-)}=\frac{i m}{2} \frac{1}{(1+ z \bar z)}  \left( \frac{d z}{z} - \frac{d \bar z}{\bar z} \right),\]
but it is not unique.  
The most general zero mode for this configuration is
\[ \widetilde \chi_+(z_1,\ldots,z_N) = \left(\prod_{i=1}^N \frac{1}{(1+ z_i \bar z_i)^{\frac{|m| -1}{2}}}\right)
\sum_{p_1,\ldots,p_N=0}^{|m|-1}
a_{[p_1 \ldots p_N]} z_1^{p_1} \cdots z_N^{p_N},\]
The degeneracy is determined by the number of components of the anti-symmetric co-efficients 
$a_{[p_1 \ldots p_N]}$, but now regularity at the S pole only requires  $0\le p \le |m|-1$,
so the degeneracy is 
\[\frac{|m|!}{N!(|m| - N)!} = \frac{\bigl((2 k+1\bigr)N - 2k)!}{ N!\bigl(2 k(N-1)\bigr)!},\]
which diverges exponentially as $N\rightarrow\infty$, for any positive $k$. 
The energy gap is lost and one cannot expect the ground state (\ref{eq:Laughlin}) to be stable under
perturbations.  The introduction of the vortices changes (\ref{eq:Laughlin}) to (\ref{eq:fractional-ground-state}) and stabilizes 
the ground state, it is a singular gauge transformation and so can change the physics.


\section { Final Remarks \label{sec:Conclusions}}


Haldane's description of the quantum Hall effect on a sphere has been developed in the context of Fermions on a compact space, allowing the Atiyah-Singer index theorem to be utilised in analysing the ground state of the Hamiltonian which necessarily requires zero modes. Electron wave-functions are cross sections of the $U(1)$ bundle associated with the monopole.
For a single electron in the field of a magnetic monopole of charge $m$ (in magnetic units with $\frac {e^2} {h}=1$) the number of zero-modes, and hence the degeneracy of the ground state, is limited by the index theorem to $|m|$. For a system of $N$ particles Fermi statistics then gives the unique ground state (\ref{eq:Haldane}) if and only if $N=|m|$
and the  filling factor is $\nu =1$.
The uniqueness, and hence stability, of the Haldane ground state wave function  for the integer quantum Hall effect (which is the same as the Laughlin ground state function except for a geometric factor) is seen to be a consequence of the index theorem which limits the dimension of the space of zero modes.

The fractional quantum Hall effect can be studied in the composite Fermion scenario by viewing a monopole of charge $m$ to be $|m|$ individual monopoles of charge $\pm 1$ and promoting some of the Dirac strings associated with these monopoles to be statistical gauge field vortices which bind to electrons, forming composite Fermions. The vortices necessarily reduce the magnitude of the background magnetic field seen by the composite fermions and the index theorem again dictates that the degeneracy of the ground state is finite.  Apart from the usual geometrical factor on the sphere the ground state wave-function is a product of a holomorphic and an anti-holomorphic field 
if and only if the vortices are of strength $2 k$ with $k$ an integer.
The ground state wave-function for a system consisting of $N$ composite fermions (\ref{eq:fractional-ground-state}) is then unique if and only if the filling factor is $\frac{1}{2 k+1}$, for large $N$. Removing the vortices by use of a singular gauge transformation then gives the Laughlin ground state for the fractional quantum Hall effect in the Laughlin series, again apart from a geometrical factor.

It would be interesting to apply similar techniques to the higher dimensional quantum Hall effect \cite{Zhang+Hu, Karabali+Nair} in which $S^2$ is replaced by $S^4$ and the spinors are cross-sections of an $SU(2)$ bundle, but we leave that to future work.



\chapter{Holographic Duality with a Dyonic Reissner-Nordstr\"om Black Hole}
\label{chp3_ADS-CFT_for_Reissner_Nordstrom_BH}

\section{Introduction}

The proposition of the Holographic principle by 't Hooft in \cite{Hooft}, and its formulation in string theory by Susskind \cite{Susskind}, has led to an abundance of research into theories that utilise the principle to tackle otherwise mathematically intractable problems using geometric techniques. The most shining example of which has been the Anti de Sitter-conformal field theory correspondence (AdS/CFT), first proposed by Maldacena in \cite{Maldacena}. In the two and half decades since Maldacena's conjecture of the correspondence a flood of research has followed and the applications have expanded to a broad range of topics. A comprehensive introduction to the general topic of this Gauge/Gravity duality can be found in \cite{Ammon2015}. The success of this correspondence in giving an avenue to examine strongly coupled conformal field theories has motivated the extension of these techniques to condensed matter systems. 

As yet there is no analytic proof of the AdS/CFT correspondence and thus the broader topic of Gauge/Gravity duality does not admit a rigorous way in which to apply its techniques to condensed matter systems. There is however a collection of prescriptive approaches which have been devised. In this chapter we will focus primarily on those methods formulated in \cite{Son, Iqbal+Liu, Iqbal+Liu2} and their applications by Liu et al. in \cite{McGreevyetal} where they used the correspondence to look for signatures of non-Fermi liquids.

Specifically we consider an asymptotically AdS space with a charged black hole, a Reissner-Nordstr\"om (RN-AdS$_{4}$) black hole with a dyonic charge, i.e. electric and magnetic charges. This is an extension of the work done in \cite{McGreevyetal} but with the modification that we use a spherical event horizon, as opposed to a planar one, and due to the inclusion of a magentic monopole we utilise the eigen-functions and eigen-values found in chapter \ref{Chpt_2} in our analysis here. In \cite{McGreevyetal} the authors mainly consider zero temperature systems for the boundary quantum field theory of their paper. We are instead interested in looking at non-zero temperature and specifically looking at the temperature for an AdS black hole where it undergoes a phase transition, i.e. where the heat capacity diverges. This is akin to what occurs in the Hawking-Page phase transition \cite{HawkPage}, though details differ in that Hawking and Page consider an AdS black hole in the presence of a photon gas which acts to stabilise the black hole as it undergoes the phase transition while we consider only the phase transition of the black hole in a vacuum.

Work on fermionic bulk fields in RN-AdS$_{4}$ to examine holographic condensed matter systems has been carried out in \cite{Ren} where the authors include a spin-orbit coupling and examine the Green's functions for the boundary theory at nonzero temperatures and densities, and find Rashba like dispersion relations but do not examine the behaviour of the boundary theory at the phase transition temperature. In \cite{Dias} the authors use the AdS/CFT correspondence to look at fermionic instabilities for RN-AdS$_{4}$ black-holes and find that their are no linear mode instabilities for fermionic fields in this space, but again their work does not include a magnetic charge nor examine the behaviour of the Green's functions for the boundary theory at the aforementioned temperature.

We replicate the results found in \cite{McGreevyetal} and then find resonance behaviour for the same system of quasiparticle peaks with our spherical event horizon. We then consider non-zero temperature, specifically looking at the values of the Green's functions at the phase transition temperature $T= T_{p}$ of the black hole. The behaviour of the Green's functions at this temperature should be indicative of a phase transition in the dual theory. We present numerical results which we believe are consistent with this understanding.

This chapter then will proceed as follows. Firstly we will give a broad introduction to the general topic of holographic duality, drawing a thread between the concepts on which it relies in \S \ref{Holog}. This will give us some intuition as to why the prescriptions we will utilise throughout the remainder of this chapter are reasonable. In \S \ref{Hol.Q.M} we will look at the applications of this approach to condensed matter systems, what is known as holographic quantum matter, and discuss the relevance of the phase transition for asymptotically AdS black holes. We will review the prescription from \cite{Iqbal+Liu} and its applications to non-Fermi liquids in \cite{McGreevyetal} \S \ref{Holo.Spinors}. In \S \ref{Results} we will present our findings.

\newpage
\section{Background to Holographic Duality} \label{Holog}

\subsection{Duality}
The concept of duality exists across mathematics and physics. Though its specific usage depends on the context there is an essential character of physical theories being dual to one another that is neatly outlined in \cite{Caste}. The particular type of duality we are concerned with is that of strongly coupled theories dual to weakly coupled ones.

An elementary example of a duality in a physical model is the relationship between the electric and magnetic fields in a vacuum. This duality arises from a symmetry of the source free Maxwell's equations
\begin{align}
\nabla \cdot \mathbf{E} &= 0 \ , \quad \nabla \times \mathbf{E}= -\frac{\partial \mathbf{B}}{\partial t} \\
\nabla \cdot \mathbf{B} &= 0 \ , \quad \nabla \times \mathbf{B}= \mu_{0}\epsilon_{0}\frac{\partial \mathbf{E}}{\partial t}.
\end{align}
The symmetry is straightforward, sending $\mathbf{E} \rightarrow \mathbf{B}$ and $\mathbf{B} \rightarrow -\mu_{0}\epsilon_{0} \mathbf{E}$ leaves Maxwell's equations invariant. This symmetry reflects a duality of the theory, whereby the electric and magnetic fields are dual to one another in this model. There is a more general form of this duality which was derived by \cite{GibRash} and reveals even further the symmetry in this theory , a concise introduction to this can be found here \cite{BaynDolan}. If a source charge for the electric field is reintroduced this symmetry is broken, unless we also introduce a magnetic charge. As was covered in chapter \ref{Chpt_2} introducing a magnetic monopole leads to Dirac's quantization condition
\begin{equation}
\frac{qm}{2\pi\hbar} = N \label{dirac.quant}
\end{equation}
where $q,m$ are the electric and magnetic charges respectively and $N \in \mathbb{Z}$. From this equation one can see \eqref{dirac.quant} that the strength of the magnetic charge is inversely proportional to the strength of the electric charge. In this way the strong coupling of the magnetic field is "dual" to the weak coupling of the electric field and vice versa. This is the most straight forward example of a duality in a theory that displays this connection between dual fields and their coupling strengths.

There are many systems that exhibit duality between weakly coupled and strongly coupled theories. One of the first examples was discovered by Kramers and Wannier in 1941 \cite{Kramer}. This duality exists between the free energy of a two-dimensional Ising model at low temperature and one at high temperature. The temperature here plays the role of the coupling in the theory and the high and low temperatures are the weak and strong couplings respectively. This duality is explicit and there is an analytic transformation to go from one theory to the other \cite{Kramer}. 

Another strong-weak duality can be found between the massive Thirring model and the sine-Gordon scalar theory. This was first discovered by Coleman \cite{Coleman} between a fermionic field in one spatial dimension and a scalar field from the sine-Gordon equation. His method of calculation involved showing that the Green's functions for both theories gave equivalent spectra. There is a concise derivation of this duality in \cite{Juricic} which shows explicitly how the Lagrangians are equivalent and how the couplings display the strong-weak duality. This particular duality sheds light on the process of bosonization of fermionic particles in 1+1 dimensions. Note there is not a dimensional reduction here, nor in the previous two examples. Thus these dualities are of a different character to the AdS/CFT correspondence, which is still a strong-weak coupling duality, but also relies on the CFT being one dimension lower than the AdS space it is dual to.

The duality at the heart of the AdS/CFT correspondence, and the first realisation of the holographic principle in \cite{Maldacena}, is the relationship between a type IIB string theory in AdS$_{5} \times \mathbf{S}^{5}$ and a $\mathcal{N} = 4$, SU($N$) super-symmetric Yang-Mills theory in four space-time dimensions, where $\mathcal{N}$ is the number of independent supersymmetries and $N$ the degree of the symmetry for the gauge fields. This is the dual CFT on the boundary of the string theory in the background AdS$_5$ space. The relevant coupling parameters for this duality on the AdS side are $l_{s}$ and $L$ , the string length and radius of curvature, and on the CFT side is the 't Hooft coupling $\lambda = g^{2}_{YM} N$ and $N$. These parameters are related via 
\begin{equation}
\left(\frac{L}{l_s}\right)^4 \propto \lambda, \quad  \left(\frac{L}{l_p}\right)^4 \propto N
\end{equation}
where $l_{p} = \sqrt{\frac{\hbar G}{c^3}}$ is the Planck length. The strongest statement we have of the AdS/CFT correspondence is that the generating functional of the field theory is equivalent to the partition function for the string theory on the boundary of the AdS space
\begin{equation}
Z_{\text{CFT}_4} = Z_{\text{AdS}_5}. \label{ADSCFT}
\end{equation}
A proof of this duality would require a complete non-perturbative description of quantized string theory in a curved space, which we do not have. Instead we must consider the limit as $\lambda \rightarrow \infty$ and $N \rightarrow \infty$ where the strings are weakly coupled and the AdS side of the duality is effectively a classical gravity theory. Equation \eqref{ADSCFT} can then be given by 
\begin{equation}
Z_{\text{CFT}_4}[{\bf{A}}] \approx e^{iS_{\text{AdS}}} \label{ADSCFT1}
\end{equation} 
Where $\bf{A}$ here are source fields for the CFT, that will depend on the behaviour of the dual bulk fields on the boundary of the AdS space, and $S_{\text{AdS}}$ is the classical gravity action. This correspondence, following a similar line of reasoning for the specific case of AdS$_5$/CFT$_4$ generalises to AdS$_{d+1}$/CFT$_{d}$. A more precise statement can be given, for example, for a bulk field $\phi(x)$ that on the boundary takes on the value $\phi_{0}(x)$, which acts as a source for an operator $\mathcal{O}$ in the dual theory,  
\begin{equation}
\left\langle \exp\left[ \int d^dx \ \phi_0 (x) \mathcal{O} (x) \right]\right\rangle_{\text{CFT}} = e^{iS_{\text{AdS}}[\phi|_{\partial \mathcal{M}}= \phi_0]} \label{ADSCFT2}
\end{equation} 
with $\mathcal{M}$ the AdS$_{d+1}$ space. An outline of this equivalence can be found in \cite{Gubser}, \cite{Witten}, where the role of bulk-boundary correspondence is explored in more depth. Equation \eqref{ADSCFT2}, along with the analogous expression for a fermionic field (discussed in \S \ref{Holo.Spinors}), allows us to compute correlation functions for strongly coupled field theories in $d$ dimensions on the boundary of a $d+1$ dimensional space time.
For a detailed introduction to these arguments and the evidence for its veracity see \cite{Ammon2015}.

The correspondence between AdS in $d+1$ dimensions and a CFT in $d$ dimensions is a realisation of the holographic principle. In the next section then we will explore the details of this principle.

\subsection{The Holographic Principle}\label{HoloPrinc} 

The holographic principle is a direct consequence of the work done on black hole thermodynamics. As was mentioned in the first chapter of this thesis, one of the successful marriages of quantum field theory (QFT) and general relativity (GR) has been within the regime of the microscopic understanding of black hole thermodynamics. From chapter \ref{Chpt_1} we discussed that the temperature of a black hole is given by its Hawking temperature. For a metric in $d$ dimensions of the form 
\begin{equation}
ds^2 = -f(r) dt^2 + \frac{dr^2}{f(r)} + r^2 d\Omega^{2}_{d-2}
\end{equation} 
with an event horizon at $r=r_{h}$ the temperature of the black hole is given by 
\begin{equation}
T_{H} = \frac{|f'(r_{h})|}{4\pi}. \label{Hawk_Rad}
\end{equation}
where we are working in units of $\hbar= c=G =1$. A natural quantity to consider, once you have a relation between particular qualities of the system and its temperature, is entropy. How does the entropy of this system relate to geometric and physical qualities of the black hole? Bekenstein \cite{Bekenstein} had first proposed a proportionality relation between the area of the event horizon, for a Schwarzschild black hole, $A$ and the entropy of the black hole $S_{BH}$, motivated by the fact that a black hole's event horizon hides information and that classically it does not decrease. Once Hawking had explained the temperature of a black hole quantum mechanically, the constants of proportionality were worked out giving
\begin{equation}
S_{BH} = \frac{k_{B}A}{4 {l_{p}}^{2}}.
\end{equation}
This relationship highlights something particular about the nature of space-time in the presence of a black hole. If we take a particular interpretation of the entropy as a measure of the information contained within a system, then this equation is telling us that what we know about the space is not proportional to the volume, as we could reasonably expect. Instead it is saying that the information about the space is contained on a two-dimensional surface in the space. 't Hooft had precisely this interpretation in \cite{Hooft} and considered that the information in a space was indeed encoded on the horizon of the black hole. He took inspiration from the developments of lattice quantum field theory and proposed that the information was encoded on the boundary in a two dimensional system of spins on a lattice, with lattice spacing $l_{p}$. This lattice of two spin-like degrees of freedom, which at each site could be spin-up or spin-down, played the role of a binary structure in which all of the information of the space was encoded. In \cite{Susskind} Susskind developed this notion even further within the context of string theory. He imagined a screen at the boundary of this space, which again featured as the binary system that encoded the information about the space. This ``screen" was composed of illuminable dots which were also spaced a distance $l_{p}$ apart. If a particular dot was illuminated it would be considered `on', and the unlit dots were considered to be in the `off' position. Developing an image of the space seems tractable from this setup but determining distances between objects in the space was more complicated. The argument Susskind puts forward to explain this is beyond what we require for this introduction, but can be found in \cite{Susskind}. The fundamental contribution Susskind makes here is to extend the understanding about the relationship between the information about the space and the event horizon. Instead of the information being \emph{on} the event horizon, it is actually encoded on the boundary of the space, like a hologram. This is the holographic principle: the details and physical information in the space is related to a holographic ``image" on the boundary of the space. Holographic duality then, which the AdS/CFT correspondence is the most successful realisation of, is the relationship between a general relativistic theory in the bulk related to an inherently quantum mechanical system on the boundary.
With this in mind we should review the geometric properties of AdS and its boundary.

\subsection{The Geometry of AdS Spacetime} \label{ADS_Geom} 

The General Relativistic origins for de Sitter (dS) and AdS space-time is in the search for solutions to the Einstein field equations that included a cosmological constant in the Einstein-Hilbert action in a vacuum, again with $G$ and $c$ set to $1$,
\begin{equation}
S = \frac{1}{16\pi}\int d^d x \sqrt{-g} \left( R - 2\Lambda \right)
\end{equation} 
where the dS and AdS solutions correspond to negative and positive cosmological constant $\Lambda$, respectively and $d$ is the dimension of the space-time. The metrics for these space-times are the solutions to the Einstein field equation that arises from varying this action with respect to the metric, giving
\begin{equation}
R_{\mu \nu} -\frac{1}{2} g_{\mu \nu} R + \Lambda g_{\mu \nu} = 0.
\end{equation}
Taking the trace of this equation with the inverse-metric one finds
\begin{align}
g^{\mu \nu} R_{\mu \nu} -\frac{1}{2} g^{\mu \nu} g_{\mu \nu} R + \Lambda g^{\mu \nu}g_{\mu \nu} &= 0 \\
R -\frac{d}{2}R + d\Lambda &= 0 \\
R &= \frac{2d}{d-2} \Lambda \label{ADS.1}
\end{align}
It's clear from \eqref{ADS.1} that the metric that would satisfy these equations would be one of constant scalar curvature. From a geometric perspective, considering spaces of constant curvature, the familiar examples are those of embeddings of surfaces in Euclidean space. The plane, sphere and hyperboloid are the typical examples of embedded surfaces in 3-dimensional Euclidean space with constant vanishing, positive and negative curvatures respectively (if homogeneity and isotropy are imposed they are the only three possibiities). These however are all Riemannian manifolds. When considering space-time manifolds we are looking at Lorentzian manifolds (or pseudo-Riemannian manifolds). Within this framework then Minkowski, dS and AdS space-time can be thought of precisely along these lines, they are the equivalent examples with vanishing, positive and negative scalar curvatures, respectively. Thus the space we are interested in, AdS, is that which has a negative cosmological constant and a constant, negative, scalar curvature. Though if this is the Lorentzian analogue of a hyperbolic plane embedded in Euclidean space we don't imagine that there is a boundary of this plane. It extends off to infinity. How then is there a boundary for our AdS space, on which we expect to find our dual CFT? If we consider, in a similar way to our Euclidean examples, that the AdS space is actually a hyperbolic hyper-surface embedded in a higher-dimensional space we can more clearly display the geometric properties of this manifold. A more in-depth treatment of this can be found in \cite{L.Sokol}

\begin{figure}[h!]
\centerline{\includegraphics[width=0.9 \textwidth]{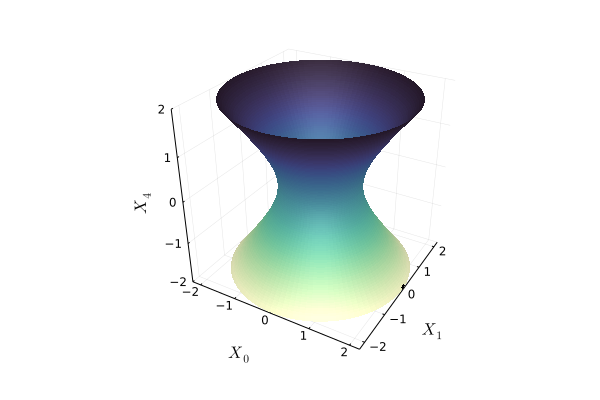}}
\caption{Here we have plotted a portion of the complete four dimensional hyper-surface from equation \eqref{ADS.3}. Specifically it is a plot of the surface $X_{4}^2 - X_{0}^2  -X_{1}^2 = -L^2$, embedded in three dimension (one space dimension and two time dimensions), with $L=1$. }
\label{AdS_Hyper_Surf.1.1}
\end{figure}

As in this chapter we are considering AdS$_{4}$, we will begin with an ambient 5-dimensional (2+3) flat Lorentzian manifold with coordinates $(X_0 ,X_1 ,X_2 ,X_3, X_4)$, with two time-like coordinates $X_0, \ X_1$. The metric on this is given by
\begin{equation}
ds^2 = -dX_{0}^2  -dX_{1}^2 + dX_{2}^2 + dX_{3}^2 + dX_{4}^2. \label{ADS.2}
\end{equation}  
Consider a 4-dimensional hyperbolic hyper-surface in this space given by the equation
\begin{equation}
X_{2}^2 + X_{3}^2 + X_{4}^2 - X_{0}^2  -X_{1}^2 = -L^2 \label{ADS.3}
\end{equation}
where $L$ is a constant and, given the above metric, all the points on this hyper-surface are equidistant from the origin. Thus sometimes the term pseudo-sphere is used to refer to such hyper-surfaces, and $L$ is known as the radius of curvature. This is also part of the reason why this particular 4-dimensional space-time is considered maximally symmetric. We can visualise the embedding of this hyper-surface in a lower dimensional representation, see figure \ref{AdS_Hyper_Surf.1.1}.

\noindent A suitable parametrization for coordinates on the complete hyper-surface are given by
\begin{align}
X_0 &= L\sin(\frac{t}{L})\cosh(\frac{\rho}{L}) \\
X_1 &= L\cos(\frac{t}{L})\cosh(\frac{\rho}{L}) \\
X_2 &= L\sinh(\frac{\rho}{L})\sin{\theta}\cos{\phi} \\
X_3 &= L\sinh(\frac{\rho}{L})\sin{\theta}\sin{\phi} \\
X_4 &= L\sinh(\frac{\rho}{L})\cos{\theta}
\end{align} 
where the coordinates take the values $-\pi L < t < \pi L$,  $\rho \ge 0$ and $\theta ,\ \phi$ are the usual spherical coordinates with $0 \le \theta \le \pi$ and $0 \le \phi \le 2\pi$. It's straightforward to check that these coordinates satisfy \eqref{ADS.3}. Now substituting them into \eqref{ADS.2} the metric on our hyper-surface is given by
\begin{equation}
ds^2 =  -\cosh^{2}(\frac{\rho}{L})dt^2 +d\rho^2 + L^2\sinh^2(\frac{\rho}{L})(d\theta^2 + \sin^{2}\theta d\phi^2) 
\end{equation}
and upon noticing the hyperbolic sine function in front of what looks like the metric on a unit 2-sphere we can spot a natural redefinition for $\rho$. Setting $r := L\sinh(\frac{\rho}{L})$ gives
\begin{equation}
ds^2 =  -\left(1 + \frac{r^2}{L^2} \right)dt^2 + \frac{1}{\left(1 + \frac{r^2}{L^2} \right)}dr^2 + r^2(d\theta^2 + \sin^{2}\theta d\phi^2) \label{ADS.5.1}
\end{equation}
and we arrive at the familiar metric, with global coordinates, for AdS$_{4}$. Typically with this metric one also takes a universal covering of the time coordinate, such that $t\in \mathbb{R}$, to avoid closed time-like curves. 

\noindent To display the relationship between the cosmological constant and $L$ we move to what are known as Poincare coordinates. They are given by the following parametrization; 
\begin{align}
X_0 &= \frac{1}{2z}(L^2 + |\vec{\tr}|^2 - \tti^2) \\
X_1 &= L\frac{\tti}{z} \\
X_2 &= L\frac{x}{z} \\
X_3 &= L\frac{y}{z} \\
X_4 &= \frac{1}{2z}(L^2 - |\vec{\tr}|^2 + \tti^2)
\end{align}
where $|\vec{\tr}|^2 = x^2 +y^2 + z^2$. The metric then becomes
\begin{equation}
ds^2 = \frac{L^2}{z^2}(-d\tti^2 + dx^2 + dy^2 + dz^2) \label{ADS.4}
\end{equation}
with an obvious coordinate singularity at $z=0$, where the metric is not defined. The above form of the metric generalises to a $d$-dimensional AdS space so working through the Christoffel symbols to calculate the Ricci scalar from the metric in $d$ dimensions one finds
\begin{equation} 
R = \frac{-d(d-1)}{L^2}
\end{equation}
which is consistent with our space-time being a hyperbolic space. Combining this with equation \eqref{ADS.1} we have 
\begin{equation} 
\Lambda = \frac{-(d-1)(d-2)}{2 L^2}.
\end{equation}
From the above metric \eqref{ADS.4} we can see that this space is conformally equivalent to Minkowski space. That is to say $g^{ADS}_{\mu \nu} = \Omega^2 g^{Mink}_{\mu \nu}$, where $\Omega$ is some scalar function of the coordinates, albeit for a segment of AdS, known as the Poincare patch. There is a hint towards the location of our conformal boundary in \eqref{ADS.5.1}. For $r \rightarrow \infty$ the time and angular components of the metric diverge but the radial component $g_{rr}\rightarrow 0$. This can be more clearly seen with the change of coordinates given by $r = L\tan(q)$, where $0 \le q < \frac{\pi}{2} $. The metric \eqref{ADS.5.1} becomes 
\begin{equation}
ds^2 =  \frac{1}{\cos^{2}(q)}\left( -dt^2 + L^{2} dq^2 + L^{2}\sin^{2}(q)(d\theta^2 + \sin^{2}\theta d\phi^2)\right)
\end{equation}
which at first does not appear to gives us more information as to the structure of the boundary of AdS. Though if we consider a metric conformally related by $\tilde{g}^{\mu \nu} = \cos^{2}(q){g}^{\mu \nu}$ we have
\begin{equation}
d\tilde{s}^2 = -dt^2 + L^{2} dq^2 + L^{2}\sin^{2}(q)(d\theta^2 + \sin^{2}\theta d\phi^2)
\end{equation}
which now includes the point $q= \frac{\pi}{2}$ and at this point we have a hyper-surface which has a metric 
\begin{equation}
d\tilde{s}^2 = -dt^2 + L^{2}(d\theta^2 + \sin^{2}\theta d\phi^2)
\end{equation}
which is the \emph{conformal spatial infinity} of AdS space-time and has the structure of $\mathbb{R}^1 \times S^2$. This establishes that for $r \rightarrow \infty $ there is a conformal boundary with some well-defined structure. With this outlined we can proceed to the further role that AdS space-time plays in the discussion of holography and specifically geometerizing the energy scale. 

\subsection{Geometerizing the Energy Scale} \label{Geom_Eng}

If we take inspiration from one of the conceptual adjustments that is required when moving from classical Newtonian mechanics to special and general relativity we can better understand the holographic quality of AdS space-time. The conceptual adjustment I am referring to is the consideration of time as a local coordinate, and no longer a universal time, in the sense of Newton, that parametrizes spatial coordinates. This observation gave rise to the progress made by Minkowski and others in understanding the geometric properties of special relativity. Thus considering not merely coordinates $(x(t),y(t),z(t))$ but $(t,x,y,z)$, where these space-time coordinates can be functions of proper time $\tau$ (or any affine parameter $\lambda$), we are afforded a new insight in to the structure of our physical reality.

Similarly, from the discussion in \S \ref{HoloPrinc} on Holography we can glean a general approach that is arising here; associating Physical quantities with geometric properties. Historically it was not the purpose of the endeavour but seems to arise naturally via the discovery of relationships between geometric and thermodynamic quantities. So what aspect of AdS space-time are we looking to ascribe a new physical meaning to? To begin with we look at a 2-dimensional dimensionless metric
\begin{equation}
ds^2 = \frac{dx^2 + dy^2}{z^2} \label{ADS.5}
\end{equation}
where here $z$ is some constant which has units of length, as do $x$ and $y$. The metric allows for the definition of distances (as well angles) on a manifold, thus in this case the numerical value of $ds^2$ remains unchanged under a different choice of units for $z,x,y$, making it scale invariant. If instead we were to consider the units of $x$ and $y$ to be fixed and varied the dimensions of $z$ we would consider this as some length scale of the metric, i.e. different choices of $z$ corresponding to different measures of distance for the space. If we were to then consider incorporating this change into the geometry of the space it would require a relatively simple adjustment to \eqref{ADS.5}
\begin{equation}
ds^2 = \frac{dx^2 + dy^2 + dz^2}{z^2}. \label{ADS.6}
\end{equation}
which is the metric for the Poincare upper half-space in 3-dimensions, with $z>0$. From this perspective we can consider a flow of the space given by \eqref{ADS.5} at different length scales. 

From here it does not require too much of an intuitive leap to see that if we consider a Minkowski metric in the numerator of \eqref{ADS.6}
\begin{equation}
ds^2 = \frac{-dt^2 + dx^2 + dy^2 + dz^2}{z^2}. 
\end{equation}
we arrive at \eqref{ADS.4} which is AdS$_{4}$ space-time in the Poincare patch, with $L = 1$. The important point here is that the length scale is encoded into $z$ and $z$ is promoted to being geometric, in the same way time becomes geometric in Minkowski space-time. With this in mind we can consider an interpretation of the radial coordinate in the broader topic of gauge/gravity duality, referred to in the literature as the holographic dimension \cite{Hartnoll}.

So far, we have only examined the geometerization of this scale from the AdS (gravity) perspective, but what does this correspond to within the broader gauge/gravity duality? On the field theory side, this flow corresponds to the flow of energy scales. In the context of a weakly coupled quantum gravity theory, or classical gravity, there is a strongly coupled QFT on the boundary. Depending on the specific QFT, this dual description can span different energy regimes. For instance, in quantum chromodynamics (QCD), the theory is strongly coupled at low energies in the infrared (IR) regime, where confinement and bound states emerge, but becomes weakly coupled at high energies in the ultraviolet (UV) regime due to asymptotic freedom. However, the precise identification of the boundary field theory in the gauge/gravity duality is not always straightforward. Regardless, given this understanding of the holographic dimension, there is a natural way to interpret the flow along the radial coordinate $z$ as a geometric representation of renormalization group (RG) flow.

We have explained above an imprecise notion of a foliation of a space in which each hyper-surface for constant $z$ has a dual QFT at different energy scales. We can think of the 'holographic' dimension of AdS space as the renormalization dimension, i.e. the limits of high energy and low energy cut-offs for the dual QFT are contained within this extra dimension and the flow along it is a geometric picture of renormalization group flow. This understanding was further articulated in \cite{Iqbal+Liu2} in the context of the universality of the hydrodynamic limit via the membrane paradigm for black holes. The membrane paradigm is a simplifying tool for calculating the thermodynamic effects of the exterior of a black hole by considering a fictitious, thin, classically radiating fluid that sits just above the event horizon \cite{Thorne}. Iqbal and Liu associated this fluid with the dual QFT at low energies, motivated by the argument that, at sufficiently long length scales, QFTs should be described by hydrodynamics \cite{0704.0240}. They used this understanding to derive equations governing the evolution of the QFT dual from the horizon to the high energy regime of the strongly coupled theory at the boundary, giving a clear presentation of this geometric renormalization flow. Crucially this also articulated how the behaviour of the dual field theory on the boundary at finite temperature was determined by the horizon's geometry and thermodynamic quantities. In the next section we will discuss the role of thermodynamics and how we can apply the AdS/CFT correspondence to condensed matter systems.

\newpage
\section{Critical Points and Conformal Symmetry}\label{Hol.Q.M}

A conformal transformation of a space is often described as preserving angles but not distances. More precisely conformal transformations are those transformations of the
metric of  a space-time $g_{\mu \nu} (x)$ that leave it invariant up to a positive, arbitrary scale factor that can depend on the coordinates of the space-time such as
\begin{equation}
g_{\mu \nu}(x) \mapsto \tilde{g}_{\mu \nu} = \left(\Omega(x)\right)^{-2}g_{\mu \nu}(x).
\end{equation}
which can be seen to preserve the causal structure of the space-time but not the length of space-time intervals. The complete set of conformal transformations of Minkowski space in $d > 2$ dimensions are translations, Lorentz transformations, dilations and the special conformal transformation. For an infinitesimal conformal transformation for the coordinates of a flat space-time, i.e. $g_{\mu \nu} = \eta_{\mu \nu}$, with $\Omega(x) = e^{-\alpha(x)}$ we have 
\begin{equation}
x^{\mu} \mapsto x^{\mu} + a^{\mu} + \omega^{\mu}_{\nu}x^{\nu} + \lambda x^{\mu} + b^{\mu} x^{\nu} x_{\nu} - 2(b^{\nu}x_{\nu})x^{\mu}. \label{CTran}
\end{equation}
where the transformations in the above equation \eqref{CTran} appear in the order listed above. The generators corresponding to these quantities are respectively
\begin{align}
P_\mu &= -i\partial_\mu, \\
J_{\mu\nu} &= i(x_\mu\partial_\nu-x_\nu\partial_\mu), \\
D &= -ix_\mu\partial^\mu, \\
K_\mu &= i(x^{\nu}x_{\nu}\partial_\mu-2x_\mu x_\nu\partial^\nu).
\end{align}
For $d=2$ dimensions the number of local conformal transformations is infinite. 

A conformally invariant field theory we have already come across is the massless Klein-Gordon equation in chapter \ref{Chpt_1}. From equation \eqref{K.G._eq} we saw that under the conformal transformation to Rindler coordinates our equations of motion remained the same. There is another feature of CFT's that we can see from the energy-momentum tensor of this field. For a Klein-Gordon field in two space time dimensions the energy-momentum tensor is
\begin{equation}
T_{\mu \nu } = \partial_{\mu} \phi \partial_{\nu} \phi - \frac{1}{2} g_{\mu \nu}\left(\partial^{\alpha} \phi \partial_{\alpha} \phi \right)
\end{equation}
and taking the trace of this gives
\begin{equation}
g^{\mu \nu} T_{\mu \nu} = 0.
\end{equation}
This is an essential property of CFTs that conformal symmetry imposes. Their energy-momentum tensors are traceless if they are invariant under scale transformations.~We can see this more generally for a field theory that is conformally invariant by considering the conserved Noether current associated with scale invariance
\begin{equation}
j^{\mu} = x^{\nu} T_{\nu}^{\mu}.
\end{equation}
By looking at the conservation equation for this current we find
\begin{align}
&\partial_{\mu}j^{\mu} = 0 \\
&\partial_{\mu}\left(x^{\nu} T_{\nu}^{\mu} \right) =0\\
&T^{\mu}_{\mu} + x^{\nu}\partial_{\mu}\left( T_{\nu}^{\mu} \right) =0 \\
&T^{\mu}_{\mu} =0,
\end{align}
where we have used the fact that $\partial_{\mu} T_{\nu}^{\mu} = 0$. Thus a scale invariant field theory has a traceless energy-momentum tensor. For more details on conformal field theories see \cite{CFTBook}.

If we focus just on dilations, $x^{\mu} \mapsto \lambda x^{\mu}$, a central quality of all CFTs is that they are scale invariant. A feature of most condensed matter systems is that they are not generally scale invariant and thus not described by CFT's. There is a saving grace however, around a quantum critical point there are condensed matter systems, or condensed matters theories (CMT), that exhibit scale invariance. That is to say, when approaching critical points the scale dependent parameters in the model become less and less relevant and the scale invariant quantities dominate. In this domain the AdS/CFT correspondence can be used to capture universal qualities of the dual boundary theory. 
In this thesis we apply the techniques of the AdS/CFT correspondence beyond CFT's, in an effort to capture signatures of CMT's in the dual theory. This approach has been used over the last two decades to analyse and provide new universality classes of condensed matter systems. Comprehensive introductions to the use of gauge/gravity duality to these systems, sometimes called the AdS/CMT correspondence, can be found in \cite{McGreevy, Hartnoll}. Our aim within this context is to investigate the impact of the phase transition of an asymptotically AdS black hole on a corresponding dual theory, that we show has a fermi surface at zero temperature.

\begin{figure}[h!]
\centerline{\includegraphics[width=0.8\linewidth]{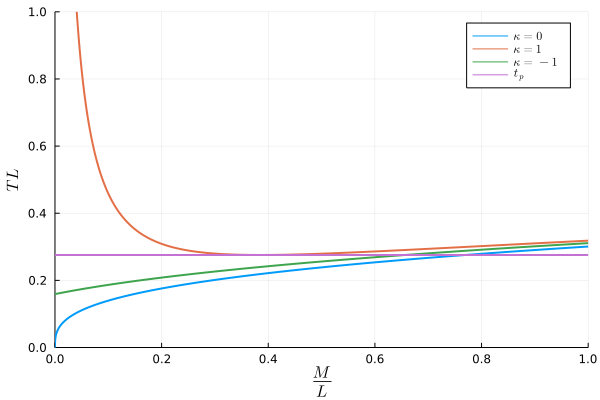}}
\caption{The dimensionless temperature of an AdS black hole as a function of the dimensionless mass for all three values of $\kp$. We have also here plotted the location of the phase transition temperature $t_{p} \approx 0.276$.}
\label{HP1}
\end{figure}

\begin{figure}[h!]
\centerline{\includegraphics[width=0.8\linewidth]{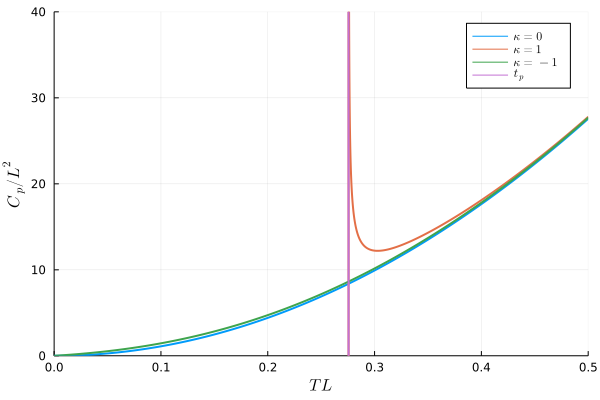}}
\caption{The dimensionless heat capacity of an AdS black hole as a function of the dimensionless temperature for all three values of $\kp$. The location of the phase transition temperature $t_{p}$, with $\kp=1$, is plotted as the vertical asymptote where $C_{p}\rightarrow \infty$.}
\label{HP2}
\end{figure}

As was discussed in \S\ref{Geom_Eng} Iqbal and Liu showed in \cite{Iqbal+Liu2} that the dual strongly coupled field theory on the boundary was determined by the properties of the bulk black hole horizon. The Hawking-Page phase transition \cite{HawkPage} for a black hole only occurs in an asymptotically AdS space-time with a spherical event horizon. An asymptotically AdS$_{4}$ space time is given by 
\begin{equation}
    ds^{2}=-f(r)^2dt^2+\frac{dr^2}{f(r)^2}+r^{2}d\Omega_{\kp}^2 \label{AsymADS}
\end{equation}
with
\begin{equation}
    f(r)^2 = \kp-\frac{2\mathcal{M}}{r} + \left( \frac{r}{L} \right)^2.
\end{equation}
and
\begin{equation}
d\Omega_{\kp}^2 = \left\{\begin{array}{lll}
d\theta^{2} + \sin^{2}(\theta)d\phi^2, &   \kp= 1 \\
d\theta^{2} + d\phi^2, &  \kp=0 \\
d\theta^{2} + \sinh^{2}(\theta)d\phi^2, & \kp= -1 \\ \end{array} \right.
\end{equation}
The $\kp$'s here determine the geometry of the event horizon at $r=r_{h}$: for $\kp=1$ the event horizon is a sphere, $\kp=0$ it is a flat plane, and $\kp=-1$ is a hyperbolic plane. At the horizon we have 
\begin{equation}
\kp-\frac{2\mathcal{M}}{r_{h}} + \left( \frac{r_{h}}{L} \right)^2 = 0 \quad \rightarrow \mathcal{M} = \frac{r_{h}}{2} \left( \kp + \frac{r_{h}^2}{L^2}\right)
\end{equation}
and the temperature for this black hole is given by equation \eqref{Hawk_Rad}
\begin{equation}
T = \frac{|f'(r_{h})|}{4\pi} = \frac{1}{2\pi}\left( \frac{\mathcal{M}}{r_{h}^2} + \frac{r_{h}}{L^2} \right) = \frac{1}{4\pi} \left( \frac{\kp}{r_{h}} + \frac{3r_{h}}{L^2} \right).
\end{equation}
If we treat $\mathcal{M}$ as the internal energy of the black hole, with the pressure assigned to the cosmological constant and is thus fixed, we can find the heat capacity given by
\begin{equation}
C_{p} = \frac{\partial \mathcal{M}}{\partial T} = \frac{\partial \mathcal{M}}{\partial r_{h}} \left(\frac{\partial T}{\partial r_{h}}\right)^{-1} =  \frac{2\pi r_{h}^2\left(\kp + 3(r_{h}^2 / L^{2})\right)}{3(r_{h}^2 / L^2) - \kp}.
\end{equation}
In dimensionless parameters
\begin{equation}
x= \frac{r_{h}}{L}, \quad \tilde{\mathcal{M}}=\frac{\mathcal{M}}{L}, \quad \tilde{T}=TL, \quad \tilde{C_{p}}= \frac{C_{p}}{L^2}
\end{equation}
the above expressions for, mass, temperature and heat capacity become
\begin{align}
\tilde{\mathcal{M}} =&\frac{x(x^2 +\kp)}{2}\\
\tilde{T} =&\frac{3x^2+\kp}{4\pi x}\\
\tilde{C_{p}} =&\frac{2\pi x^2 (3x^2 + \kp)}{3x^2 -\kp} \label{cp}.
\end{align}
It is clear from the above expression for the heat capacity that it can only diverge, i.e. undergo a phase transition, when $\kappa =1$. Plotting the heat capacity as a function of the temperature and mass, figures \ref{HP1} and \ref{HP2}, we can see this behaviour clearly. When $\kp=1$ there is a phase transition temperature $T_{p} = t_{p}/L= \frac{\sqrt{3}}{2\pi L}$ where $C_{p} \rightarrow \infty$.

 This behaviour does not occur for the other values of $\kp$. These figures show that for $t=t_{p}$ the heat capacity diverges and as can be seen from equation \eqref{cp}, the heat capacity is negative in the region of $t<t_{p}$. This indicates that the black hole undergoes a phase transition from a stable state to an unstable one where it radiates away its energy. This is the essence of the Hawking-Page phase transition but in \cite{HawkPage} they introduce a photon gas to show that the AdS black hole can achieve stable thermal equilibrium with the radiation. For our purpose though, as  a phase transition is present for just an asymptotically AdS black hole, it is sufficient to analyse the dual theory at the equivalent phase transition temperature for a RN-AdS$_{4}$ black hole without further complications. In \S\ref{Results} a nearly identical calculation will follow for a RN-AdS$_{4}$ black hole which will allow us to look for signatures of a phase transition in our dual field theory on the boundary at this temperature. Before this though we will briefly review the prescriptive approach we will be using to obtain our results, as outlined in \cite{Son, Iqbal+Liu}.

\newpage

\section{AdS/CFT Prescription for Spinors}\label{Holo.Spinors}
As has been noted often in the literature, though the explicit examples of holographic duality require string theory as the starting point \cite{Maldacena, Ammon2015, Hartnoll}, the techniques involved are quite general and only require us to use concepts from general relativity and quantum field theory. The list of techniques and prescriptions are vast and could not be meaningfully covered in this chapter; we will instead briefly review the prescription outlined in \cite{Iqbal+Liu} for calculating real-time retarded Green's functions for a fermionic operator $\mathcal{O}$ that is dual to a spinor field in a background asymptotic AdS$_{d+1}$ space time. As in equation \eqref{AsymADS} the metric
\begin{equation}
    ds^{2}=-f(r)^2dt^2+\frac{dr^2}{f(r)^2}+r^{2}d\Omega_{d-1}^2 \label{AsymADS1}
\end{equation} 
has the behaviour near the boundary 
\begin{equation}
g_{tt}, g^{rr}, g_{ii} \approx r^2, \quad r\rightarrow \infty
\end{equation}
where the subscript $i$ denotes the angular components of the metric. For the time being we do not specify the function $f(r)$ in the metric, just that it has the above boundary behaviour. 

In Lorentzian signature there is not an explicit way to arrive at the Green's functions from the derivatives of an action principle. The prescription instead begins with an Euclidean action and then analytically continues the results to Lorentzian signature. To illustrate the prescription we will begin with a massless scalar field action with the above metric
\begin{equation}
S = -\frac{1}{2}\int d^{d+1}x\sqrt{-g}(\nabla_{\mu} \phi \nabla^{\mu} \phi) 
\end{equation}
which when analytically continued to Euclidean signature via
\begin{equation}
t \rightarrow -i\tau
\end{equation}
gives
\begin{equation}
S_{E} = \frac{1}{2}\int d^{d+1}x\sqrt{g}(\nabla_{\mu} \phi \nabla^{\mu} \phi). 
\end{equation}
We can then use the AdS/CFT prescription from equation \eqref{ADSCFT2}, with $S_{\text{AdS}}$ the bulk action $S_{E}$ evaluated at the boundary value $\phi_{0}$ for the field $\phi$. From \eqref{ADSCFT2} we can find one point functions for $\mathcal{O}$
\begin{equation}
\langle \mathcal{O}(x) \rangle_{\phi_0} = -\frac{\delta S_{\rm{AdS}}}{\delta \phi_0 (x)} = - \lim_{r\to\infty}\Pi_E (r,x)|_{\phi} \label{OP1}
\end{equation}
where $\Pi_{E}$ is the canonical momentum conjugate to $\phi$ with respect to foliation in the $r$-direction and $x$ is shorthand for all coordinates in the space except $r$. The canonical momentum is defined in this way as the $r$ coordinate plays the role of the energy scale in the dual field theory, i.e. the dynamics are described in terms of the bulk field evolving according to $r$. The Fourier transform of this gives
\begin{equation}
\langle \mathcal{O}(\omega_E, \vec{k}) \rangle_{\phi_0}  = -\lim_{r\to\infty}\Pi_E (r,\omega_E, \vec{k})|_{ \phi} \label{Op1.1}
\end{equation}
where $\omega_{E}$ and $ \vec{k}$ are boundary momentum space coordinates. The frequency is distinguished here for Euclidean signature as it is connected with the time coordinate and thus is analytically continued by $\omega \rightarrow i\omega_{E}$. The Green's function then is given in Euclidean signature as 
\begin{equation}
G_E(\omega_E, \vec{k}) = -\lim_{r\to\infty}\frac{\Pi_E (r,\omega_E, \vec k)|_{ \phi}}{\phi (r,\omega_E,\vec{k})} \biggr|_{\phi_0 =0} \label{Green1.1}
\end{equation}
where the notation of $\phi_{0} = 0$ indicates that in evaluating the ratio above we take a linear approximation of \eqref{Op1.1} and thus the part of \eqref{Green1.1} that is independent of $\phi_{0}$. From here, to obtain the Green's function in Lorentzian signature, we analytically continue the bulk field $\phi$ to Lorentz signature. The relationship between the fields in momentum space is given by
\begin{equation}
\phi(r, \omega_{E}, \vec{k}) \biggr|_{\omega_{E} = -i\omega} = \phi_{R}(r, \omega, \vec{k})
\end{equation}
and this allows us to write down the retarded Green's function 
\begin{equation}
G(\omega, \vec{k}) = \lim_{r\to\infty}\frac{\Pi (r,\omega, \vec k)|_{ \phi_{R}}}{\phi_{R} (r,\omega,\vec{k})}
\end{equation}
with the condition that the field $\phi_{R}$ satifies the boundary conditions; $\lim_{r\rightarrow\infty} \phi_{R} = \phi_{0} $, $\phi_{0}\ne 0$ and is in-falling at the horizon, i.e. selecting only the solutions that have ingoing modes of the bulk field at the horizon. For the fermionic theory then the set up is similar.

Given a boundary theory fermionic operator $\mathcal{O}$ that is dual to a spinor field $\psi$ in an AdS background we can compute two point functions for this dual operator just by considering the quadratic part of the action for $\psi$ in the bulk 
\begin{equation}
S[\psi] = \int d^{d+1}x\sqrt{-g} i \left( \bar{\psi} \Dslash \psi - M \bar{\psi}\psi   \right) + S_{bd}
\end{equation}
where $S_{bd}$ are boundary terms that ensure a well defined variational principle for the action \cite{Henneaux}. The Dirac operator $\Dslash$ is 
\begin{equation}
\gamma^{\mu}D_{\mu}= \gamma^{a}(e^{-1})^{\mu}_{a}(\partial_{\mu}+\frac{1}{4}\gamma^{bc}(\omega_{bc})_{\mu})
\end{equation}
and $\bar{\psi} = \psi^{\dagger}\gamma^t$. The $\gamma$'s are gamma matrices in $d+1$ dimensions, where for our results $d=3$. In the next section we will consider a $U(1)$ gauge potential that features in the Dirac operator above, but for the purposes of introducing this prescription we leave it out. The $e$'s are components of orthonormal one forms and are related to the metric by 
\begin{equation}
    \eta_{ab}e^{a}_{\mu}e^{b}_{\nu}=g_{\mu\nu}
\end{equation}
where the Roman letters denote frame indices while Greek letters denote coordinate indices. The $\omega$'s are the matrix elements of spin connection one forms and we can find these using Cartan's first structure equation (assuming zero torsion)
\begin{equation}
    de^a= -\omega^{a}{}_{b}\wedge e^{b}.
\end{equation}
as in chapter \ref{Chpt_2}. We then analytically continue to Euclidean signature via
\begin{equation}
\gamma^{t} \rightarrow -i\gamma^{\tau}
\end{equation} 
which gives the Euclidean action
\begin{equation}
S[\psi] = -\int d^{d+1}x\sqrt{g} \left( \bar{\psi} \Dslash \psi - M \bar{\psi}\psi   \right) + S_{bd}.
\end{equation}
The AdS/CFT prescription for spinors is then 
\begin{equation}
\left\langle \exp\left[ \int d^dx \ \left( \bar{\psi}_0 \mathcal{O} + \bar{\mathcal{O}}\psi_{0} \right)\right]\right\rangle_{\text{CFT}} = e^{-S_{\text{AdS}}[\bar{\psi}_{0},\psi_0]}
\end{equation} 
where $\psi_{0}$ is the value of the spinor field at the boundary $r\rightarrow \infty$. The prescription requires that we find a solution $\psi = \psi_{+} + \psi_{-}$, to the equations of motion with in-falling boundary conditions at the horizon. Where $\psi_{\pm}$ are related to each other via
\begin{equation}
\psi_{\pm} = \frac{1}{2}(1\pm\gamma^{r}) \psi
\end{equation}
The canonical momentum conjugate to $\psi_{\pm}$ is given via
\begin{equation} 
\Pi_{\pm} =\frac{\delta S_{E}}{\delta \psi_{\pm}} =\mp \sqrt{\frac{g}{g_{rr}}} \bar{\psi}_{\mp}
\end{equation}
where again some care must be taken when varying the action and choosing appropriate boundary terms, see appendix C of \cite{Iqbal+Liu}. Then expanding the solution near the boundary $r \rightarrow \infty$ (with $L=1$) such that 
\begin{equation}
\psi_{+}= Ar^{M-\frac{d}{2}} + Br^{-M-\frac{d}{2} -1}, \quad \psi_{-}= Cr^{M-\frac{d}{2} -1} + Dr^{-M-\frac{d}{2}}.
\end{equation}
and putting this back in to the dirac equation for $r \rightarrow \infty$ yields the relationship between the coefficients to be
\begin{equation}
C = \frac{i\gamma^{\mu}k_{\mu}}{2M-1} A, \quad B = \frac{i\gamma^{\mu}k_{\mu}}{2M+1} D, \quad k_{\mu} = (-\omega,\vec{k})
\end{equation}
and the canonical momenta are given by 
\begin{equation}
\Pi_{+} = -\bar{C}r^{M+\frac{d}{2} -1} - \bar{D}r^{-M+\frac{d}{2}}, \quad \Pi_{-}= \bar{A}r^{M+\frac{d}{2}} + \bar{B}r^{-M+\frac{d}{2} -1}, \quad r\rightarrow \infty.
\end{equation}
It's clear that at the boundary the term $A$ is dominant and should thus be identified with the source for the boundary operator, i.e.
\begin{equation}
\lim_{r \rightarrow \infty}r^{-M+\frac{d}{2}}\psi_{+} = A =\psi_{0}.
\end{equation}
and in an analogous way to \eqref{OP1} we find that the expectation value for $\bar{\mathcal{O}}$ is given by the canonical momentum
\begin{equation}
\langle \bar{\mathcal{O}} \rangle_{\psi_{0}} = -\lim_{r \rightarrow \infty}r^{M-\frac{d}{2}}\Pi_{+} = \bar{D}
\end{equation}
where we extract the finite term in the limit above. To obtain the Green's functions we cannot take a ratio of these two quantities as they are spinors so instead we find a matrix $\mathcal{S}$ whereby
\begin{equation}
D(k_{\mu}) = \mathcal{S} A(k_{\mu}).
\end{equation}
The boundary Euclidean Green's function then is given by
\begin{equation}
G_{E}(k_{\mu}) = \mathcal{S}(k_{\mu})\gamma^{\tau}
\end{equation}
which when we analytically continue to Lorentzian signature gives us
\begin{equation}
G_{R}(k_{\mu}) = i\mathcal{S}(k_{\mu}) \gamma^{t}.
\end{equation}
This will be the central focus of the next section, where we begin by briefly laying out the approach taken in \cite{McGreevyetal}.

\newpage
\section{Holographic Condensed Matter}\label{Results}

\subsection{Dual Green's Functions for a Fermionic Field in Reissner-Nordstr\"om-AdS$_{4}$}
In this section we will be looking at a charged fermionic field in a RN-AdS$_{4}$ space-time that is dual to a charged fermionic operator on the boundary of this space, as in \cite{McGreevyetal}. The fermionic action in the bulk is given above in \S \ref{Hol.Q.M} as
\begin{equation}
S[\psi] = \int d^{4}x\sqrt{-g} i \left( \bar{\Psi} \Dslash \Psi - M \bar{\Psi}\Psi   \right) + S_{bd} \label{SpinAc}
\end{equation}
with $\Psi$ a four component spinor with charge $e$ which on the boundary is the charge of the dual fermionic operator $\mathcal{O}$. An RN-AdS$_{4}$ metric is that which satisfies the Einstein equations given by varying the action
\begin{equation}
S_{\text{Grav}} = \frac{1}{16\pi}\int d^4 x \sqrt{-g} \left( R - 2\Lambda \right) -\frac{1}{4} \int d^4 x \sqrt{-g} F^{ab}F_{ab}
\end{equation}
where the electromagnetic potential for a dyonic charge is given by
\begin{equation}
A = q\left(\frac{1}{r_h} - \frac 1 r \right)dt + \tm(\pm 1 - cos(\theta))d\phi .
\end{equation}
The monopole charge $\tm$ is half the monopole charge used in chapter \ref{Chpt_2} that we introduce for the sake of neatness of the following equations. The components of the electromagnetic tensor are found from taking the exterior derivative of the potential $F=dA$. It's straight forward then to calculate that
\begin{equation}
F^{ab}F_{ab} = -\frac{(q^2 - \tm^2)}{r^4}.
\end{equation}
The inclusion of these charges amounts to a modification of the metric we used in \S \ref{Hol.Q.M} by introducing a dyonic charge $Q$ on the black hole, such that 
\begin{equation}
ds^2 =-f(r)^2 dt^2+\frac{dr^2}{f(r)^2}+r^{2}d\Omega_{k}^2 
\end{equation}
has the function in the metric given by
\begin{equation}
f(r)^2 = \kappa-\frac{2\mathcal{M}}{r} + \frac{Q^2}{r^2} + \left( \frac{r}{L} \right)^2.
\end{equation}
where $\kappa$ here can take the values $0,1$, and $Q$ is given by
\begin{equation}
Q^2 = 4\pi(q^2 + \tm^2)
\end{equation}
where the factor of $4\pi$ comes from solving the Einstein equations. At the black hole horizon radius $r_{h}$ we have
\begin{equation}
  \kp -  \frac{2\mathcal{M}}{r_{h}} + \left(\frac{Q}{r_{h}}\right)^2 + \left( \frac{r_{h}}{L}\right)^2  = 0 \rightarrow \mathcal{M} = \frac{1}{2}\left(\kp r_h + \frac{Q^2}{r_{h}} + \frac{r_{h}^3}{L^2}  \right)
\end{equation}
which allows us to write the above function $f(r)$ in a convenient form
\begin{align}
    f(r)^2 &= \kp (1-\frac{r_{h}}{r}) + Q^2(\frac{1}{r^2} - \frac{1}{r_{h} r}) + \frac{1}{L^2}(r^2 - \frac{r_{h}^3}{r}) \\
    &= \kp (1-\frac{r_{h}}{r}) + \frac{Q^2}{r_{h}^2}((\frac{r_{h}}{r})^2 - \frac{r_{h}}{r}) + \frac{r_{h}^2}{L^2}((\frac{r}{r_{h}})^2 - \frac{r_{h}}{r})
\end{align}
From this we can calculate the Hawking temperature 
\begin{equation}
    T_{H} = \frac{|(f(r_{h})^2)'|}{4\pi} = \frac{1}{4\pi r_{h}}(\kp + \frac{3 r_{h}^2}{L^2} - \frac{Q^2}{r_{h}^2}) \label{Hawktemp1}
\end{equation}
which we can see for non-negative temperatures $Q^2 \le r_{h}^2(\kp + \frac{3 r_{h}^2}{L^2})$ and at equality the temperature is zero. We will discuss the temperature of  the phase transition that occurs for this system later in this section. In \cite{McGreevyetal} they have $\kp=0$ and focus mainly on zero temperature, we will vary these quantities in our analysis.

The effect of introducing a charge $Q$ in the metric above, and with it $U(1)$ gauge fields $A_{\mu}$ for both $q$ and $\tm$, on the dual CFT is that the boundary theory now has a global $U(1)$ symmetry. The question that McGreevey et al. then ask is: if there is strongly coupled boundary field theory with a finite $U(1)$ charge density does it contain Fermi surfaces, specifically of the kind found by the author of \cite{Senthil}? In their analysis, which is common in the literature of AdS/CMT \cite{Ammon2015}, they interpret $q$ as the chemical potential of the boundary theory and the product of $eq$ from the gauge potential in the covariant derivatives of the dirac operator 
\begin{equation}
D_{\mu} =  \partial_{\mu}+\frac{1}{4}\gamma^{bc}(\omega_{bc})_{\mu} + ie A_{\mu}
\end{equation}
as the effective chemical potential of the system. They pursue this search for Fermi surfaces by primarily looking at the spectral functions of the fermionic operators from the boundary theory. The spectral function of an operator is proportional to the imaginary part of the Green's function $G_{R}$ associated with that operator and is a measure of the density of states that couple to the operator. Using the prescription described in \S \ref{Hol.Q.M} we can find these Green's functions numerically.

In \cite{McGreevyetal} the authors find strong indications of the existence of a Fermi surface of a non-Fermi liquid that comes in the form of quasi-particle peaks in the spectral functions. They determine that this system is a non-Fermi liquid based on the scaling behaviour of the spectral functions. They find this Fermi surface for black hole temperature $T = 0$ (i.e. for $Q = \frac{\sqrt{3}r_{h}^2}{L}$), where increasing the temperature appears to flatten out the peaks. In this section we extend this analysis to a spherical horizon $\kp=1$ and non-zero temperature, specifically around $T=T_{p}$ where the heat capacity diverges. To begin though we will find the Green's functions used to observe this behaviour. 

\noindent The equations of motion from the action in \eqref{SpinAc} give the Dirac equation
\begin{equation}
(\Dslash -M)\Psi =0 
\end{equation}
where the Dirac operator in a curved space-time is given as in \S \ref{Hol.Q.M} but with the inclusion of the gauge potentials for the electric and magnetic fields
\begin{equation}
\Dslash = \gamma^{\mu}D_{\mu}= \gamma^{a}(e^{-1})^{\mu}_{a}(\partial_{\mu}+\frac{1}{4}\gamma^{bc}(\omega_{bc})_{\mu} + ie A_{\mu}).
\end{equation}
The orthonormal one-forms are
\begin{align}
    e^{a}&= e^{a}{}_{\mu}d x^{\mu}\\ 
    e^{0}&= f d t, \ e^{1}= \frac{dr}{f}, \ e^{2}= rd\theta, \ e^{3}= r\sin\theta d\phi
\end{align}
and using Cartan's first structure equation
\begin{equation}
    de^a= -\omega^{a}{}_{b}\wedge e^{b}
\end{equation}
we can find the connection one-forms by taking the exterior derivative of the $e$'s above and doing some rearranging
\begin{align}
    de^{0}&= f' e^1\wedge e^0 \\
    de^{1}&= 0\\ 
    de^{2}&= \frac{f}{r}e^1\wedge e^2\\ 
    de^{3}&= \frac{f}{r}e^1\wedge e^3 + \frac{\cot\theta}{r}e^2\wedge e^3.
\end{align}
The connection one-forms are 
\begin{align*}
    \omega_{0 1} &= -f' e^{0} = -f f' dt \\
    \omega_{1 2} &= -\frac{f}{r} e^{2} = -f d\theta  \\
    \omega_{1 3} &= -\frac{f}{r} e^{3} = -f \sin\theta d\phi\\
    \omega_{2 3} &= -\frac{\cot\theta}{r} e^{3} = -\cos\theta d\phi.
\end{align*}
The Dirac equation reads 
\begin{align}
    \Dslash \Psi -M\Psi &= \gamma^{a}(e^{-1})^{\mu}{}_{a} D_{\mu} \Psi -M\Psi = 0\nonumber \\
    &=\left(\gamma^{0}(e^{-1})^{t}{}_{0}D_{t} + \gamma^{1}(e^{-1})^{r}{}_{1}D_{r} + \gamma^{2}(e^{-1})^{\theta}{}_{2}D_{\theta} + \gamma^{3}(e^{-1})^{\phi}{}_{3}D_{\phi}\right)\Psi - M\Psi\nonumber \\
                        &=\left(\frac{\gamma^{0}}{f}D_{t} + \gamma^{1}f D_{r} + \frac{\gamma^{2}}{r}D_{\theta} + \frac{\gamma^{3}}{r\sin\theta}D_{\phi}\right)\Psi - M\Psi
\end{align}
with
\begin{align}
  D_t &= \partial_t - \frac 1 2 \gamma^{01} f f'+ i e A_t = -i \omega - \frac 1 2 \gamma^{01} f f'+ i e A_t, \\
  D_r&=\partial_r,\\
  D_\theta&=\partial_\theta - \frac 1 2 \gamma^{1 2} f,\\
  D_\phi&=\partial_\phi  -\frac 1 2 \gamma^{1 3} f \sin\theta -\frac 1 2 \gamma^{2 3}(-\cos\theta)  +ie A_{\phi }.
\end{align}
In \cite{McGreevyetal} the authors use the gamma matrix basis given by,
\begin{equation} 
\gamma^0 = \begin{pmatrix} 0 & i\sigma^{2} \\ i\sigma^{2} & 0 \end{pmatrix}, \quad
  \gamma^1 = \begin{pmatrix} {\bf 1} & 0 \\ 0 & -{\bf 1} \end{pmatrix}, \quad
  \gamma^2 = \begin{pmatrix} 0 & \sigma^1 \\ \sigma^1 & 0 \end{pmatrix}, \quad
  \gamma^3 = \begin{pmatrix} 0 & \sigma^3 \\ \sigma^3 & 0 \end{pmatrix},\label{MajoranaGam}\end{equation}
such that the dual boundary theory then has a valid gamma matrix basis for the (2+1) dimensional fermionic theory, and the boundary dual spinors transform as Dirac spinors. Beginning with a spinor of the form 
\begin{equation}
\psi = \frac{e^{-i\omega t + ik_{i}x^{i}}}{(f(r))^{\frac{1}{2}}}\begin{pmatrix} \phi_{+} \\ \phi_{-}\end{pmatrix}
\end{equation}
where the $k_{i}$'s are planar wave numbers for the non-radial coordinates, with $\kp =0$. They then set $k_{2} =0$ as the system is rotationally symmetric in the transverse coordinates and only $k_{1}$ features in their equations. For our purposes the eigenvalues on the sphere $\lambda$ will be equivalent to their $k_{1}$ for $\kp=0$ in the metric. The $\phi_{\pm}$ are two spinors that they find to have asymptotic behaviour at the boundary given by
\begin{equation}
\phi_{+} = Ar^{M} + Br^{-M-1}, \quad \phi_{-} = Cr^{M-1} + Dr^{-M}
\end{equation}
where the coefficients here are related as they are in \S \ref{Hol.Q.M}. They further specify the components of the two component spinors 
\begin{equation}
\phi_{\pm} = \begin{pmatrix} y_{\pm} \\ z_{\pm} \end{pmatrix}
\end{equation}
to arrive at two sets of decoupled equations from the Dirac equation. Following the prescription given in the aforementioned section they arrive at an expression for the Green's function that is diagonal in their basis
\begin{equation} 
G_R = \lim_{r\rightarrow \infty} r^{2 M} \begin{pmatrix} \xi_+ & 0 \\ 0 & \xi_- \end{pmatrix} \label{Green}
\end{equation}
with
\begin{equation}
\xi_{+} = \frac{iy_{-}}{z_{+}}, \quad \xi_{-} = -\frac{iz_{-}}{y_{+}}
\end{equation}
and the $\xi$'s obey a flow equation, as can be found in \cite{Iqbal+Liu2}, that comes from substituting these ratios in to the equations of motion
\begin{equation}
f(r)\partial_{r}\xi_{\pm} = -2M\xi_{\pm} \mp \left(\frac{k_{1}}{r} \mp \frac{\omega - e A_{t}}{f(r)} \right) \pm \left(\frac{k_{1}}{r} \pm \frac{\omega - e A_{t}}{f(r)} \right)\xi_{\pm}^2 
\end{equation}

We can arrive at exactly this flow equation and thus be able to extract the boundary values of these quantities for the Green's functions numerically but to use the eigen-values and eigen-functions for the Dirac Operator on the sphere, denoted $\Dslash_{S^2}$, we start by using the gamma matrix basis
\begin{equation} 
\gamma^0 = \begin{pmatrix} 0 & {\bf 1} \\ -{\bf 1} & 0 \end{pmatrix}, \quad
  \gamma^1 = \begin{pmatrix} 0 & \sigma_3 \\ \sigma_3 & 0 \end{pmatrix}, \quad
  \gamma^2 = \begin{pmatrix} 0 & \sigma_1 \\ \sigma_1 & 0 \end{pmatrix}, \quad
  \gamma^3 = \begin{pmatrix} 0 & \sigma_2 \\ \sigma_2 & 0 \end{pmatrix},\label{Chiral}\end{equation}
which gives the Dirac operator as
    \begin{equation}
    \Dslash =
      \begin{pmatrix}
      0 & \frac{i(-\omega +A_t)}{f} +f\sigma_3 \nabla_r +\frac 1 r \Dslash_{S^2} \\
        \frac{-i(-\omega +A_t)}{f} +f\sigma_3 \nabla_r +\frac 1 r \Dslash_{S^2} & 0
      \end{pmatrix}
      \end{equation}
    where
    \begin{equation}
    \nabla_r=\partial_r + \frac 1 r + \frac{f'} {2 f}.
    \end{equation}
With a spinor of the form
    \begin{equation}
    \Psi = \begin{pmatrix} F(r)_+ \eta_+ +F(r)_- \eta_- \\
      G(r)_+ \eta_+ +G(r)_- \eta_-  \end{pmatrix}
    \end{equation}
with $\eta_{\pm}$ eigen-functions on the sphere such that
\begin{equation}
\Dslash_{S^2} \eta_\pm= \mp i  \lambda \eta_\pm, \qquad \sigma_3 \eta_\pm = \eta_\mp.
\end{equation}
Substituting this in to the Dirac equation with the above form of $\Dslash$ we get four equations,
  \begin{align}
i\left(    \frac{(-\omega+ e A_t)}{f}   -\frac{\lambda}{r}\right)G_+ + f \nabla_r G_- 
   & =M F_+\\
i\left( \frac{(-\omega+ e A_t)}{f}+\frac{\lambda}{r}\right)G_-  + f \nabla_r  G_+ 
     & =M F_- \\
-i\left(\frac{(-\omega+ e A_t)}{f}  +\frac{\lambda}{r}\right) F_++ f \nabla_r F_- 
      &=M G_+\\
-i\left( \frac{(-\omega+ e A_t)}{f} -\frac{\lambda}{r} \right)F_- + f \nabla_r F_+ 
      &=M G_-.
  \end{align}
for $\lambda\ne 0$,  and $\eta_+$ and $\eta_-$ being linearly independent. Defining
\begin{equation} 
\widetilde F_\pm = r \sqrt{f} F_\pm, \qquad \widetilde G_\pm = r \sqrt{f} G_\pm
\end{equation}
and
\beq K_\pm(r) = \frac{\pm(\omega - e A_t)}{f} - \frac{\lambda}{r}\eeq
these simplify to   
  \begin{align}
  i K_-\tilde{G}_+ + f \partial_r \tilde{G}_- 
   & =M \tilde{F}_+ \label{GplusFplus} \\
-i K_+\tilde{G}_-  + f \partial_r  \tilde{G}_+ 
     & =M \tilde{F}_- \label{GminusFminus}\\
i K_+ \tilde{F}_++ f \partial_r \tilde{F}_- 
      &=M \tilde{G}_+ \label{FplusGplus} \\
-i K_-\tilde{F}_- + f \partial_r \tilde{F}_+ 
      &=M \tilde{G}_-. \label{FminusGminus}
  \end{align}
Re-arranging these by taking linear combinations (\ref{GplusFplus})$\pm$(\ref{FminusGminus}) and (\ref{GminusFminus})$\pm$(\ref{FplusGplus}) amounts to returning to the gamma matrix basis \eqref{MajoranaGam} used by \cite{McGreevyetal}. This gives
  \begin{align}
    (f\partial_r \mp M)(\tilde{G}_- \pm \tilde{F}_+)&=-i K_-(\tilde{G}_+ \mp \tilde{F}_-)\\
    (f\partial_r \mp M)(\tilde{G}_+ \pm \tilde{F}_-)&= i K_+(\tilde{G}_- \mp \tilde{F}_+)
    \end{align}
which we can identify with the components of $\phi_{\pm}$ by
\begin{equation} 
y_+ = a(\tF_+ + \tG_-), \ y_- = b(\tF_+ - \tG_-), \ z_+ = b(\tG_+ + \tF_-),
    \ z_- = a(\tG_+ - \tF_-),\label{ypmzpm}
\end{equation}
where $a$ and $b$ are arbitrary non-zero constants. We then arrive at the equations
\begin{align}
(f\partial_r \mp M) y_\pm &= \mp i K_- z_\mp \label{dy}\\
(f\partial_r \mp M) z_\pm &= \mp i K_+ y_\mp \label{dz}
\end{align}
which are identical to the equations of motion in the basis used by \cite{McGreevyetal} with $k_{1}$ replaced by $-\lambda$. Following the same identification of $\xi_{\pm}$
    \begin{align}
            \xi_+ & = \frac{i y_-}{z_+}= \frac{i(\tF_+ - \tG_-)}{(\tF_- +\tG_+)}\\
     \xi_- &= -\frac{i z_-}{y_+}=
              \frac{i(\tF_- -\tG_+)}{(\tF_+ +\tG_-)}.
    \end{align}

we arrive at the flow equation 
\begin{equation} 
f(r) \partial_r \xi_\pm = -2 M \xi_\pm \mp K_\mp(r) \pm K_\pm(r) \xi_\pm^2\label{plusminusODE}.
\end{equation}

To obtain the boundary conditions necessary for the numerical calculations it is convenient to use $z = \frac{r_{h}}{r}$. The function in the metric can then be written as
\begin{equation}
 f(z)^2= \frac{(1-z)}{z^2}\left(\kp z^2 - \frac{Q}{r_{h}}^2z^3 + \frac{r_{h}^2}{L^2} (1+z+z^2)\right). \label{eq.25}
\end{equation}
The boundary is now at $z=0$ and the horizon is at $z=1$. The derivative of $\xi_{\pm}$ with respect to $z$ is given by
\begin{equation}
    \partial_{r} \xi_{\pm} = -(\frac{z^2}{r_h})\partial_{z}\xi_{\pm}
\end{equation}
and our flow equation is given by 
\begin{equation}
f(z)(-\frac{z^2}{r_{h}})\partial_{z}\xi_{\pm}(z)= -2 M \xi_\pm \mp K_\mp(z) \pm K_\pm(z) \xi_{\pm}^{2}.
\end{equation} 
To simplify things further $f(z)$ can be written as $f(z)^2 = \frac{\alpha^2 (1-z)}{z^2}h(z)^2$, as in equation \eqref{eq.25}. The function $h(z)^2$ is
\begin{equation}
h(z)^2 = \frac{\kp z^2}{\alpha^2} - \frac{{Q}^2 z^3}{\alpha^4 L^2} + 1+z+z^2
\end{equation}
where we have introduced the parameter $\alpha = \frac{r_{h}}{L}$. This allows us to write our flow equation as 
\begin{multline}
(1-z)h(z)^2 \partial_{z}\xi_{\pm} = \frac{2ML(1-z)^{\frac{1}{2}}}{z}h(z)\xi_{\pm} \pm \frac{\lambda}{\alpha} (1-z)^{\frac{1}{2}}h(z)(1-\xi_{\pm}^2)  \\- (\frac{\omega L}{\alpha} - \frac{e q}{\alpha^2}(1-z))(1+\xi_{\pm}^2) \label{eq.26}
\end{multline}
Setting $L=1$ and redefining $\tilde{\omega} = \frac{\omega}{\alpha}$, $\tilde{\lambda} = \frac{\lambda}{\alpha}$ and $\frac{e q}{\alpha^2} \rightarrow \tq$ we have
\begin{equation}
(1-z)h^2 \partial_{z}\xi_{\pm} = \frac{2M(1-z)^{\frac{1}{2}}}{z}h\xi_{\pm} \pm \tilde{\lambda} (1-z)^{\frac{1}{2}}h(1-\xi_{\pm}^2) - (\tilde{\omega} - \tilde{q}(1-z))(1+\xi_{\pm}^2)
\end{equation}
We need to analyse the behaviour of this equation near the horizon so we choose coordinates $z = 1 - \epsilon^2$, and $h(z)^2$ becomes
\begin{align}
h(z)^2 &= \left(\frac{\kp}{\alpha^2} - \frac{{Q}^2}{\alpha^4} + 3 + \mathcal{O}(\epsilon^2) \right) \\
h(z)^2 &= \tilde{h}^2 + \mathcal{O}(\epsilon^2)
\end{align}
Which, to lowest order is related to the Hawking temperature by
\begin{equation}
\frac{4\pi T_{H}}{r_{h}} = \left(3+ \frac{\kp}{\alpha^2} - \frac{Q^2}{\alpha^4} \right) = \tilde{h}^2 \quad \rightarrow \quad T_{H} = \frac{\alpha}{4 \pi} \left(3+ \frac{\kp}{\alpha^2} - \frac{Q^2}{\alpha^4}\right)
\end{equation}
The derivative of $\xi_{\pm}$ becomes $\partial_{z}\xi_{\pm} = -\frac{1}{2\epsilon}\frac{\partial\xi_{\pm}}{\partial\epsilon}$ so our near horizon equation looks like
\begin{align}
(-\frac{\epsilon}{2})h(z)^2 \frac{\partial \xi_{\pm}}{\partial \epsilon} &\approx \frac{2M \epsilon}{(1-\epsilon^2)}h\xi_{\pm} \pm \tilde{\lambda} \epsilon h(1-\xi_{\pm}^2) - (\tilde{\omega} - \epsilon^2 \tq )(1+\xi_{\pm}^2) \\
-\frac{\epsilon}{2} h(z)^2 \frac{\partial \xi_{\pm}}{\partial \epsilon} &= (2M \epsilon)h\xi_{\pm} \pm \tilde{\lambda} \epsilon h(1-\xi_{\pm}^2) - \tilde{\omega}(1+\xi_{\pm}^2) + \mathcal{O}(\epsilon^2).
\end{align}
The infalling boundary condition stated in \S \ref{Holo.Spinors} from \cite{Iqbal+Liu} imposes regularity for these solutions on the horizon i.e. at $z=1$, provided $\omega \ne 0$, $\xi_{\pm}(1) = i$. Taylor expanding the function $\xi_{\pm}$ we have
\begin{align}
    \xi_{\pm}(1- \epsilon^2) &= \xi_{\pm}(1) + ((1-\epsilon^2) - 1)(\frac{-1}{2\epsilon})\frac{\partial \xi_{\pm}}{\partial \epsilon}(1) + \mathcal{O}(\epsilon^2) \\
    \xi_{\pm}(1- \epsilon^2) &= i + \frac{\epsilon}{2}\frac{\partial \xi_{\pm}}{\partial \epsilon}(1) + \mathcal{O}(\epsilon^2) \\
    (\xi_{\pm}(1- \epsilon^2))^2 &= -1 + i\epsilon\frac{\partial \xi_{\pm}}{\partial \epsilon}(1) + \mathcal{O}(\epsilon^2) \\
    1+(\xi_{\pm}(1- \epsilon^2))^2 &= i\epsilon\frac{\partial \xi_{\pm}}{\partial \epsilon}(1) + \mathcal{O}(\epsilon^2) \\
    1- (\xi_{\pm}(1- \epsilon^2))^2 &=  2- i\epsilon\frac{\partial \xi_{\pm}}{\partial \epsilon}(1) + \mathcal{O}(\epsilon^2)
\end{align}
Close to the horizon then we have
\begin{align}
(-\frac{\epsilon}{2})\tilde{h}^2 \frac{\partial \xi_{\pm}}{\partial \epsilon} &= (2M)\tilde{h}(i\epsilon + \frac{\epsilon^2}{2}\frac{\partial \xi_{\pm}}{\partial \epsilon}) \pm \tilde{\lambda} \tilde{h}(2\epsilon - i\epsilon^2\frac{\partial \xi_{\pm}}{\partial \epsilon}) - \tilde{\omega}(i\epsilon\frac{\partial \xi_{\pm}}{\partial \epsilon}) + \mathcal{O}(\epsilon^3)\\
(-\frac{\epsilon \tilde{h}^2}{2})\frac{\partial \xi_{\pm}}{\partial \epsilon} &= 2\tilde{h} \epsilon (iM \pm \tilde{\lambda}) - i\tilde{\omega} \epsilon \frac{\partial \xi_{\pm}}{\partial \epsilon} + \mathcal{O}(\epsilon^2)
\end{align}
which when looking at the lowest power of $\epsilon$ gives
\begin{equation}
\frac{\partial \xi_{\pm}}{\partial \epsilon}(1) = \frac{2\tilde{h}(iM \pm \tilde{\lambda})}{i\tilde{\omega} -\frac{\tilde{h}^2}{2}}
\end{equation}
and our horizon boundary condition then is $\xi_{\pm}(1 - \epsilon^2) = i + \epsilon \frac{\partial \xi_{\pm}}{\partial \epsilon}(1)  $ for $\omega \ne 0$. Numerical solutions for $\epsilon =0$ are problematic as our flow equation is singular there. Thus it is necessary to start at some finite, but suitably small, $\epsilon$, which should allow us to get reasonable solutions from our computational calculations.
 
The behaviour of this flow equation at the AdS boundary is also worth going through. We employ a similar approach as above and delicately examine equation \eqref{eq.26} as we go further out to the boundary at $z=0$. We can't just sub in this value because the term with the mass diverges. If we multiply \eqref{eq.26} through by $z$ and then set it to zero we get

\begin{equation}
\xi_{\pm}(0) = 0
\end{equation}
provided $\partial_{z} \xi_{\pm}$ is finite at the boundary. This gives us little information about the behaviour of our functions as we approach the boundary. If we rearrange equation \eqref{eq.26} by dividing through by the factor in front of the $\partial_{z} \xi_{\pm}$ we have 
\begin{equation}
\partial_{z} \xi_{\pm} = \frac{2M}{z(1-z)^{\frac{1}{2}}h(z)}\xi_{\pm} \pm \frac{\tilde{\lambda}}{(1-z)^{\frac{1}{2}}h(z)}(1-\xi_{\pm}^{2}) - \frac{\tilde{\omega} -\tilde{q}(1-z)}{(1-z)h(z)^2}(1+\xi_{\pm}^{2}).
\end{equation}
If we first look at the behaviour of $h^{2}(z)$ close to zero. For $z\ll 1$ we have
\begin{align}
h(z)^2 &= \frac{\kp z^2}{\alpha^2} - \frac{{Q}^2 z^3}{\alpha^4 L^2} + 1+z+z^2 \\
&= 1 + z + \mathcal{O}(z^2) \\
\intertext{so near the boundary at $z=0$ we have}
h(z)^2 &\approx 1 + z \\
h(z) &= (1+ z + \mathcal{O}(z^2))^\frac{1}{2} \approx 1 + \frac{z}{2} \\
(1-z)^{\frac{1}{2}} &\approx 1 + \frac{z}{2}. 
\end{align}
Now we can look at our non-linear ODE near the boundary given these reasonable approximations for small $z$. 
\begin{align}
\partial_{z} \xi_{\pm} &= \frac{2M}{z(1-\frac{z}{2})(1+\frac{z}{2})}\xi_{\pm} \pm \frac{\tilde{\lambda}}{(1-\frac{z}{2})(1+\frac{z}{2})}(1-\xi_{\pm}^{2}) - \frac{\tilde{\omega} -\tilde{q}(1-z)}{(1-z)(1+z)}(1+\xi_{\pm}^{2})\\
\partial_{z} \xi_{\pm} &= \frac{2M}{z(1-\frac{z^2}{4})}\xi_{\pm} \pm \frac{\tilde{\lambda}}{(1-\frac{z^2}{4})}(1-\xi_{\pm}^{2}) - \left(\frac{\tilde{\omega}}{1-z^2} -\frac{\tilde{q}}{(1+z)}\right)(1+\xi_{\pm}^{2}).
\end{align}
Using the expression for a geometric series $\frac{1}{1-z} = 1-z+ z^2 -z^3 ...$ for $z < 1$ and neglecting terms of order $z^2$ or higher we have
\begin{equation}
\partial_{z} \xi_{\pm} = 2M\xi_{\pm}(\frac{1}{z}+\frac{z}{4}) \pm \tilde{\lambda} - \tilde{\omega} + \tilde{q}(1-z)
\end{equation}
which upon going nearer the boundary again and ignoring terms of order $z$ we have 
\begin{equation}
\partial_{z} \xi_{\pm} = \frac{2M}{z}\xi_{\pm} \pm \tilde{\lambda} - \tilde{\omega} + \tilde{q}.
\end{equation}
Now we can tackle this differential equation by using an integrating factor. This gives 
\begin{equation}
\frac{d}{dz}\left(z^{-2M}\xi_{\pm}\right)=\left[\pm \tilde{\lambda} - \tilde{\omega} + \tilde{q}\right]z^{-2M}
\end{equation} 
where integrating and simplifying gives 
\begin{equation}
\xi_{\pm} = \left(\frac{\pm \tilde{\lambda} - \tilde{\omega} +\tilde{q}}{1-2M}\right)z +c_{\pm}z^{2M}.
\end{equation}
So as $z \rightarrow 0$ we have that $\xi_{\pm}$ must fall off like $z^{2M}$, where $0\leq M < \frac{1}{2}$. This bound on $M$ is mentioned in \cite{Iqbal+Liu} at the end of their prescription for spinors as the condition that ensures the fermionic fields are normalizable. Note that the expression for the Green's function $G_{R}$ above has a factor in front that requires us to extract the finite terms in the limit $r \rightarrow \infty$. In terms of $z$ this is given by
\begin{equation} 
G_R = \lim_{z\rightarrow 0} z^{-2 M} \begin{pmatrix} \xi_+ & 0 \\ 0 & \xi_- \end{pmatrix} \label{Green}
\end{equation}
and the finite part we extract is the integration constant $c_{\pm}$. 

\begin{figure}[h]
\includegraphics[scale=0.3]{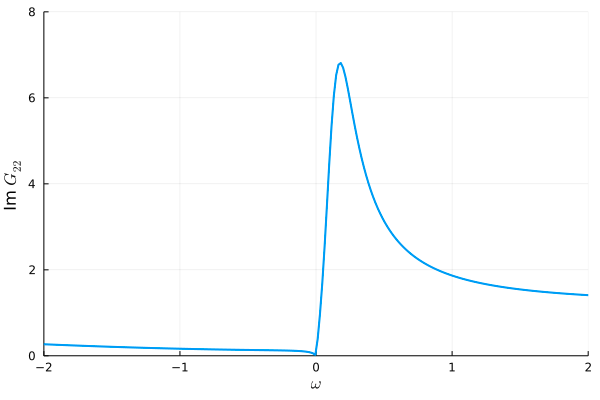} 
\includegraphics[scale=0.3]{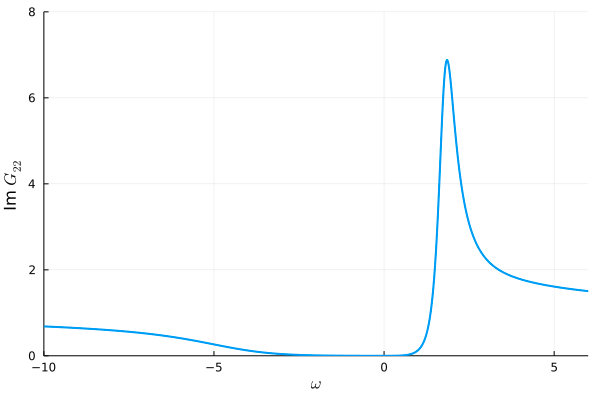}
\caption{Shown is Im$G_{22}(\omega)$ for the values $\lambda = 1.2$ (left) and $\lambda=3$ (right), with $\kp = 0$, $\alpha = 1$, $Q^2= 3$ and thus $T_{H} =0$. These bumps show the region in which the spectral functions deviate from the vaccuum behaviour, which is Im$G_{ii} \rightarrow 1$ as $\omega \rightarrow \pm \infty$. These finite peaks exist past the region where the fermi surface is, shown in figure \ref{Surfk0}.}
\label{ImG22k0}
\end{figure}

\begin{figure}[h]
\includegraphics[scale=0.33]{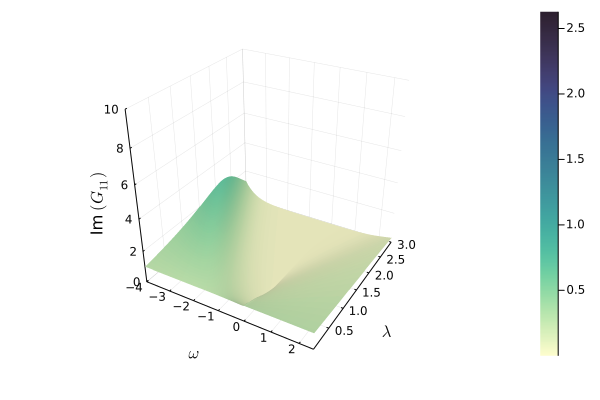} 
\includegraphics[scale=0.33]{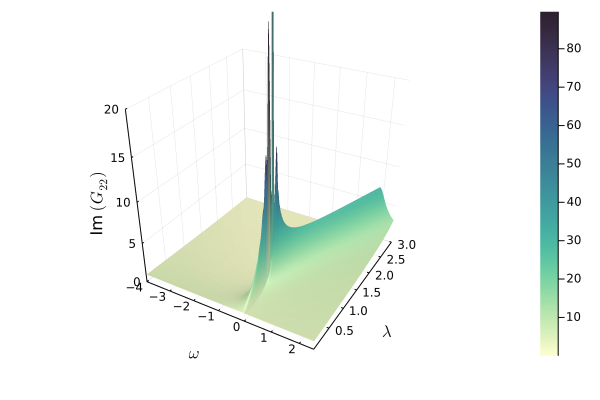}
\caption{This graph shows the qualitative behaviour indicative of a fermi surface as a pole in the dispersion relation in Im$(G_{22})$. It appears as we approach $\omega \rightarrow 0^{-}$ and at the value $\lambda_{f_0} \approx 0.9185$. The oscillations that can be seen in the right figure above are a feature of instabilities in the numerical calculations near the pole, they are not physical.}
\label{Surfk0}
\end{figure}

There are certain immediate qualities of the components of the Green's function that we can exploit to narrow the range of our numerical calculations. It is a straight forward substitution to see that
\begin{align}
G_{11}(\omega,\lambda) &= G_{22}(\omega, -\lambda) \\
G_{11}(\omega,\lambda, -e) &= -G_{22}(-\omega, \lambda, e)
\end{align} 
so it is the sufficient to focus on $\lambda, e \ge 0$. Our numerical calculations are done using the Julia programming language and a good first check of our code is to try reproduce the results found in \cite{McGreevyetal}. Our code can be found at the address in the bibliography \cite{zenodocodo}. Looking at the equation \eqref{eq.26} analytically we examine its behaviour by starting at the horizon $z=1$ and then moving off to the boundary $z=0$. As stated above, computationally these values are not viable as \eqref{eq.26} is singular at these values. Instead we must be careful to choose appropriately small values of $\epsilon$ in the range of $z$ as we integrate from $z=1- \epsilon^2$ to $z=\epsilon^2$ so that we capture as close to the true behaviour as possible of the retarded Green's functions of our dual QFT. Setting $\tm=M=\kp=0$ we find our numerics are in good agreement with results found in \cite{McGreevyetal}. Figures \ref{ImG22k0} and \ref{Surfk0} show the behaviour of the spectral function (imaginary parts of the green's function) and are calculated for zero temperature, with $Q^2 = 3$, $\alpha = 1$ and the charge of the fermion set to $e=-1$. These results match exactly with those in \cite{McGreevyetal} and can be checked against their figures fig.1 and fig.2. From the right surface plot in figure \ref{Surfk0} we see there exists this quasi-particle-like peak for Im$G_{22}(\omega\rightarrow 0^{-}, \lambda_{f_{0}})$ that is identified as fermi surface for the dual fermionic field theory in \cite{McGreevyetal}. Our notation with the subscript $f_{0}$ here for the location of the fermi surface $\lambda_{f_{0}}$ is to denote that this is the fermi surface for the case where $\kp=0$.

With these results we are confident in the accuracy of our numerical calculations and can proceed to varying these parameters for different horizon geometry's and temperatures.

\newpage
\subsection{Spectral Functions for a Spherical Event Horizon}\label{Results1}

We have previously examined the thermodynamics of an asymptotically AdS$_{4}$ black hole for the case where $\kp=1$ in \S \ref{Hol.Q.M}, though with no charge on the black hole. We saw that once there is a spherical event horizon the heat capacity can diverge i.e. a phase transition can occur for the black hole. Thus examining the behaviour of our dual field theory's Green's functions by varying the temperature of the black hole will be part of the focus of this section but before looking at this there is interesting behaviour in merely changing the geometry of the event horizon from $\kp =0$ to $\kp =1$ and keeping the temperature at $T_{H} =0$. We focus primarily on the massless case with magnetic charge $\tm =0$, which means $\lambda = n$ as we have a discrete spectrum of eigenvalues on the sphere.

Crucially what plays a role in the case of $\kp = 1$ is that the horizon radius $r_{h}$ is now a parameter and must be considered in the numerics. In \cite{McGreevyetal} with $\kp = 0$, their results are all scaled by the horizon radius and it thus doesn't feature as a parameter that needs to be chosen.

\subsubsection{Zero Temperature}
\begin{figure}[h]
\includegraphics[scale=0.33]{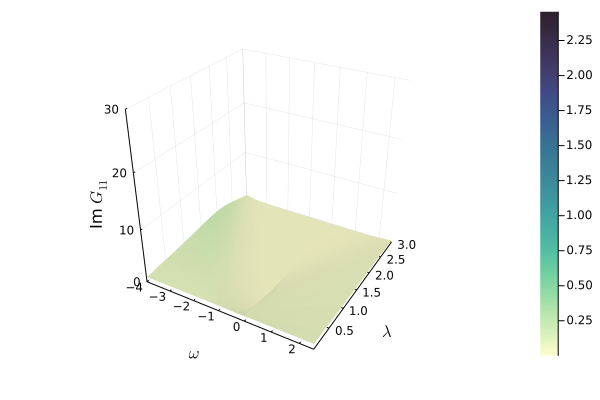}
\includegraphics[scale=0.33]{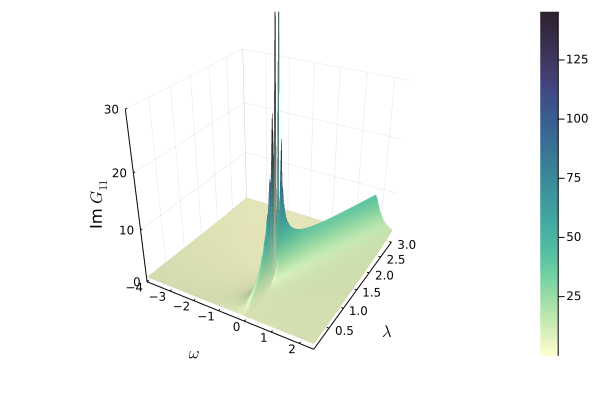}
\caption{Here we have plotted the dispersion relations for Im$G_{11}(\omega,\lambda)$ and Im$G_{22}(\omega,\lambda)$, with $\kp = 1$, $\alpha = 1$, $Q^2= 4$ and thus $T_{H} =0$. The surface for Im$G_{22}(\omega,\lambda)$ captures the behaviour of the infinite peak that indicates the presence of the fermi surface at $\lambda_{f_1}$.}
\label{dispersion1}
\end{figure}

\begin{figure}[h]
\includegraphics[scale=0.21]{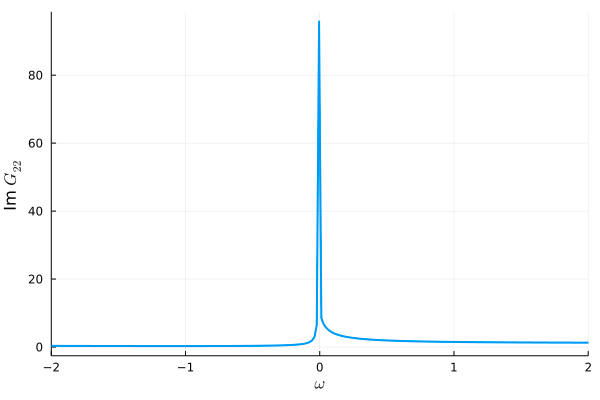} 
\includegraphics[scale=0.21]{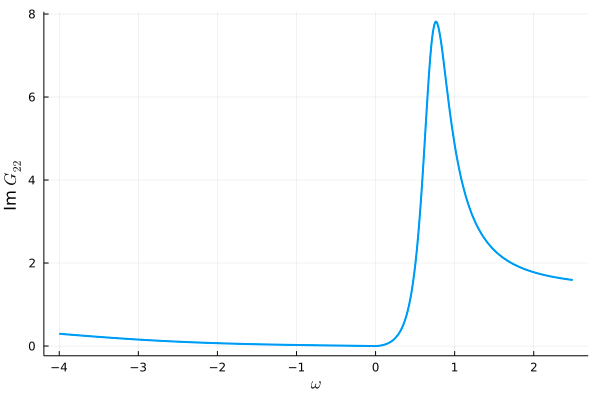}
\includegraphics[scale=0.21]{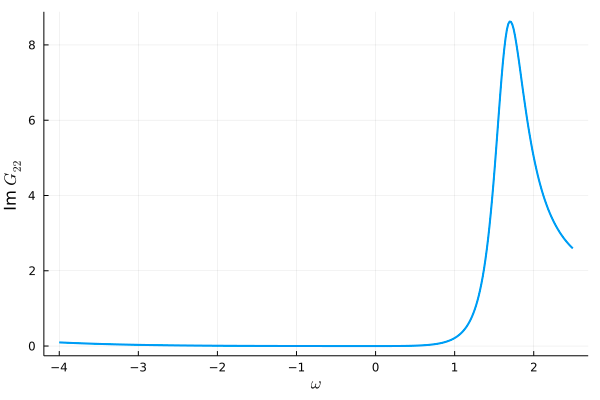}

\caption{From left to right we have plot the  Im$G_{22}$ against $\omega$ for discrete values of $\lambda = 1,2,3$, again with $\kp = 1$, $\alpha = 1$, $Q^2= 4$ and thus $T_{H} =0$. The general behaviour matches with the figures from the previous section but the peaks deviate further from the vacuum behaviour of the spectral functions for the same ranges of $\omega$.}
\label{peaks1}
\end{figure}

For the zero temperature case we look to make our analysis as similar as possible to that of \cite{McGreevyetal} by setting $r_{h}=L=1$, and only changing $\kp$ from $0$ to $1$. From equation \eqref{Hawktemp1} zero temperature for this event horizon geometry is reached when $Q^2=4\pi q^2 = 4$. At this temperature and charge we find that we recover qualitatively similar behaviour as in \cite{McGreevyetal} but that the peaks deviate further from the vaccuum. We find that we still have an infinite spike, indicative of a fermi surface, in the same region of $\omega \rightarrow 0^{-}$ but at a higher value of $\lambda_{f_1} \approx 1.0499$ which can be seen in figure \ref{dispersion1}. Note we have plotted these surfaces for a continous range of $\lambda$ but only the discrete values are physically allowed, examples of which are plotted in \ref{peaks1}. The fermi surface now lies between the discrete eigenvalues on the sphere. A possible explanation as to why the fermi surface shifts could be to do with the change in chemical potential at $T_{H}=0$ from $\kp =0$ to $\kp=1$. The chemical potential with zero magnetic monopole is $q = \frac{Q}{2\sqrt{\pi}}$. At zero temperature for $\kappa =0$, $q_{0}=\sqrt{\frac{3}{4\pi}}$. For $\kappa = 1$ the chemical potential is $q_{1}=\frac{1}{\sqrt{\pi}}$. For a non-relativistic ideal fermi gas of mass $M$ the relationship between the Fermi energy (which is the chemical potential at zero temperature) and the Fermi wave-number is  
\begin{equation}
\epsilon_{f} \propto \frac{\lambda_{f}^2}{M}.
\end{equation}
For the massless case there is an obvious problem with this so if we instead assume the relation where 
\begin{equation}
\epsilon_{f}\propto \sqrt{\lambda_{f}^2 - M^2}
\end{equation}
then for zero mass we have a linear proportionality relation between the Fermi energy and the values of $\lambda = \lambda_{f}$. Looking at the difference in ratios of the $\lambda_{f}$'s to the chemical potentials we find
\begin{equation}
\frac{q_1}{q_0} - \frac{\lambda_{f_{1}}}{\lambda_{f_{0}}} \approx 0.01.
\end{equation} 
Thus it seems a linear proportionality relation between these quantities is reasonable and it is viable that the switch from $\kp =0$ to $\kp =1 $ mainly contributes to increasing the chemical potential which shifts the fermi surface up to a higher value of $\lambda$, though further numerical investigation of the spectral functions would be required to confirm this.

\newpage 

\subsubsection{Phase Transition Temperature}
\begin{figure}[h]
\includegraphics[scale=0.6]{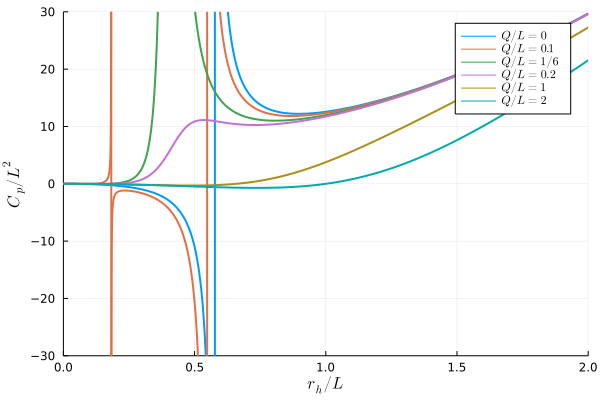}
\caption{The dimensionless heat capacity as a function of $\alpha$ plotted for various values of $Q/L$ labelled in the graph. We can see for values of $Q/L > 1/6$ there is no phase transition. At the maximum value for $Q/L = 1/6$ the heat capacity is always positive for any value of $r_{h}/L$. As $Q/L$ is lowered to zero the heat capacity changes sign on either side of the asymptotes (plotted in the same colour as their corresponding graphs). These are the regions in which the black hole phase transition is unstable. There is another flip in sign for $C_{p}/L^2$ in the range $1/6 >Q/L > 0$ which may indicate that the black hole undergoes another phase transition and becomes stable again before going to zero.}
\label{CPRN}
\end{figure}

\begin{figure}[h!]
\includegraphics[scale=0.6]{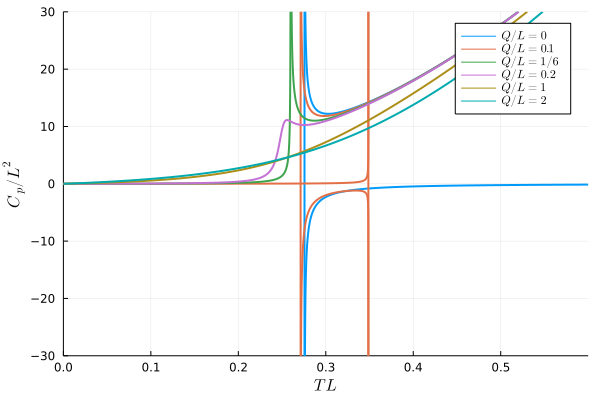}
\caption{The dimensionless heat capacity $C_{p}/L^2$ as a function of $TL$ plotted for the same range of values as figure \ref{CPRN}. We can see similar behaviour again for the heat capacity as it diverges as the temperature is lowered, again only becoming negative when $1/6 >Q/L \ge 0$.}
\label{CPRN1}
\end{figure}

As was discussed in \cite{McGreevyetal}, for $\kappa = 0$, increasing the temperature of the system appears to flatten out the peaks and the spike at the fermi surface becomes smooth. There is no specifically interesting temperature apart from $T_{H}=0$ when $\kappa =0$. When $\kappa = 1$ however there is a specific non-zero temperature that seems natural to examine and that is the temperature at which the phase transition for the black hole occurs. Following a similar approach from \S\ref{Hol.Q.M} we begin with the mass of the black hole from above as
\begin{equation}
\mathcal{M} = \frac{r_h}{2}\left(\kp  + \frac{Q^2}{r_{h}^2} + \frac{r_{h}^2}{L^2}  \right)
\end{equation}
the temperature is 
\begin{equation}
T = \frac{1}{4\pi r_{h}}(\kp + \frac{3 r_{h}^2}{L^2} - \frac{Q^2}{r_{h}^2})
\end{equation}
where we drop the subscript $H$ from the temperature from here on. Taking the mass to be the internal energy again, the heat capacity is then given by 
\begin{equation} 
C_{p} = \frac{\partial \mathcal{M}}{ \partial T} = \frac{2\pi r_{h}^2 \left( 3\frac{r_{h}^4}{L^2} + r_{h}^2 -Q^2 \right)}{3\frac{r_{h}^4}{L^2} - r_{h}^2 + 3 Q^2}.
\end{equation}
In dimensionless parameters
\begin{equation}
\alpha = \frac{r_{h}}{L}, \quad \tQ = \frac{Q}{L}, \quad \tilde{\mathcal{M}}=\frac{\mathcal{M}}{L}, \quad \widetilde{T}=TL, \quad \tilde{C_{p}}= \frac{C_{p}}{L^2}
\end{equation}
the above expressions for, mass, temperature and heat capacity become
\begin{align}
\tilde{\mathcal{M}} =&\frac{\alpha^4 +\kp \alpha^2 + \tQ^2}{2\alpha}\\
\widetilde{T} =&\frac{3\alpha^4+\kp \alpha^2 -\tQ^2}{4\pi \alpha^3}\\
\tilde{C_{p}} =&\frac{2\pi \alpha^2 (3\alpha^4 + \kp \alpha^2 - \tQ^2)}{3\alpha^4 -\kp \alpha^2 + 3\tQ^2} \\ \label{cp1}
=&\frac{8\pi^2 \alpha^5 \widetilde{T}}{3\alpha^4 -\kp \alpha^2 + 3\tQ^2}.
\end{align}
The heat capacity diverges at 
\begin{equation}
\alpha_{\pm}^2 = \frac{1}{6}\left(1 \pm \sqrt{1-36\tQ^2}  \right)
\end{equation}
and there is an upper bound on the charge of the black hole where a phase transition can occur which is at $\tQ^2 = \frac{1}{36}$. The numerator of the heat capacity is positive provided the temperature remains positive. The denominator needs to be handled more carefully. For non-zero temperature we can fix the horizon radius by focusing on the region of the temperature where the heat capacity diverges and thus tie $\alpha_{\pm}$ to the charge on the black hole. For $\alpha = \alpha_{\pm}$ the temperature is given by 
\begin{equation}
\widetilde{T}_{\pm} = \frac{2\alpha_{\pm}^2 -4\tQ^2}{4\pi \alpha_{\pm}^3}
\end{equation} 
which is positive for $0\le \tQ^2 \le \frac{1}{36}$. At the upper bound of $\tQ^2$ there is only one value of the horizon radius $\alpha_{0}^2 = \frac{1}{6}$ for the phase transition to occur. The temperature at the phase transition with this charge is $\widetilde{T}_{p} = \frac{\sqrt{6}}{3\pi}$. The above range for $\tQ^2$ is sufficient for the numerator of $\tilde{C_{p}}$ to be positive so that we are in a stable region for the heat capacity but the denomimator changes sign as we get to lower values of the horizon radius as can be seen from figures \ref{CPRN} and \ref{CPRN1}. We have to be careful then in our numerics when choosing a value for the horizon radius so that we are in the region where the system is stable and at the phase transition $C_{p} \rightarrow +\infty$.

\begin{figure}[h]
\includegraphics[scale=0.33]{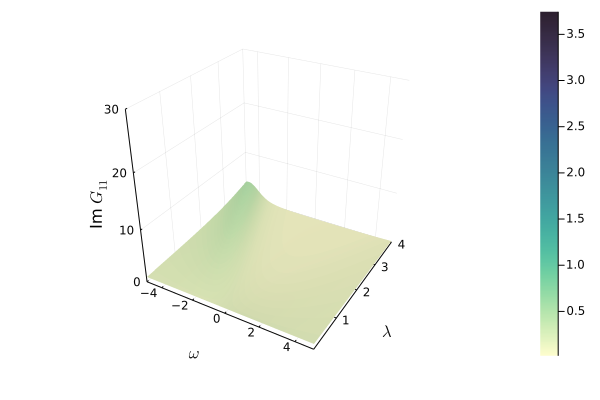}
\includegraphics[scale=0.33]{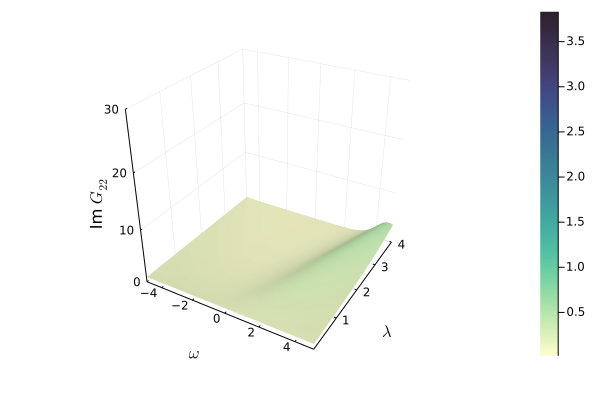}
\includegraphics[scale=0.33]{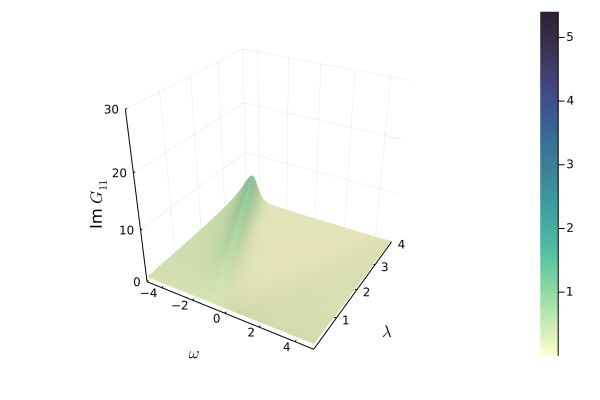}
\includegraphics[scale=0.33]{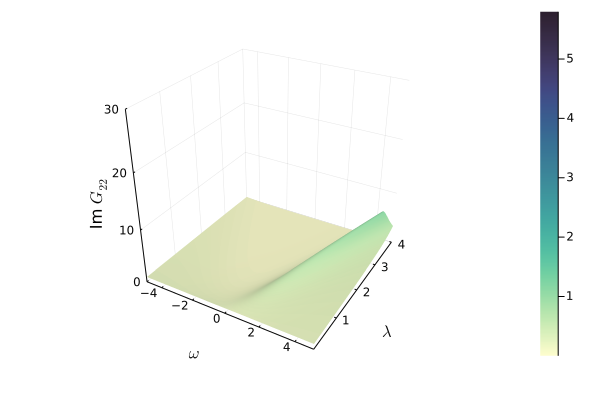}
\includegraphics[scale=0.33]{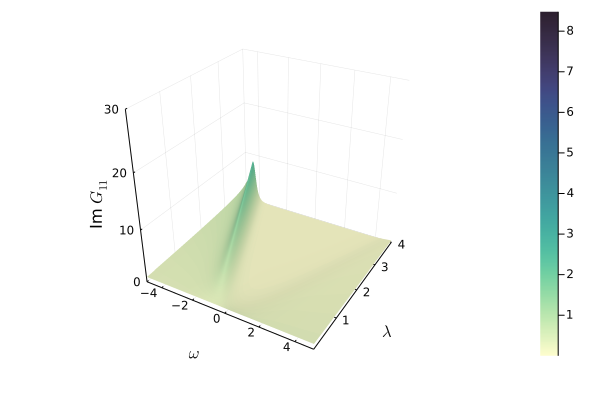}
\includegraphics[scale=0.33]{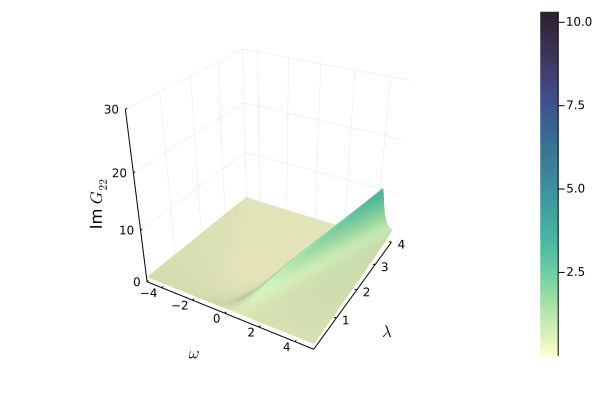}
\includegraphics[scale=0.33]{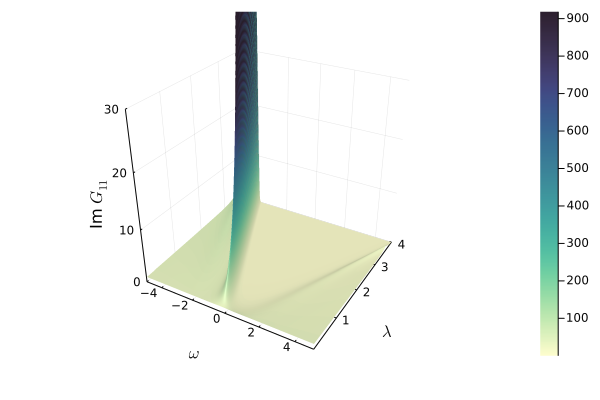}
\includegraphics[scale=0.33]{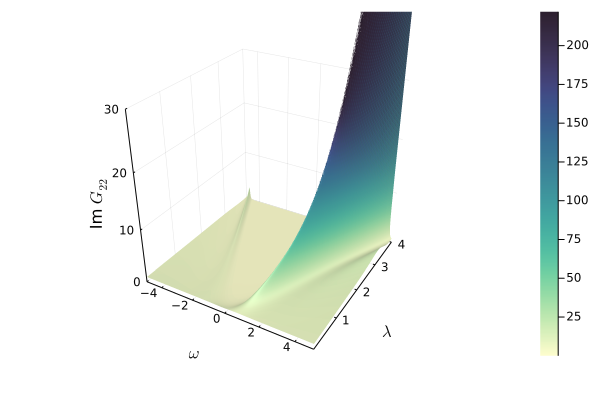}
\caption{Here we have plotted the dispersion relation for both components of the Green's function for $\tQ =\frac{1}{6}$ with progressively lower values of $\alpha$. From top to bottom we have $\alpha=\alpha_0 + 2$, $\alpha=\alpha_0 + 1$, $\alpha=\alpha_0 + 0.5$, $\alpha=\alpha_0$. We see that a smooth ridge appears as we approach the value of $\alpha$ where the phase transition occurs for both Im$G_{11}$ with $\omega<0$ and Im$G_{22}$ with $\omega>0$, and then there is a clear divergence at the phase transition.}
\label{ImG_Tp}
\end{figure}

\newpage
If we consider the maximum value for $\tQ^2 = \frac{1}{36}$, which is the minimum temperature for the upper sign of $\alpha = \alpha_{+}$, there is only one pole for the heat capacity and it only diverges positively (see the green graph in figure \ref{CPRN}). In figure \ref{ImG_Tp} we set $L=1$ and start at some large value for the horizon radius (which is some higher temperature $T>T_{p}$) and lower the value of the horizon radius and temperature to the region where the phase transition occurs, at $\alpha=\alpha_{0}$. We find that we get a clear divergence of the spectral functions in this region. This is telling us that at the phase transition for the black hole the density of states that are coupling to the fermionic operator diverges. What is occuring here is not clear as our boundary theory is some strongly coupled $(2+1)$ dimensional femionic field theory. A possible explanation for what is being captured here could be a deconfining phase transition like that proposed by the authors in \cite{QED31} for $(2+1)$ QED at finite temperature. They argue that a Berezinskii–Kosterlitz–Thouless (BKT) like transition occurs at the temperature
\begin{equation}
T_{c} = \frac{\tilde{g}^2}{8\pi\left(1-\frac{\tilde{g}^2}{12\pi \tilde{m}}\right)}
\end{equation} 
where $\tilde{g}$ is the gauge coupling and $\tilde{m}$ is the mass of the fermions. This expression for the critical temperature is valid only for small values of $\frac{T}{\tilde{m}}$ and $\frac{\tilde{g}^2}{\tilde{m}}$. Lattice numerical simulation were carried out in \cite{QED32} which found evidence of the existence of this phase transition. Though this is a possible candidate for what is occurring in our system, it is not without flaws: notably, this phase transition requires a large mass compared to the gauge coupling. Moreover, BKT like transitions primarily affect vortex configurations and transport properties rather than directly influencing the density of states. To further qualify this result requires more exploration of the parameter space which we leave to future work.

\chapter{Conclusion}
\label{Conclusion}

The general theme and purpose of this thesis was to demonstrate the efficacy of analysing quantum systems within specific geometric settings to simplify the calculation of meaningful quantities. We considered three distinct quantum systems, structuring the thesis such that each chapter focused on one of these systems. Given the unique nature of each system, a brief overview of the necessary background was provided at the beginning of each chapter to establish the foundational concepts required for the analysis.

In chapter \ref{Chpt_1} we addressed Unruh radiation \cite{Unruh1976}, specifically in the context of an accelerated two state atom. Detecting Unruh radiation presents significant experimental challenges, as achieving a temperature of $T \sim 1 $K requires the proper acceleration of the observer to be $a \sim 10^{20}$m$/$s$^{2}$. In \cite{Svidzinsky2018}, the authors attempted to tackle this challenge, via perturbative methods, by examining a two state atom coupled to a quantized scalar field in its ground state, being linearly accelerated towards a mirror with the goal of stimulating photon emission at lower accelerations. We extended this analysis along a similar line of reasoning by investigating a more general scenario where a variety of accelerated trajectories were taken into account. 

\noindent Specifically, in section \ref{free}, by modelling a two-state atom undergoing simple harmonic motion in free space, we showed that it is possible to achieve frequencies within the bounds of experimental realization. However, this result is not without its limitations as we are employing perturbative methods slightly outside the realm of their applicability. To overcome this we could look to employ non-perturbative numerical simulations to determine the extent to which higher-order or non-linear factors contribute. Another aspect we could expand on is the simplicity of the model used; we could incorporate a broader directional dependence of photon emission, rather than restricting it to a single axis, which could have the effect of capturing an increased likelihood of photon production.

The focus of chapter \ref{Chpt_2}, was the application of the Atiyah-Singer index theorem \cite{Atiyah} to the quantum Hall effect within a spherical geometry, using the Dirac operator for non-interacting fermions in the presence of a background magnetic field. This magnetic field originates from a Wu-Yang magnetic monopole \cite{Wu+Yang} positioned at the center of the sphere. We derived wave functions for higher Landau levels as cross-sections of a non-trivial \( U(1) \) bundle, where the zero-point energy vanishes, ensuring no perturbations can lower the energy further. Importantly, the Atiyah-Singer index theorem constrains the degeneracy of the ground state. Building upon this, we investigated the fractional quantum Hall effect through the composite fermion model. Here, vortices in the statistical gauge field were introduced by promoting Dirac strings, associated with the monopole field, to physical vortices. A unique ground state was obtained only when these vortices carried an even number of flux units, effectively counteracting the background field and reducing the effective field experienced by the composite fermions. This approach yielded a unique gapped ground state and, in the limit of large particle numbers, reproduced fractional filling factors of the form \( \nu = \frac{1}{2k+1} \).

\noindent In addition to the spherical geometry considered here, future studies could explore how different manifolds affect the application of the index theorem and the resulting physics of the quantum Hall effect. The curvature and topology of alternative manifolds might lead to new insights into the structure of Landau levels and their degeneracies. Furthermore, extending these methods to examine particles with different spins---such as higher-spin fermions or bosonic fields---may reveal novel quantum phases and richer topological structures.

Finally, chapter \ref{chp3_ADS-CFT_for_Reissner_Nordstrom_BH} focused on the application of techniques derived from the AdS/CFT correspondence to condensed matter systems, an area of research sometimes referred to as the AdS/CMT correspondence. We reviewed the holographic prescription for spinors from \cite{Iqbal+Liu} and its application in \cite{McGreevyetal}, which investigates signatures of non-Fermi liquid behaviour in the spectral functions of a fermionic operator dual to a bulk fermionic field in an asymptotically AdS Reissner-Nordstr\"om (RN-AdS$_4$) background with a \( U(1) \) gauge field. While their work focused on a flat event horizon and zero black hole temperature (corresponding to zero temperature of the dual boundary system), we extended this analysis to consider spherical event horizons and non-zero temperatures, specifically at the phase transition temperature of the black hole.

\noindent At zero temperature, the introduction of a spherical event horizon appears to shift the location of the Fermi surface in the boundary theory, which we attribute primarily to the change in chemical potential induced by the transition from a flat to a spherical event horizon. We also examined the behavior of the boundary theory at non-zero temperatures, particularly at the black hole's phase transition temperature. Our findings suggest indications of a phase transition in the boundary theory, consistent with a \( (2+1) \)-dimensional \( U(1) \) fermionic theory, also known as QED$_3$. These results still require further investigation and the broader parameter space needs to be explored more to be sure of our interpretation.

\noindent Due to the nature of the AdS/CMT correspondence we are restricted in our ability to exactly determine the dual field theory we are describing. One direction for further study is to include a non-zero magnetic charge in the numerical calculations, which may reveal additional features of the dual fermionic theory. Additionally, investigating the scaling behaviour and scaling exponents of the Green’s functions, as is carried out in \cite{McGreevyetal}, could better characterize the nature of the dual boundary theory, potentially offering deeper insights into its critical phenomena and phase structure.

In summary, this thesis demonstrated that geometric and general relativistic techniques can provide powerful tools for simplifying the study of quantum systems and uncovering novel physical phenomena.

\setcounter{secnumdepth}{-1}
\chapter{Appendix: Vortices on a Sphere} 
\label{appendixA}

A vortex of strength $v$ at the north-pole of the sphere, $z=0$, with flux $2\pi v$ out of the sphere, is described by a magnetic potential
\begin{equation}
a=\frac{v}{2 i} \left[ \frac{d z}{z} 
- \frac{d \bar z}{\bar z}\right].\label{eq:vortex-N}
\end{equation}
Then $d a=0$ provided $z\ne 0$ but, if we isolate the N-pole by surrounding it by a small circle $S^1_\epsilon$ of radius $\epsilon$ 
centred on $z=0$, 
\begin{equation}
\int_{S^1_\epsilon} a =
 \frac{v}{2 i}\int_{S^1_\epsilon} \frac{d z}{ z} 
-\frac{v}{2 i}\int_{S^1_\epsilon} \frac{d \bar z}{\bar z}
=   \pi  v  - \pi (- v)= 2\pi v.\label{eq:vortex-S1}
\end{equation}
Thus (\ref{eq:vortex-N}) describes a point vortex of strength $v$ at the N-pole,
$f = da$ is a $\delta$-function at $z=0$,
which can be represented by
\[\partial_{\bar z}\left(\frac{1}{z}\right)=
  \partial_{z}\left(\frac{1}{\bar z}\right) = 2\pi \delta(z).\] 
However this potential also gives an anti-vortex, of strength $-v$, at the S pole:
the anti-podal point is given by $z \rightarrow \frac{1}{\bar z}$, which sends $a\rightarrow -a$. This is perhaps clearer using polar co-ordinates, $(\theta,\phi)$, 
in which 
\[ a = v d \phi\]
which represents an infinite straight flux tube in 3-dimensions, threading through both the N-pole and the S-pole of the sphere.  The total flux through the sphere 
arising from $f = da$ is zero.

\begin{figure}[h]
\centerline{\includegraphics{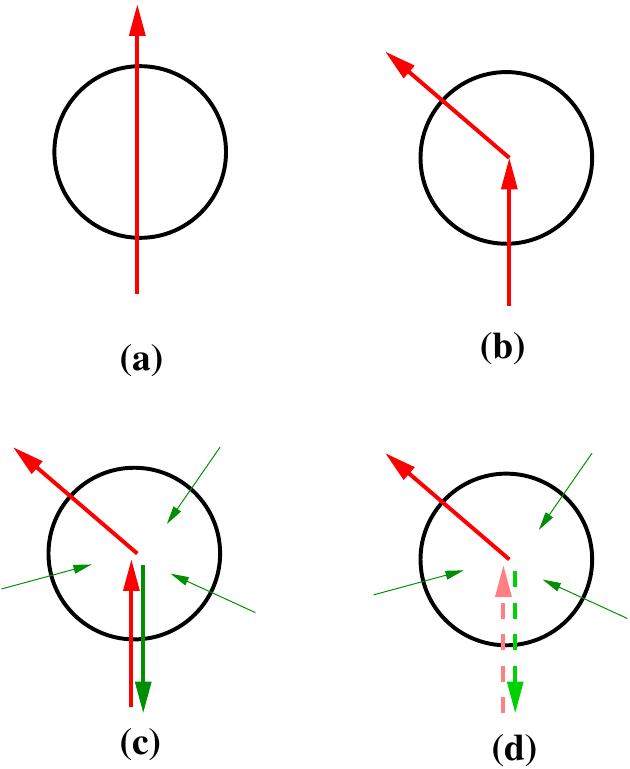}}
\caption{A vortex threading the sphere. (a) shows the simple vortex in (\ref{eq:vortex-N}) piercing the sphere at the north and south poles; 
(b) shows the vortex in (\ref{eq:vortex-A}), piercing the sphere at $z_1$ and the south pole; (c) shows the combination of the vortex in (b) combined with a Dirac monopole of charge $-1$ uniformly distributed on the sphere together with its accompanying string through the south pole; (d) the Dirac string and the vortex through the south pole cancel leaving a uniform monopole field with a vortex at $z_1$.  The total flux through the sphere in (d) is zero --- the Dirac string has been moved  from the south pole to the point $z_1$.}
\label{HallVortex1.1}
\end{figure}

The position of the vortex through the N-pole can be moved around by using
\begin{equation}
a=\frac{v}{2 i} \left[ \frac{d z}{(z-z_1)} 
- \frac{d \bar z}{(\bar z - \bar z_1)}\right],\label{eq:vortex-A}
\end{equation}
representing a vortex of strength $v$ through the point $z_1$, but there is still
an anti-vortex through the S-pole for any finite $z_1$.
However the vortex at the S-pole can be removed by adding a uniform magnetic field with a semi-infinite solenoid threading the S pole and terminating at the centre of the sphere,
\begin{equation}
a=\frac{v}{2 i} \left[ \frac{d z}{(z-z_1)} 
- \frac{d \bar z}{(\bar z - \bar z_1)}\right] 
+\frac{v}{2 i}\left(\frac{z d \bar z - \bar z d z}{1+z \bar z}\right),\label{eq:solenoid}
\end{equation}
giving the field strength
\[ f=d a = i\left(2 \pi  v \delta(z_1)  + \frac{v}{2}\frac{1}{(1+ z \bar z)^2}\right) d z \wedge d \bar z \] 
This is perfectly regular at the S pole and represents a magnetic monopole of charge $-v$ at the centre of the sphere together with a point vortex of strength $v$ at $z_1$,
the total flux is zero (see figure \ref{HallVortex1.1}). It is actually like a Dirac monopole with its accompanying
string threading the sphere at $z_1$, but a Dirac string is a gauge artifact, a vortex is not.

If there are $N-1$ vortices all of the same strength $v$ positioned at $z_j$ then the fields are simply added:
\begin{equation}a=\frac{v}{2 i} \sum_{j=1}^{N-1}\left( \frac{d z }{z-z_j}
-  \frac{d \bar z}{\bar z-\bar z_j}\right)
-\frac{i (N-1) v}{2}\left(\frac{z d \bar z - \bar z d z}{1+z \bar z}\right).\label{eq:a}
\end{equation}
The corresponding field strength is
\[ f = d a =i\left( 2 \pi v \sum_{j=1}^{N-1}\delta(z-z_j) -
\frac{(N-1)v}{(1 + z \bar z)^2} \right)  dz \wedge \bar dz\]
and
\[ \int_{S^2} f =0.\]
If in addition a background monopole field with charge $m'$ is present then the total gauge potential on the northern hemisphere is
\begin{equation} A^{(+)}  = \frac{v}{2 i} \sum_{j=1}^{N-1}\left( \frac{d z }{z-z_j}
-  \frac{d \bar z}{\bar z-\bar z_j}\right)
+ \frac{i m}{2}\left(\frac{z d \bar z - \bar z d z}{1+z \bar z}\right),\label{eq:A+}
\end{equation}
where $m=m' - (N-1)v$,
and the field strength is
\[ F = d A^{(+)} = i\left(2 \pi v \sum_{j=1}^{N-1} \delta(z-z_j) + 
\frac{m}{2 (1+z \bar z)^2}\right)dz \wedge d \bar z.\]
On the southern hemisphere we take the potential to be
\begin{align} A^{(-)} 
&= \frac{v}{2 i} \sum_{j=1}^{N-1}\left( \frac{d z }{z-z_j}
-  \frac{d \bar z}{\bar z-\bar z_j}\right)
- \frac{i (N-1)v (z d \bar z - \bar z d z)}{2(1+z \bar z)}
+\frac{i m'}{2} \frac{1}{(1+ z \bar z)}  \left( \frac{d z}{z} - \frac{d \bar z}{\bar z} \right)\nonumber\\
&= \frac{v}{2 i} \sum_{j=1}^{N-1}\left( \frac{d z }{z-z_j}
-  \frac{d \bar z}{\bar z-\bar z_j}\right)
+ \frac{i}{2} \left( \frac{m}{(1+ z \bar z)}  +  (N-1) v \right) 
  \left( \frac{d z}{z} - \frac{d \bar z}{\bar z} \right),\label{eq:A-}
\end{align}
which is perfectly well defined as $|z|\rightarrow \infty$. Again
\[ F = d A^{(-)} = i\left(2 \pi v \sum_{j=1}^{N-1} \delta(z-z_j) + 
\frac{m}{2 (1+z \bar z)^2} \right)dz \wedge d \bar z.\]
The total flux is
\[ \int_{S^2} F= 2\pi \bigl[m + (N-1)v \bigr],\]
see figure 2.

\begin{figure}[h]
\centerline{\includegraphics{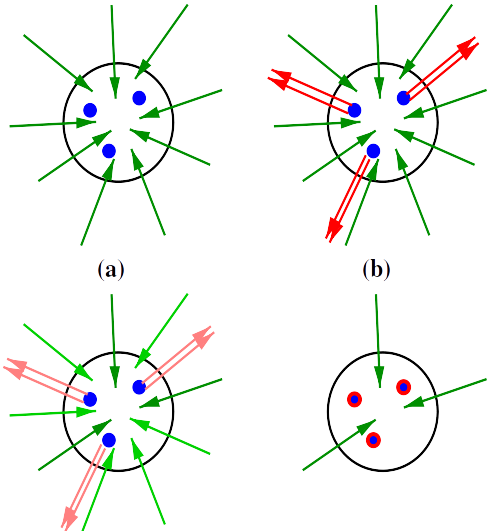}}
\caption{Composite fermions. (a) represents three electrons in a uniform background flux with total magnetic charge $m=-9$, giving filling factor $\frac{1}{3}$;
(b) six Dirac strings are promoted to be real vortices and attached to the electrons in pairs; (c) the total magnetic flux is now $m'=-3$; (d) the resulting configuration consists of three composite fermions in a field of strength $-3$ giving an effective filling factor 1.}
\end{figure}

What we have done here is taken $|m|$ monopoles each of charge $\pm 1$
(depending on the sign of $m$) and promoted the Dirac strings on 
$N-1$ of them to be real vortices at $z_i$, but leaving $|m'|$ of them as Wu-Yang monopoles, for which the Dirac string is a gauge artifact.
The configuration is indistinguishable from that of a monopole of charge $m$
together with $N-1$ vortices.


\backmatter
\addcontentsline{toc}{chapter}{Bibliography}
\bibliography{KitaevModelBib}
\bibliographystyle{unsrt}
\end{document}